\documentclass{article} %
\usepackage{iclr2027_conference,times}

\usepackage{amsmath,amsfonts,bm}

\def\eqref#1{equation~\ref{#1}}

\def\1{\bm{1}}

\DeclareMathAlphabet{\mathsfit}{\encodingdefault}{\sfdefault}{m}{sl}
\SetMathAlphabet{\mathsfit}{bold}{\encodingdefault}{\sfdefault}{bx}{n}

\usepackage{hyperref}
\usepackage{url}
\usepackage{amsmath,amssymb,booktabs,graphicx,microtype,multirow}
\usepackage{enumitem}
\usepackage{tikz}
\usetikzlibrary{arrows.meta,positioning,fit,shapes.geometric}

\definecolor{solo}{HTML}{4C78A8}
\definecolor{social}{HTML}{E69F00}
\definecolor{evolve}{HTML}{008F7A}
\definecolor{softgray}{HTML}{F2F3F5}

\graphicspath{{figures/}}
\title{From Solo to Social Learning: Characterizing Recursive Social Improvement in LLMs}

\author{Kunal Jha\textsuperscript{1}
\And
Max Kleiman-Weiner\textsuperscript{1,*}
\And
Natasha Jaques\textsuperscript{1,*}
}

\usepackage{comment}
\newtoggle{comment}
\toggletrue{comment}
\usepackage{etoolbox}

\iclrfinalcopy %
\begin{document}

\maketitle
\begingroup
\renewcommand{\thefootnote}{\arabic{footnote}}
\footnotetext[1]{Department of Computer Science,
University of Washington, Seattle, WA 98195, USA. Correspondence to
\texttt{kjha@uw.edu}. *Equal Advising}
\endgroup
\begin{abstract}
Large language models (LLMs) can now improve themselves by revising the instructions they follow, and LLM agents are increasingly orchestrated to work together on complex problems. However, self-improvement methods typically optimize one system at a time, and multi-agent frameworks often have every model work toward a shared goal. We ask a different question. When each agent pursues its own reward, can self-improving LLMs learn from one another well enough to improve the whole population? We call this capability \textit{recursive social improvement}. We study populations that revise skill files and choose whether, when, and whom to copy from. Independent search, learning from peers, and acting all share one token budget. In controlled environments, established social-learning algorithms benefit from peers, but three LLMs do not. They earn less reward per token than solo learners, and explore too narrowly or run out of tokens before acting. We then let the models write and revise their own skills. Observing peers changes how they improve, helping one model find useful skills sooner and another spend less on private search. Neither, however, outperforms independent learners at the same cost. Skills are copied, revised, and passed on, so one discovery can seed further search. Yet these exchanges concentrate the population around fewer independent discoveries. Together, these results show that LLMs can make learning more efficient by copying from peers, but not yet more effective.
\end{abstract}

\section{Introduction}
\vspace{-0.5em}

Large language models (LLMs) are now the core reasoning component of many automated systems, such as coding assistants. The performance of these systems depends not only on the model's weights but also on the instructions and procedures the model follows, such as \emph{skill files}: written procedures an agent reads when solving a task \citep{lin2026harnessupdatingharnessbenefit}. Because skill files are text supplied at inference time, they can be revised, copied, and shared without changing the model.

Existing methods for improving these instructions, including prompt optimization, reflection, and evolutionary search, typically focus on a single system \citep{fernando2023promptbreederselfreferentialselfimprovementprompt,khattab2023dspycompilingdeclarativelanguage,shinn2023reflexionlanguageagentsverbal,yuksekgonul2024textgradautomaticdifferentiationtext,agrawal2026gepareflectivepromptevolution}. The system tries candidate procedures and retains what works. Each attempt costs tokens and gives only a noisy estimate of a procedure's value, so the system faces a dilemma between exploration and exploitation: spend tokens searching for a better procedure, or use the best one found so far.

When many self-improving agents work side by side, each can also learn from what the others have found. This is not hypothetical: agents can already download skills and scaffolding that other agents use from GitHub \citep{gavrielc2026nanocoai}, and a large body of work on multi-agent systems has LLMs coordinate and build on one another's work \citep{du2023improvingfactualityreasoninglanguage,li2023camelcommunicativeagentsmind,hong2024metagptmetaprogrammingmultiagent,zhuge2024languageagentsoptimizablegraphs,feng2025heterogeneousswarmsjointlyoptimizing,park2026scalingdiscoverytesttimecommunication}, most prominently \citet{openaiNavierStokesMillennium}, who coordinated roughly 10,000 agents on their solution to the Navier--Stokes problem. This prior work on swarms asks how to make multiple LLMs work together on a single task. We instead ask what happens when many self-interested LLM agents optimize their own separate tasks but can see what their peers produce.

\begin{figure}[!tbp]
\centering
\includegraphics[width=0.9\linewidth]{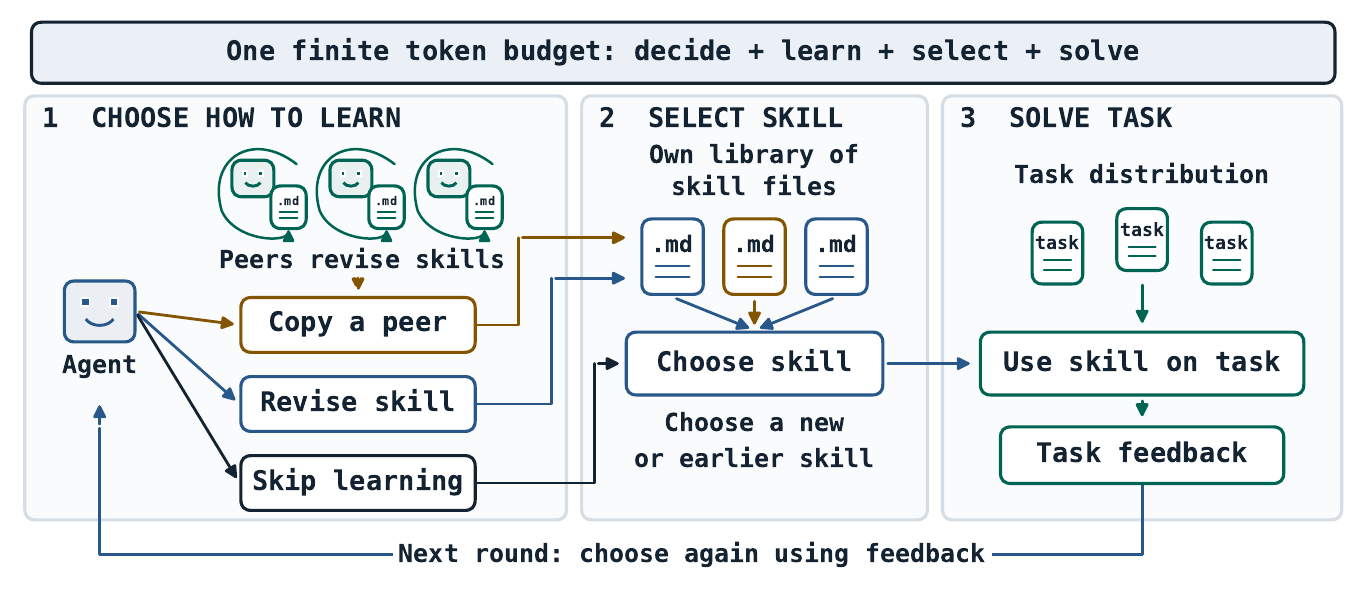}
\caption{Recursive social improvement. An agent can copy a peer's skill file, revise one of its own, or skip learning. Copying and revision add alternatives to its library of reusable task-solving instructions. It then chooses any saved skill, including an earlier version, and uses it to solve a task. Feedback informs the next round. The entire process shares one finite token budget. Peers follow the same loop independently, each pursuing its own task reward.}
\label{fig:overview}
\end{figure}

This setting poses a tension long studied in behavioral ecology, cultural evolution, and economics \citep{grossmanImpossibility,barnard1981producers,Boyd1988-ki,rogers1988social,laland2004social}: learning from others is valuable only if someone searches independently, yet each individual would prefer that someone else pay the cost. As more agents reuse, fewer explore, depleting the discoveries that make social learning useful. Learning from the wrong peer can also be worse than learning alone \citep{lazer2007network,Barkoczi_2016,Derex2016-vv}. We call the prospect of resolving this tension without central coordination \emph{recursive social improvement}: agents improve their own procedures, learn from others' improvements, and revise what they acquire, so that one agent's discovery becomes the starting point for another's search (Figure~\ref{fig:overview}).

We study populations of agents that each solve their own tasks, are rewarded only on those tasks, and can see the skills their peers use. Each round, an agent chooses whether to search privately, observe a peer, or act, all from the same token budget. We count every token spent by every agent, including the search behind each copied skill, and ask whether a social population outperforms independent learners that spend the same amount. We borrow the design of classic social-learning tournaments \citep{laland2004social,Rendell2010-ro,Song2016-wf,Kendal2018-is}, but our agents allocate their own computation and share skills that can be revised. We begin in a controlled environment where the value of each option is known, then move to populations that write, revise, and exchange skill files for instruction-following tasks. Our contributions are:

\begin{enumerate}[leftmargin=*,topsep=0pt, itemsep=0pt]
\item \textbf{A formulation of recursive social improvement.} We define a reinforcement learning problem in which each self-interested agent decides, under one token budget, whether to revise its own procedures, copy a peer's, or act. We distinguish efficient social learning (more performance per token) from effective social learning (outperforming independent learners at equal cost).
\item \textbf{Useful social information is not enough.} A hand-designed learning policy gains a social advantage in our controlled environment, but every LLM tested is 30--52\% less efficient when observing peers than solo populations of the same model. Copied options usually beat what the recipient found alone, yet social populations explore less, and neither reserving tokens for acting nor requiring private search closes this gap.
\item \textbf{Observing peers changes learning speed and cost, but not performance at matched cost.} With peer access, GLM-5.3-Flash averages 1.3 percentage points higher held-out accuracy across learning than solo populations of the same model, while GPT-OSS-120B uses a third fewer learning tokens. This makes GPT-OSS more efficient, but at matched learning tokens neither model shows a clear advantage over independent learners. Skills are copied and revised again, yet by the end most agents' skills trace back to only one or two of the population's first revisions.
\end{enumerate}

Overall, LLM agents build on their peers' skills, but when LLMs control social learning, populations do not outperform independent learners that spend the same number of tokens. As more self-interested agents improve side by side, using peers' discoveries well becomes a capability in its own right. Future work should study how to train or incentivize agents to balance independent exploration with observation, so populations improve faster and further than they would through independent search.

\vspace{-0.5em}
\section{Related Work}
\vspace{-0.5em}

\textbf{Self-improving LLM agents.} Prompt optimization, reflection, and evolutionary search find better instructions without changing model weights, and related work evolves agent designs, improvement strategies, and reusable skill libraries \citep{wang2023voyageropenendedembodiedagent,fernando2023promptbreederselfreferentialselfimprovementprompt,khattab2023dspycompilingdeclarativelanguage,shinn2023reflexionlanguageagentsverbal,yuksekgonul2024textgradautomaticdifferentiationtext,guo2025evopromptconnectingllmsevolutionary,hu2025automateddesignagenticsystems,agrawal2026gepareflectivepromptevolution,lin2026harnessupdatingharnessbenefit,zhang2026darwingodelmachineopenended,zhang2026hyperagents}. These methods improve one system at a time. We ask what happens when many such systems improve side by side and can see one another's work.

\textbf{Multi-agent collaboration and swarms.} A growing body of work builds multi-LLM systems that pursue a shared objective through assigned roles, debate, output aggregation, message passing, shared blackboards, or joint optimization of the system's weights and topology \citep{du2023improvingfactualityreasoninglanguage,li2023camelcommunicativeagentsmind,hong2024metagptmetaprogrammingmultiagent,li2024agentsneed,qian2024chatdevcommunicativeagentssoftware,wang2024mixtureofagentsenhanceslargelanguage,zhuge2024languageagentsoptimizablegraphs,feng2025heterogeneousswarmsjointlyoptimizing,feng2025modelswarmscollaborativesearch,liu2026messagepassingenablesefficient,park2026scalingdiscoverytesttimecommunication, qi2026economymindsemergingmultiagent,salemi2026llmbasedmultiagentblackboardinformation}. Studies of these systems ask when coordination beats a strong single agent, why it fails, and how malicious contributors affect it \citep{cemri2025multiagentllmsystemsfail,kim2026sciencescalingagentsystems,yang2026usmeasuringmitigatingmalicious}; others transfer knowledge across agents through generated instructions or shared repositories of trajectories \citep{mohtashami2024sociallearningcollaborativelearning,kim2026multiagenttransactivememory}. In all of these, sharing is part of the system design and agents work toward a common goal. In our setting, no agent is told to collaborate, share, or diversify: each maximizes its own reward, reuse is a choice it makes and pays for, and any population-level benefit must emerge from those individual choices.

\textbf{Social learning, cultural evolution, and games.} The tension between producing and reusing information has been studied in producer--scrounger games, cultural evolution, and studies of how copying depends on network structure, task structure, and source choice \citep{barnard1981producers,Boyd1988-ki,rogers1988social,laland2004social,lazer2007network,Barkoczi_2016,Derex2016-vv,Kendal2018-is,toyokawa2019social}. Social-learning tournaments and bandit algorithms formalize discovery, observation, and exploitation as a repeated choice among options \citep{berry1997bandits,Auer2002-ha,kleinberg2008multiarmedbanditsmetricspaces,Rendell2010-ro,bubeck2011xarmedbandits,Yoshida_2016,kolla2016collaborativelearningstochasticbandits,Nakayama2017-wp,landgren2019distributedcooperativedecisionmakingmultiarmed,chawla2024gossipinginserteliminatealgorithmmultiagent}, while reinforcement learning agents and LLM populations can be trained or observed to use social cues, accumulate knowledge across generations, and develop collective biases \citep{ndousse2021emergentsociallearningmultiagent,cook2024artificialgenerationalintelligencecultural,perez2024culturalevolutionpopulationslarge,piatti2024cooperatecollapseemergencesustainable,anthropicPatternsProblems,el2026physicsagentsstatisticalmechanics,Flint2026-lv,jha2026tapesstrongcoevolutioncomputation}. We bring this tradition to self-improving LLM agents which share revisable procedures and must divide one token budget between learning from others and acting.

\textbf{Costly information and the allocation of computation.} Economic models study how social information changes the incentive to acquire private information, including the result that freely observable information undermines that incentive, and how agents decide whether and whom to observe; human foraging studies likewise show that attending to others competes with independent search \citep{grossmanImpossibility,Burguet2000-ez,Kultti2007-wy,Mueller-Frank2016-ah,Song2016-wf,Wu2021-js}. We ask how LLM agents make these decisions when choosing a peer and interpreting its skill is paid from the same token budget used to act, and how those choices affect each agent's performance and the population's.

\vspace{-0.5em}
\section{Recursive Social Improvement as a reinforcement learning problem}
\label{sec:formulation}
\vspace{-0.5em}

We consider a population of $N$ agents, indexed by $i$, that learn over rounds $t=1,\dots,T$. Each agent solves its own tasks $x\sim\mathcal D_i$ and is rewarded only on those tasks. To solve a task, agent $i$ executes a \emph{skill} $\sigma$ from its library $\Sigma_{i,t}$ and receives reward $r_i(\sigma,x)$, with expected value $\mu_i(\sigma)=\mathbb E_{x\sim\mathcal D_i}[r_i(\sigma,x)]$. A skill may be as simple as an arm of a bandit or as rich as a written skill file. Each agent follows its own policy $\pi_i$, which conditions on the agent's history of actions, rewards, and observations, and we write $\boldsymbol\pi=(\pi_1,\dots,\pi_N)$ for the population. In our experiments, a policy is either a fixed-weight LLM that learns in-context \citep{moeini2025surveyincontextreinforcementlearning} or an algorithmic baseline.

\textbf{A two-stage action space.} Each round, $\pi_i$ makes two choices in order. It first selects a learning action that changes the agent's library. The agent can \emph{evolve} its skills by revising one of them using reward feedback from its own tasks and adding the result to its library, \emph{observe} a peer $j$ of its choosing, which adds the skill $\sigma_{j,t-1}$ that $j$ executed in the previous round to its library and reveals the reward $r_{j,t-1}$ it received, or skip learning. The policy then selects a skill $\sigma_{i,t}\in\Sigma_{i,t}$ and executes it to receive reward $r_{i,t}$, with $r_{i,t}=0$ for a missing or invalid action. Solo agents cannot observe. This structure follows the social-learning tournament of \citet{Rendell2010-ro}, except that observation costs tokens rather than a round's payoff. This makes the cost of observing endogenous, as it is for deployed LLM agents: it depends on how much the agent reasons about whether and whom to copy, rather than being a fixed penalty set by the environment. Because peers keep learning, the value of observing a given peer changes over time, so each agent faces a non-stationary reinforcement learning problem.

\textbf{Learning competes with acting.} Each round, agent $i$ receives the same allowance $B$ and spends
\begin{equation}
c_{i,t}=c^{\mathrm{plan}}_{i,t}+c^{\mathrm{learn}}_{i,t}+c^{\mathrm{select}}_{i,t}+c^{\mathrm{exec}}_{i,t}\leq B
\label{eq:token-decomposition}
\end{equation}
on choosing its learning action, carrying it out, choosing a skill, and executing it, so a long decision can leave too little budget to act. The population's cumulative cost is $C_t(\boldsymbol\pi)=\sum_{i}\sum_{s\leq t}c_{i,s}$. It counts every token spent by every agent, so the agent that discovers a skill pays for that search once, and peers who copy it pay only to acquire and use it.

\textbf{Measuring improvement at matched cost.} We measure the population's performance at round $t$ as the expected reward of the skills its agents execute, $J_t(\boldsymbol\pi)=\frac{1}{N}\sum_i\mu_i(\sigma_{i,t})$, and its performance at cost $C$ as $J(\boldsymbol\pi;C)=\mathbb E[J_{t_C}(\boldsymbol\pi)]$, where $t_C$ is the last round before the population's cumulative cost exceeds $C$. Let $\boldsymbol\pi^{\mathrm{soc}}$ be a population that can observe peers and $\boldsymbol\pi^{\mathrm{solo}}$ the same population with observation disabled. The social advantage at matched cost is
\begin{equation}
\Delta J(C)=J(\boldsymbol\pi^{\mathrm{soc}};C)-J(\boldsymbol\pi^{\mathrm{solo}};C).
\label{eq:matched-compute}
\end{equation}

\textbf{What counts as recursive social improvement.} We identify \emph{recursive transmission} when a skill revised by one agent is copied by another and revised again. We say a population exhibits \emph{recursive social improvement} when it shows recursive transmission and $\Delta J(C)>0$ over the costs both populations reach. $\Delta J(C)$ compares the social population with $N$ independent learners at the same total cost, so both populations search in parallel and copied skills are paid for by whoever discovered them. A positive $\Delta J(C)$ therefore cannot come from dividing work among more agents. It requires exchanges between agents to produce more than the same agents would alone. Without this requirement, recursive social improvement would describe parallel search that shares its results cheaply. We also report whether this occurs by measuring performance per token over a run,
\begin{equation}
E(\boldsymbol\pi)=\frac{\sum_{t\leq T}J_t(\boldsymbol\pi)}{C_T(\boldsymbol\pi)/N}.
\label{eq:efficiency}
\end{equation}
However, because $C_T(\boldsymbol\pi^{\mathrm{soc}})$ can differ from $C_T(\boldsymbol\pi^{\mathrm{solo}})$, a social population can raise this ratio simply by stopping search once returns diminish. This is more efficient, but it does not tell us whether the social population learns more effectively than independent learners (Appendix~\ref{app:metric-motivation}). Thus, we focus on $\Delta J(C)$ as the measure of whether a group of agents performs as more than the sum of its parts.

\vspace{-0.5em}
\section{Understanding LLM social learning in a controlled environment}
\label{sec:framework}
\vspace{-0.5em}

Before asking whether agents can build on one another's improvements, we first ask a simpler question: do agents benefit from one another's discoveries when those discoveries cannot be revised?
\begin{enumerate}[label=RQ\arabic*:,leftmargin=*,itemsep=2pt]
\item Does the ability to observe peers improve reward at matched token cost, cumulative reward, or reward per token when useful discoveries are available?
\item If observing peers does not help, why not?
\end{enumerate}

\textbf{Environment.} Our environment adapts the social-learning tournament of \citet{Rendell2010-ro}, a standard testbed for comparing individual and social learning strategies. Each round of that tournament, an agent chooses to \textsc{Innovate} (learn a new behavior's payoff), \textsc{Observe} (see a behavior a peer is performing), or \textsc{Exploit} (earn a known behavior's payoff), and the strongest strategies relied heavily on copying. We make two changes. First, as in a standard bandit, every pull both earns a payoff and reveals information about the pulled arm, so there is no separate \textsc{Innovate} move. Second, observation costs tokens rather than a round's payoff: each round, an agent may observe a peer's previous arm and payoff and then pulls an arm, with both drawing on the same budget of $B=4{,}096$ completion tokens. Unused tokens expire, and a missing or invalid pull earns no reward. The environment is a 25-armed bandit with fixed mean payoffs and noisy rewards, played by ten agents for 100 rounds. We use eight environment seeds for each of Qwen3-14B \citep{qwen3technicalreport}, Ministral-3-14B-Reasoning-2512 \citep{liu2026ministral3}, and GPT-OSS-20B \citep{openai2025gptoss120bgptoss20bmodel}, and report SE across population seeds (Appendices~\ref{app:r1-methods} and~\ref{app:reporting}).

\textbf{Algorithmic baselines.} To check that the environment rewards social learning at all, we compare three policies that use no LLM. The \emph{oracle} always pulls the arm with the highest mean payoff. \emph{Solo UCB} uses the upper confidence bound rule \citep{Auer2002-ha}: it adds to each arm's estimated payoff a bonus that is larger for arms it has tried less often, then pulls the arm with the highest total. \emph{Hierarchical UCB} applies the same rule at two levels. It first treats each source, either itself or one of its peers, as an arm and chooses one. Choosing a peer adds that peer's last arm and payoff to the agent's estimates, while choosing itself gathers no social information. A small prior biases these agents towards individual exploration initially. It then applies solo UCB to its updated estimates to choose an arm. Because choosing whom to learn from is separate from choosing what to execute, we reuse this source rule in Section~\ref{sec:r4-methods}, where choosing oneself means revising privately (Appendix~\ref{app:source-ucb}).

\begin{figure}[!tbp]
\centering
\includegraphics[width=\linewidth]{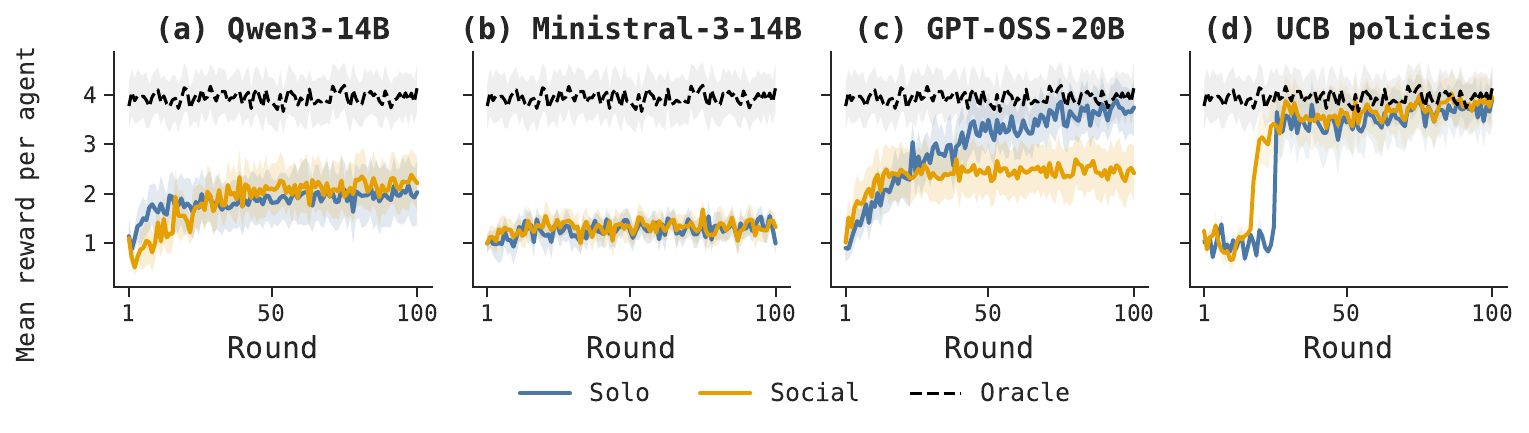}
\caption{Mean reward per agent at each round. Panels (a)--(c) show Qwen3-14B, Ministral-3-14B, and GPT-OSS-20B; panel (d) shows solo UCB and hierarchical social UCB, which use no LLM inference. All panels use the same eight environment seeds. Blue denotes solo and orange social; the black dashed line shows the oracle. Shading shows seed SE, and missing pulls earn zero.}
\label{fig:r1-reward}
\end{figure}

\textbf{An algorithmic social learner benefits from observing peers (RQ1).} Hierarchical UCB earns $25.68\pm2.87$ more cumulative reward than solo UCB over 100 rounds ($p<0.001$), about $0.26$ per round, on the same environments used for the LLM experiments (Figure~\ref{fig:r1-reward}d). The gain comes mainly from finding good arms sooner: solo UCB nearly reaches the oracle by the end of learning, but takes longer to get there (Appendix Figure~\ref{fig:r1-regret}d). The gain is nevertheless consistent: it holds across 100 simulated environments and when payoffs change every 25 rounds, and weakens only when they change every 10 rounds (Appendix Figures~\ref{fig:r1-precision} and~\ref{fig:r1-ucb-volatility}). The environment therefore contains social information that a well-designed policy can use, so if LLM agents fail to benefit from their peers, the failure lies in how they use that information.

\textbf{LLM social learners show no clear advantage over solo learners at matched cost and use tokens less efficiently (RQ1).} At matched completion-token expenditure, the social-minus-solo difference $\Delta J(C)$ is $+0.14\pm0.33$ for Qwen, $+0.37\pm0.34$ for Ministral, and $-1.24\pm0.45$ for GPT-OSS (Appendix Figure~\ref{fig:r1-matched-tokens}a--c). None is significant after correcting for these three comparisons, and including prompt tokens gives the same directions. Cumulative reward over 100 rounds shows the same pattern, changing by $+1.5\%$ for Qwen, $+4.0\%$ for Ministral, and $-20.7\%$ for GPT-OSS relative to the corresponding solo mean (Figure~\ref{fig:r1-reward}a--c). All three models also earn less reward per token when they can observe peers, with drops of $29.6\pm13.4\%$ for Qwen, $41.8\pm11.6\%$ for Ministral, and $51.9\pm7.8\%$ for GPT-OSS (Appendix Figure~\ref{fig:r1-token-efficiency}). The drop is significant for GPT-OSS after correcting for the ten comparisons in Appendix Table~\ref{tab:r1-key} ($p<0.01$): its social agents earn about half as much reward per token as its solo agents. For Qwen and Ministral, mean reward changes little, so their lower estimated efficiency comes mainly from spending more tokens, although neither drop is significant after correction. Qwen's small social gap should also not be read as strong performance, since its social agents remain below GPT-OSS's solo agents (Figure~\ref{fig:r1-reward}a,c).

\begin{figure}[!tbp]
\centering
\begin{minipage}[t]{0.40\linewidth}
\centering
\includegraphics[width=\linewidth]{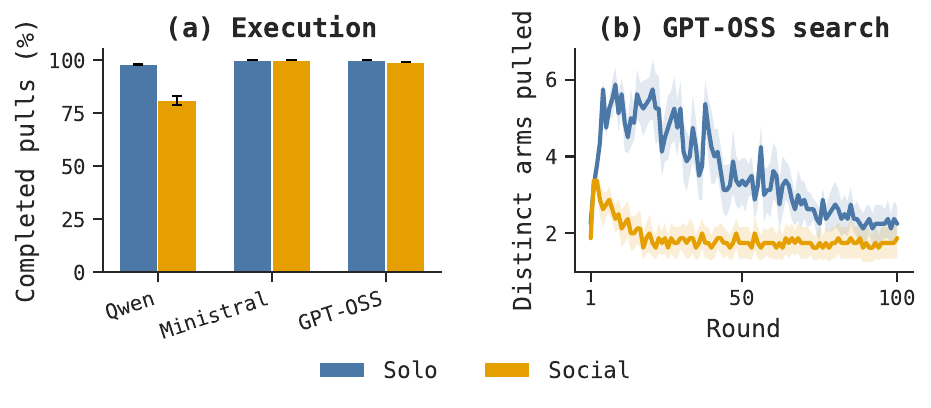}
\caption{Diagnosing social learning failures. (a) Fraction of rounds with a completed pull. (b) Distinct arms pulled by GPT-OSS populations over time. Blue denotes solo and orange social; bars and shading show SE over eight seeds.}
\label{fig:r1-diagnosis}
\end{minipage}\hfill
\begin{minipage}[t]{0.57\linewidth}
\centering
\includegraphics[width=\linewidth]{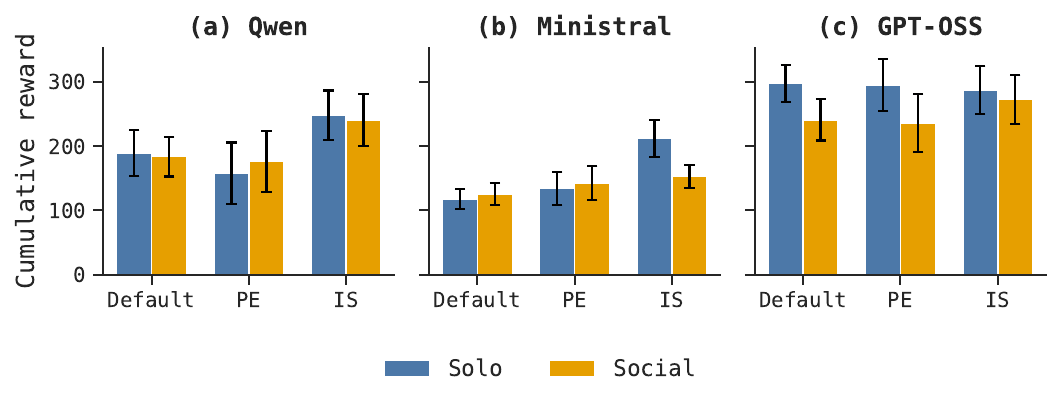}
\caption{Cumulative reward per agent through round 100 under default, protected-execution (PE), and independent-search (IS) conditions, for equally treated solo and social populations. Default pools 16 runs over the same eight environment seeds; PE and IS use eight runs each. Bars show means and SE.}
\label{fig:r1-interventions}
\end{minipage}
\end{figure}

\textbf{Copying helps the individual but narrows the population's search (RQ2).} The failure is not that agents copy poor arms. For GPT-OSS, a copied arm has a higher mean payoff than the best arm the recipient had found on its own in $93.2\%$ of comparable pulls (Appendix Figure~\ref{fig:r1-copy}a). Instead, the trajectories of social populations point to two problems. The first is under-exploration. Once agents can see good arms that peers are pulling, they collapse onto those arms and stop trying the rest of the bandit. GPT-OSS social populations quickly use far fewer distinct arms than solo populations (Figure~\ref{fig:r1-diagnosis}b), and Qwen social populations also use fewer for much of the run (Appendix Figure~\ref{fig:r1-time}d--e). In a bandit with effectively infinite arms, the ability to observe others reduces the number of independently discovered arms for all three models (Appendix Figure~\ref{fig:r2-main}b). This is the tension identified by \citet{rogers1988social} and later social-learning tournaments \citep{Rendell2010-ro}: copying helps the individual who copies, but if everyone copies, no one discovers the better arms that would make copying worthwhile. The second problem is token management. Agents that reason at length about whether and whom to observe can run out of budget before pulling an arm, earning nothing that round. Qwen's social agents complete a pull in $81.0\%$ of rounds, compared with $98.1\%$ for solo agents (Figure~\ref{fig:r1-diagnosis}a; Appendix Figure~\ref{fig:r1-completion}g--i).

\textbf{Fixing exploration and execution does not make social learners outperform solo learners (RQ2).} If these two problems explain the entire social-solo gap, fixing them should let social learners catch up. We test each with a targeted intervention that adds no tokens and is applied equally to solo and social populations. \emph{Protected execution} (PE) reserves 2,048 of each agent's 4,096 tokens for the pull. \emph{Independent search} (IS) blocks observation on a randomly assigned 15\% of agent-rounds and requires the agent to pull an arm it has neither pulled nor observed, whenever such an arm remains. IS therefore forces independent exploration rather than merely prompting the model to explore.

Both interventions fix the problem they target. PE raises the rate at which Qwen's social agents complete a pull by $19.53\pm2.24$ percentage points ($p<0.001$). IS increases the number of distinct arms GPT-OSS's social agents select by $1.27\pm0.23$ ($p<0.01$), with a similar but nonsignificant increase for Qwen (Appendix Figure~\ref{fig:r1-mechanisms}a,c). However, neither intervention makes social learners reliably better than equally treated solo learners (Figure~\ref{fig:r1-interventions}). Under PE, Qwen and Ministral show small social gains, but GPT-OSS social learning still underperforms solo learning. Under IS, Qwen's reward improves but its social agents remain slightly below its solo agents, and Ministral improves in both conditions but more for solo agents (Appendix~\ref{app:ministral-is}).

Effective social learning therefore requires more than access to useful information, independent exploration, or enough tokens to act. It also requires weighing options gathered from peers against those found privately to balance exploration and exploitation, and we find no reliable social advantage in cumulative reward, reward per token, or the new matched-cost comparison. Appendices~\ref{app:r2} and~\ref{app:r3} find similar patterns in a bandit with effectively infinite arms and on reasoning tasks with a fixed set of skill files. We next ask what changes when agents can revise and exchange the skills themselves.

\vspace{-0.5em}
\section{Can Self-Improving LLM Agents Benefit from One Another?}
\label{sec:r4-methods}
\vspace{-0.5em}

In Section~\ref{sec:framework}, agents can share what they find but cannot create better skills. We now let agents write, revise, copy, and reuse skill files, so that each agent's library $\Sigma_{i,t}$ from Section~\ref{sec:formulation} can grow over time.
\begin{enumerate}[label=RQ\arabic*:,leftmargin=*,itemsep=2pt,start=3]
\item When agents can evolve their own skill files, does observing their peers' skills help them improve sooner, spend less computation, or reach higher accuracy at the same cost?
\item Can an algorithmic rule for choosing whom to observe improve accuracy or efficiency, and does adapting that choice beat choosing peers at random?
\end{enumerate}

\textbf{Setup.} Asking whether agents build on one another's improvements requires a setting where private improvement is possible. We use IFBench \citep{pyatkin2025generalizingverifiableinstructionfollowing}, an instruction-following benchmark in which each answer must satisfy several automatically verified constraints. We use two models, GPT-OSS-120B \citep{openai2025gptoss120bgptoss20bmodel} and GLM-5.3-Flash \citep{glm5team2026glm5vibecodingagentic}, for which we confirmed single-agent skill evolution improves held-out accuracy (Appendix Figure~\ref{fig:r4-offline}a--b). They are larger than the bandit models, since writing a useful revision takes more capability than choosing an arm. They are also inexpensive enough to run many seeded populations and fall in the intermediate capability range where evolved instructions help most \citep{lin2026harnessupdatingharnessbenefit}. We use OpenEvolve \citep{openevolve} to propose revisions to a skill file and score them on the training set. For each model, we run six independent populations of five agents in every condition. After an initialization round, agents learn for 49 rounds on the same 100 training examples, and we evaluate their skills on 200 held-out examples. Only skill files change; model weights and decision prompts stay fixed. Training reward is the fraction of constraints an answer satisfies. We measure performance $J_t$ as the stricter held-out accuracy, the fraction of answers that satisfy every constraint, and estimate $\Delta J(C)$ by interpolating between saved checkpoints (Appendix~\ref{app:matched-spending}).

\begin{figure}[!t]
\centering
\includegraphics[width=.9\linewidth]{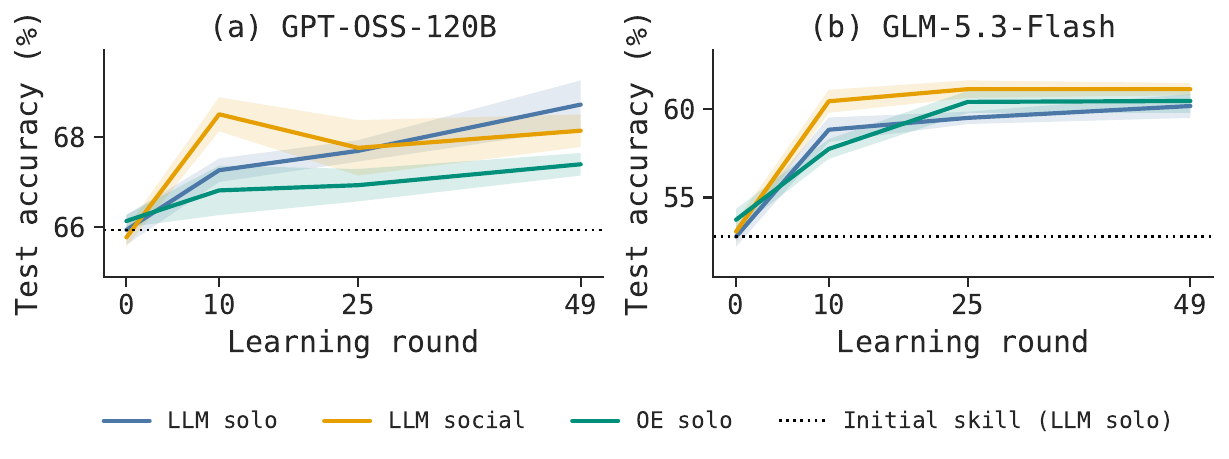}
\caption{Held-out whole-answer accuracy during skill evolution. Panels (a) and (b) compare LLM-controlled solo and social learning with full solo OpenEvolve for each model. The dotted reference is the LLM-solo initial skill. Lines show means and shaded bands show SE across six population seeds.}
\label{fig:r4-endogenous}
\end{figure}

\begin{table*}[!b]
\centering
\small
\setlength{\tabcolsep}{4pt}
\begin{tabular}{lrrrr}
\toprule
& \multicolumn{3}{c}{LLM social $-$ LLM solo accuracy (pp)} & \\
\cmidrule(lr){2-4}
Model & Trajectory & Final & $\Delta J(C)$ & Observe rate (social) \\
\midrule
GPT-OSS-120B & $+0.19\pm0.59$ & $-0.58\pm0.88$ & $+0.14\pm0.35$ & $35.92\pm1.40\%$ \\
GLM-5.3-Flash & $+1.31\pm0.61$ & $+0.95\pm1.15$ & $+0.98\pm0.85$ & $8.37\pm0.90\%$ \\
\bottomrule
\end{tabular}
\caption{Social minus solo held-out whole-answer accuracy, in percentage points. \emph{Trajectory} averages rounds 0, 10, 25, and 49; \emph{Final} is round 49; $\Delta J(C)$ compares equal population-wide learning tokens (prompts and completions, excluding hidden evaluation). \emph{Observe rate} is the fraction of learning rounds social agents spend observing. Trajectory and final SE account for both population seeds and held-out examples (Appendix~\ref{app:test-sampling}); other SE are across six seeds.}
\label{tab:r4-performance}
\end{table*}

Each round follows the setup described in Section~\ref{sec:formulation}. An agent first chooses to \textsc{Revise} a skill with OpenEvolve, \textsc{Evaluate} a skill on the training set to better estimate its score, or \textsc{Keep} its library unchanged. A social agent can also \textsc{Observe} a peer, which adds the peer's most recently executed skill and its training score to the library. The agent then selects any skill in its library and executes it on a training task. Copied skills can be revised and passed on again. Agents see a short summary of every skill they hold, but only their latest observation in full. The per-round budget is large enough to evaluate a skill on all 100 training examples and still choose, revise, and execute (Appendix~\ref{app:budget-audit}). Agents never see peers' held-out accuracy, share answers, or receive a joint reward. All agents train on the same examples, so an agent that copies a peer's skill gains the benefit of that peer's search but sees no new training data.

\textbf{Baselines.} All conditions use OpenEvolve to write and score revisions; they differ in who decides what to do with them. In \emph{LLM solo} and \emph{LLM social}, the agent's LLM decides when to revise, which skill to revise, and which skill to execute, and the two conditions differ only in whether the agent can observe peers. In \emph{full solo OpenEvolve}, OpenEvolve's archive makes these decisions instead, keeping alternative skills in separate subpopulations and executing the highest-scoring one. This shows how well an agent improves alone without an LLM making decisions. For RQ4, a \emph{UCB} controller uses the source rule from Section~\ref{sec:framework} to choose each round between revising its own skills and copying a particular peer, and a \emph{uniform} controller makes that choice at random. In both, OpenEvolve's archive then selects the skill to execute. We compare LLM social with LLM solo, and the two controllers with full solo OpenEvolve.

\textbf{Neither LLM shows a clear accuracy advantage at matched learning computation (RQ3).} At equal population-wide learning tokens, social populations score $+0.14\pm0.35$ percentage points above solo for GPT-OSS and $+0.98\pm0.85$ for GLM; neither difference is statistically significant (Table~\ref{tab:r4-performance}; Appendix Figure~\ref{fig:r4-matched-spending}). Thus, neither model meets our criterion for recursive social improvement. However, observing peers changes how quickly each model improves and how much it spends.

\textbf{GLM improves sooner with social access, while GPT-OSS spends less (RQ3).} Averaged over learning, GLM's social populations have $1.31\pm0.61$ points higher held-out accuracy than its solo populations, but their final accuracy is close to solo (Figure~\ref{fig:r4-endogenous}b; Table~\ref{tab:r4-performance}). Observing peers therefore helps GLM reach useful skills sooner, but solo learning catches up by the end, and GLM shows no clear reduction in tokens or cost. GPT-OSS follows a different pattern. Observing peers does not improve its accuracy during learning or at the end. Instead, its social agents observe in place of many private revisions and use $17.34\pm2.08$ million fewer learning tokens than solo populations, and they cost significantly less in dollars even when held-out evaluation is included (Appendix Table~\ref{tab:r4-resources}). Part of this saving comes from the design itself, since observing a skill avoids the 100-example evaluation that every new revision requires. Lower cost alone therefore does not show that agents are learning from their peers more effectively.

\begin{figure}[!t]
\centering
\includegraphics[width=.9\linewidth]{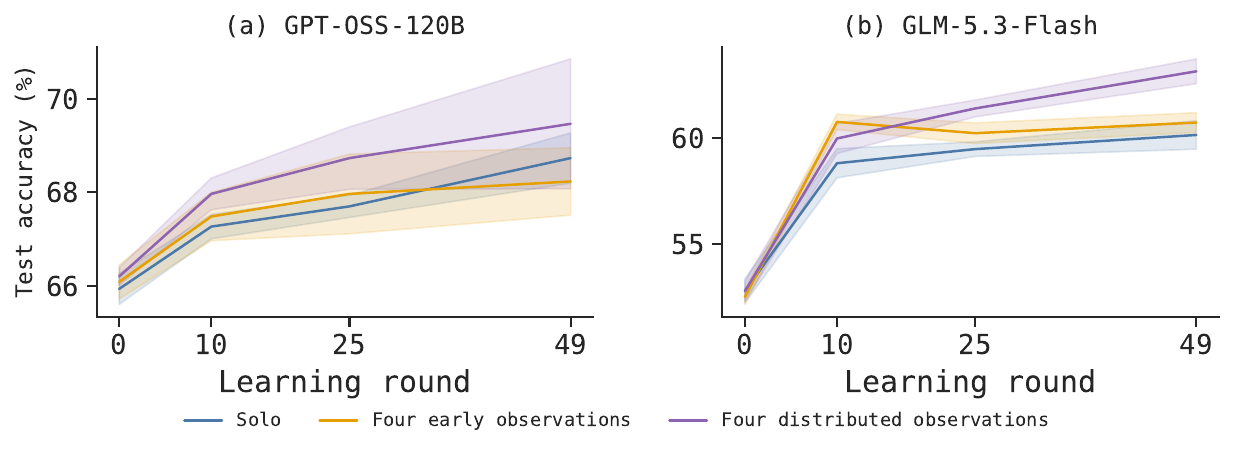}
\caption{The same number of observations has different effects depending on timing. Panels (a)--(b) compare four assigned early observations with four distributed observations and solo learning. Models choose whom to observe and which skill to execute. Shading is SE across six population seeds. The optional early-access control and paired statistics appear in Appendix~\ref{app:timing-plan}.}
\label{fig:r4-assigned-timing}
\end{figure}

\begin{figure}[!b]
\centering
\includegraphics[width=\linewidth]{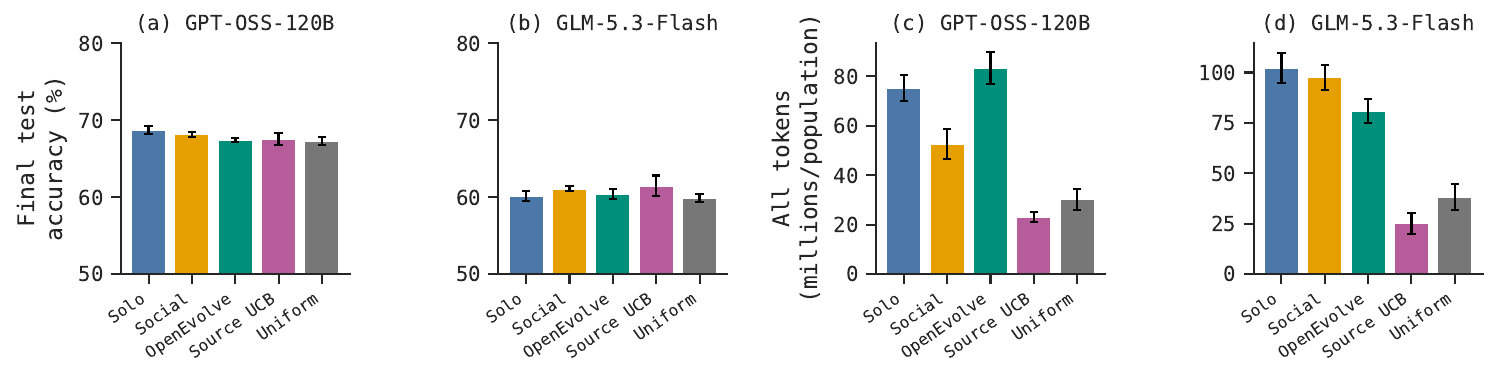}
\caption{Final held-out accuracy (a--b) and end-to-end token use, including held-out evaluation (c--d), per five-agent population. OpenEvolve denotes full solo OpenEvolve. Bars: six-seed means $\pm$ SE.}
\label{fig:r4-controller}
\end{figure}

\textbf{GLM observes rarely and early, while GPT-OSS observes often and throughout (RQ3).} GLM's social agents observe in $8.37\%$ of rounds, mostly early in learning. GPT-OSS's agents observe in $35.92\%$ of rounds, keep observing throughout the run, and make fewer private revisions as a result (Table~\ref{tab:r4-performance}; Appendix Figures~\ref{fig:r4-timing} and~\ref{fig:r4-search}). This frequent copying explains GPT-OSS's savings, but it also resembles the loss of independent exploration we saw in the bandit. GLM's pattern suggests a different hypothesis: perhaps a small amount of early observation is what makes it improve sooner.

\textbf{Spreading observations across learning works better than observing early (RQ3).} To test this hypothesis, we assign each agent exactly four observations, either all within rounds 1--10 (early) or one in each quarter of learning (distributed). Agents still choose whom to observe and which skill to execute. For GLM, distributed observation finishes $2.42\pm0.46$ points above early observation ($p<0.01$) and $3.00\pm1.13$ points above solo. GPT-OSS shows a similar improvement, but the effect is not statistically significant (Figure~\ref{fig:r4-assigned-timing}; Appendix Figure~\ref{fig:r4-timing-paired}). Early observation therefore does not explain GLM's advantage. Later observations are more likely to find skills that peers have already revised: for GLM, $95.83\%$ of copied skills differ from the shared initial file under distributed observation, compared with $79.17\%$ under early observation (Appendix Figure~\ref{fig:r4-copy-timing-behavior}c--d). Waiting gives peers time to produce something worth copying.

\textbf{Skills are revised and passed on, but they come from fewer independent discoveries (RQ3).} We trace the history of each executed skill through the recorded revisions and observations to test for recursive transmission (Section~\ref{sec:formulation}); copying the shared initial file does not count. At the final round, $83.3\pm16.7\%$ of GPT-OSS skills and $56.7\pm6.1\%$ of GLM skills show recursive transmission (Appendix Figure~\ref{fig:r4-lineages}c--d), so one agent's improvement does become the starting point for another agent's search. However, social populations build on fewer independent discoveries. We group skills into lineages, where a lineage contains every skill descended from one first revision of the shared initial file. At the end of a GPT-OSS social run, the five executed skills descend from only one lineage on average, compared with $4.83\pm0.17$ under solo learning. GLM social runs keep $2.33\pm0.21$ lineages, compared with five under solo learning (Appendix Figure~\ref{fig:r4-lineages}e--f). Skills in the same lineage can still differ, but social populations build almost all of their progress on a few early discoveries.

\textbf{Algorithmic social controllers reduce search costs without a clear benefit from adaptive source choice (RQ4).} Compared with full solo OpenEvolve, the UCB and uniform controllers use 76--78\% fewer learning tokens, and their final whole-answer accuracy stays within $1.03$ points (Figure~\ref{fig:r4-controller}; Appendix Figures~\ref{fig:r4-paired} and~\ref{fig:r4-costs}). At matched learning tokens, UCB and uniform score $+0.79\pm0.76$ and $+0.58\pm0.90$ points above full solo OpenEvolve for GPT-OSS, and $+3.27\pm1.77$ and $+1.56\pm0.36$ points above for GLM. The GLM uniform contrast is significant under a nominal paired test ($p<0.01$; Appendix Figure~\ref{fig:r4-efficiency-tokens}). When a simple rule decides when to copy, reusing peers' skills reaches similar accuracy for much less computation than searching alone, although part of this saving comes from avoiding the evaluation a revision requires. We find no clear advantage from choosing sources adaptively rather than uniformly (Appendix Figures~\ref{fig:r4-paired} and~\ref{fig:r4-efficiency-tokens}). One possible reason is that copying quickly spreads a few high-scoring skills, leaving UCB peers with nearly identical training scores to choose among (Appendix Figure~\ref{fig:r4-peer-spread}). Choosing whom to observe well may require a policy that estimates what a peer is likely to discover next, rather than relying only on its current training score.

\vspace{-0.5em}
\section{Discussion}
\vspace{-0.5em}

We studied whether self-improving LLM agents can benefit from what other self-interested agents discover, separating a population's performance per token from its performance relative to independent learners at the same cost. \textbf{First, useful social information is not enough.} In the controlled environment, an algorithmic social learner benefits from observing its peers, but no LLM does. LLM social learners explore too little and often run out of tokens before acting, yet fixing either problem does not make them outperform solo learners. \textbf{Second, observing peers changes how LLM populations learn, but not how well.} GLM reaches higher accuracy sooner, while GPT-OSS spends less, which makes it more efficient but not more effective. Neither outperforms independent learners at the same cost, whereas simple algorithmic rules for when to copy reach similar accuracy for far less computation. Agents do revise and pass on copied skills, so the gap lies in how LLMs decide when and from whom to learn.

\textbf{Limitations.} We primarily study three models in the bandit experiments and two models on a single instruction-following benchmark, so our findings may not generalize to other models or tasks. Because peers were always listed in the same order, we cannot separate a preference for particular peers from an effect of list position. We compare populations at the number of tokens both actually spent, rather than at a budget fixed in advance. Agents also keep only their latest observation in full, draw tasks from the same distribution, use a single model within each population, and cannot revise the prompts that guide their decisions. Relaxing any of these may change behavior.

\section*{Acknowledgements}
This research was supported by the UW-Amazon Science Gift Hub, UW-Tsukuba Amazon NVIDIA Cross Pacific AI Initiative (XPAI), Sony Research Award, Tinker Research Grants, Character.AI, DoorDash, Open Philanthropy, Coefficient Giving, Toyota Research Institute, the Schmidt AI2050 Fellows program, the Cooperative AI Foundation, the Foresight Institute, Jacobs CIFAR Research Fellowship, Templeton World Charity Foundation (https://doi.org/10.54224/34843), and the NSF CISE RI program, award \#2550849. This work was supported in part by Advanced Micro Devices, Inc. under the AMD University Program’s AI \& HPC Cluster. This material is based upon work supported by the Defense Advanced Research Projects Agency and the Air Force Research Laboratory, contract number(s): FA8650-23-C-7316. Any opinions, findings and conclusions, or recommendations expressed in this material are those of the author(s) and do not necessarily reflect the views of AFRL or DARPA.

\section*{AI Use}

We used generative AI tools to assist with implementing and debugging experimental and analysis code, including the synthetic bandit simulations and the OpenEvolve port; processing results and producing figures; finding potentially relevant references; and drafting and editing portions of the manuscript. The authors originated the research questions and hypotheses, selected the benchmarks and models, developed the formulation and its connection to the social-learning literature, designed the experimental conditions and causal controls, determined the interpretation and limitations of the results, as well as wrote and edited the draft. Generative AI was not used to develop theoretical results or proofs, conduct translation or qualitative analysis, or perform survey, interview, or transcription tasks, which are not applicable to this work. All suggested references were manually verified against their original sources. AI-assisted code was reviewed and tested, figures were checked against the underlying seed-level results, and the manuscript was manually revised. We take responsibility for the final content of this work, including all text, claims, code, and artifacts produced with the aid of generative AI.

\section*{Reproducibility Statement}

Sections~\ref{sec:formulation}--\ref{sec:r4-methods} describe the formalization, environments, models, comparison conditions, data splits, and evaluation metrics. The appendix specifies the bandit distributions and algorithmic baselines, skill-evolution budgets and accounting, uncertainty procedures, implementation checks, and additional results. We plan on releasing the experimental and analysis code, configurations, and seed-level summaries used to produce the figures and tables. We preserve matched seeds across primary comparisons and aggregate agents within independent populations before estimating uncertainty.

\bibliography{iclr2027_conference}
\bibliographystyle{iclr2027_conference}

\appendix

\section{Finite-bandit implementation and additional analyses}
\label{app:r1-methods}

\textbf{Environment and model details.}

\begin{figure}[!tbp]
  \centering
  \includegraphics[width=.78\linewidth]{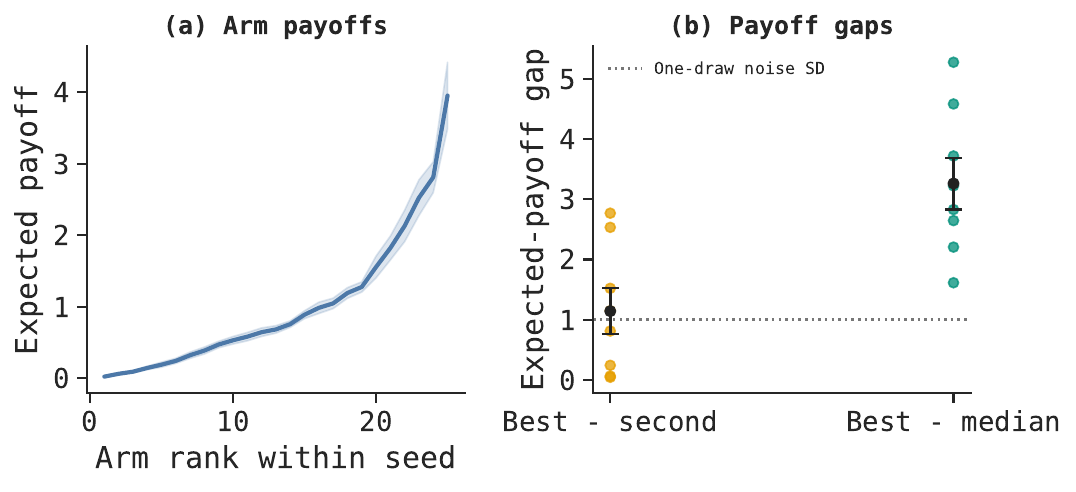}
  \caption{Audit of the stationary finite-bandit payoff distribution over eight matched
  seeds. Panel (a) shows arm means sorted within seed, averaged with seed SE.
  Panel (b) shows the best-minus-second and best-minus-median gaps for every seed;
  the dotted line is the standard deviation of one reward draw.}
  \label{fig:r1-arm-audit}
\end{figure}

The stationary environment draws 25 arm means independently from an exponential distribution with mean one. Rewards are deterministic counterfactual Gaussian draws keyed by seed, round, agent, and arm. This lets matched policies face the same potential outcome for the same pull. The eight environment seeds are shared across models and access conditions. Figure \ref{fig:r1-arm-audit}a shows the sorted mean profile over these seeds. The best arm is separated from the median by much more than one reward-noise standard deviation on average, while the best versus second-best gap is sometimes smaller, as shown in Figure~\ref{fig:r1-arm-audit}b. The environment therefore contains both meaningful strategy differences and nontrivial uncertainty near the optimum. Solo agents see their own history of pulled arms and payoffs. Social agents can additionally observe a peer's arm and payoff from the previous round, but not an action the peer has not yet completed.

The model snapshots are Qwen3-14B, openai/gpt-oss-20b, and Ministral-3-14B-Reasoning-2512. We use each model's supported reasoning template. GPT-OSS-20B uses temperature 1.0 with medium reasoning effort; the other models use the fixed V3 decoding profiles saved with every run. A strict parser accepts only the actions exposed by the condition. Parser or budget failures are scored as missing actions rather than silently dropped.

\textbf{Reward at matched token cost.} We estimate $J_t$ using the realized reward averaged over all ten agents in a round, with missing or invalid pulls assigned zero. This is per-round performance, not cumulative reward or reward divided by tokens. For each model and matched seed $s$, we choose $C_s^*=\min(C_{T,s}^{\mathrm{soc}},C_{T,s}^{\mathrm{solo}})$ and linearly interpolate each population's per-round reward against its cumulative tokens at $C_s^*$. We then subtract solo from social and report the mean paired difference with SE across the eight seeds. Thus, each pair is compared at the same expenditure, although that expenditure can differ between seeds. All pairs remain within their observed cost ranges; no trajectory is extrapolated. We use charged completion tokens for the primary bandit comparison, consistent with its budget and efficiency accounting, and check prompt-plus-completion tokens separately. Section~\ref{sec:r4-methods} uses prompt-plus-completion learning tokens. Because the formulation defines $t_C$ as the last completed round within budget, we also report that discrete comparison as a sensitivity check rather than treating interpolation as an exact observation at $C_s^*$.

\begin{figure}[!tbp]
\centering
\includegraphics[width=\linewidth]{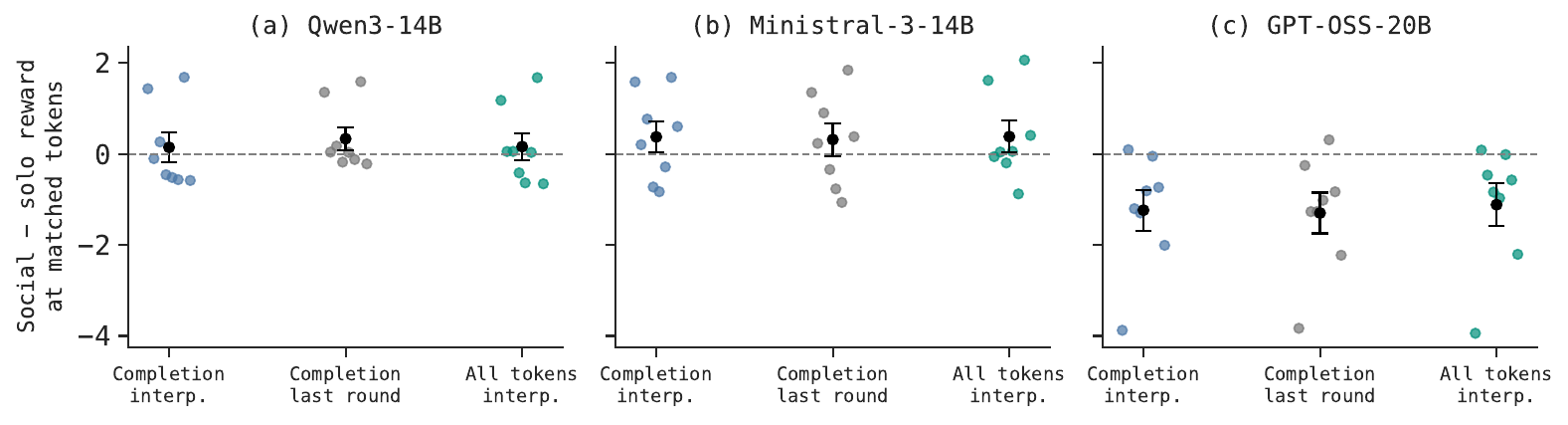}
\caption{Realized per-round social-minus-solo reward at matched token expenditure. Panels (a)--(c) show Qwen, Ministral, and GPT-OSS. Each dot is one paired environment seed; black markers and bars show mean $\pm$ SE across eight seeds. The primary comparison interpolates against completion tokens; alternatives use the last completed round or interpolate against prompt-plus-completion tokens. Missing pulls earn zero.}
\label{fig:r1-matched-tokens}
\end{figure}

Using the last completed round gives paired differences of $+0.33\pm0.25$, $+0.32\pm0.36$, and $-1.30\pm0.45$ for Qwen, Ministral, and GPT-OSS, respectively; interpolating against prompt-plus-completion tokens gives $+0.16\pm0.30$, $+0.38\pm0.35$, and $-1.12\pm0.48$ (Figure~\ref{fig:r1-matched-tokens}a--c). These checks retain the primary estimates' directions. The endpoint is chosen from expenditures, not rewards, and the analysis summarizes that shared endpoint rather than establishing an advantage or its absence at every budget.

\begin{figure}[!tbp]
\centering
\includegraphics[width=0.9\linewidth]{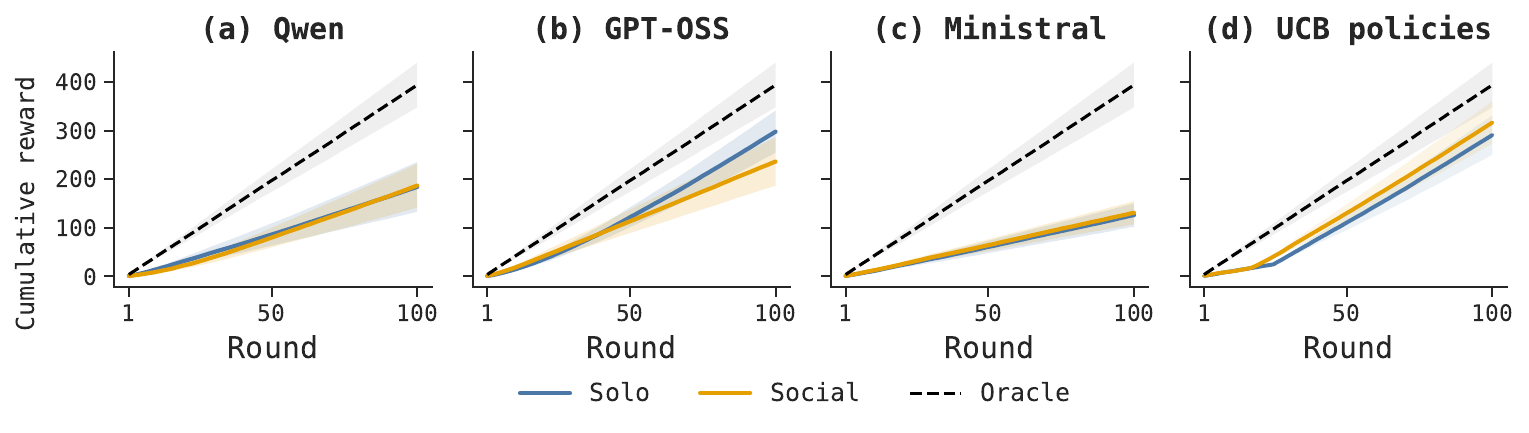}
\caption{Cumulative reward per agent over the same eight environment seeds in all panels. Panels (a)--(c) show the three LLM families; (d) shows solo and hierarchical social UCB. Every panel includes the same oracle, shown with a black dashed line and light-gray SE shading. Colored shading is seed SE.}
\label{fig:r1-cumulative}
\end{figure}

\begin{figure}[!tbp]
\centering
\includegraphics[width=.70\linewidth]{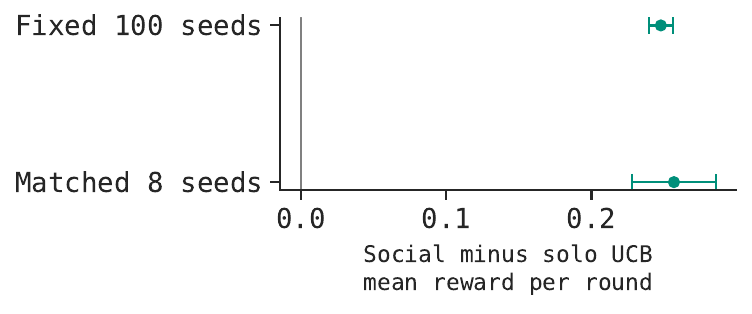}
\caption{Precision sensitivity for hierarchical versus solo UCB. The original eight environments are a subset of the fixed 100-seed extension. Paired mean differences and SE quantify uncertainty across environments, not additional information available to agents.}
\label{fig:r1-precision}
\end{figure}

\textbf{Cumulative reward and simulation precision.} Figure~\ref{fig:r1-cumulative} integrates the per-round rewards shown in the main text, using the same eight environments for every policy. Figure~\ref{fig:r1-precision} compares the eight algorithmic seeds with the fixed extension to 100. The larger set was chosen before examining its effect estimate; simulation was not stopped when a significance threshold was crossed. It improves precision for the paired algorithmic contrast, but its raw reward levels should not be directly compared with the LLMs' eight-environment average. Both the UCB curves and the shared oracle in the main figure therefore use the matched subset.

\textbf{Cumulative regret.}

We compute cumulative expected regret from the known arm means, assigning expected payoff zero to an invalid pull and averaging agents within each population. Reward and regret share the same oracle reference and are complementary views of the same learning process.

\begin{figure}[!tbp]
  \centering
  \includegraphics[width=.94\linewidth]{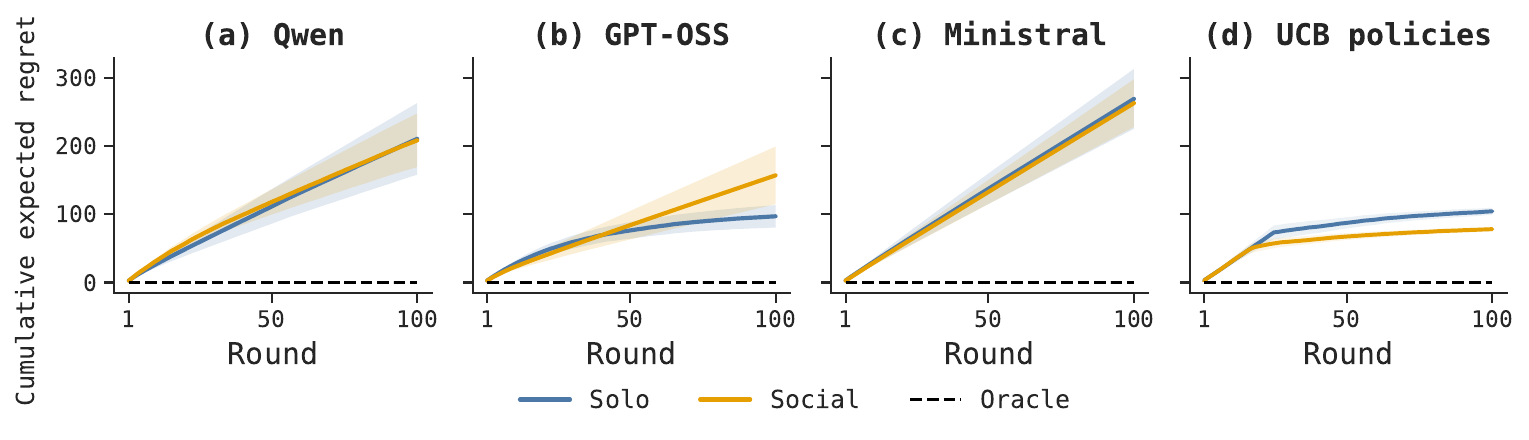}
  \caption{Cumulative expected regret for the policies in
  Figure~\ref{fig:r1-reward}. For an invalid or missing pull, the selected arm
  has expected payoff zero. Lines average agents within population seed and
  bands are seed SE over the same eight environments for all policies.}
  \label{fig:r1-regret}
\end{figure}

Figure~\ref{fig:r1-regret} is the expected-regret counterpart of the main reward figure. GPT-OSS-20B accumulates more regret with social access than alone, while Qwen and Ministral have overlapping trajectories, as shown in Figure~\ref{fig:r1-regret}a--c. Hierarchical social UCB accumulates less regret than standard solo UCB, as shown in Figure~\ref{fig:r1-regret}d. Reward and regret are not independent replications of the finding; they use the same actions and oracle means. At round 100, solo UCB's expected regret is $0.25$ reward per pull, compared with $0.11$ for hierarchical UCB and zero for the oracle.

\textbf{Copy quality and population dynamics.}

\begin{figure}[!tbp]
  \centering
  \includegraphics[width=.80\linewidth]{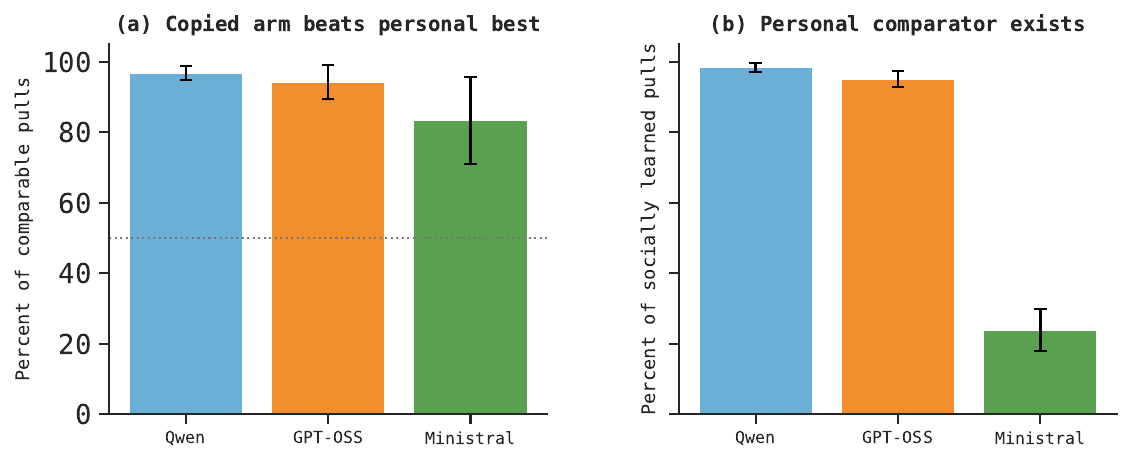}
  \caption{Local copy quality under the primary expiring-budget condition.
  Panel (a) shows the percentage of socially learned pulls whose arm mean exceeds
  the best arm the copier had personally discovered. Panel (b) shows the share
  of socially learned pulls for which that personal comparator exists. Bars are
  seed means with seed SE over eight populations.}
  \label{fig:r1-copy}
\end{figure}

\begin{figure}[!tbp]
  \centering
  \includegraphics[width=.93\linewidth]{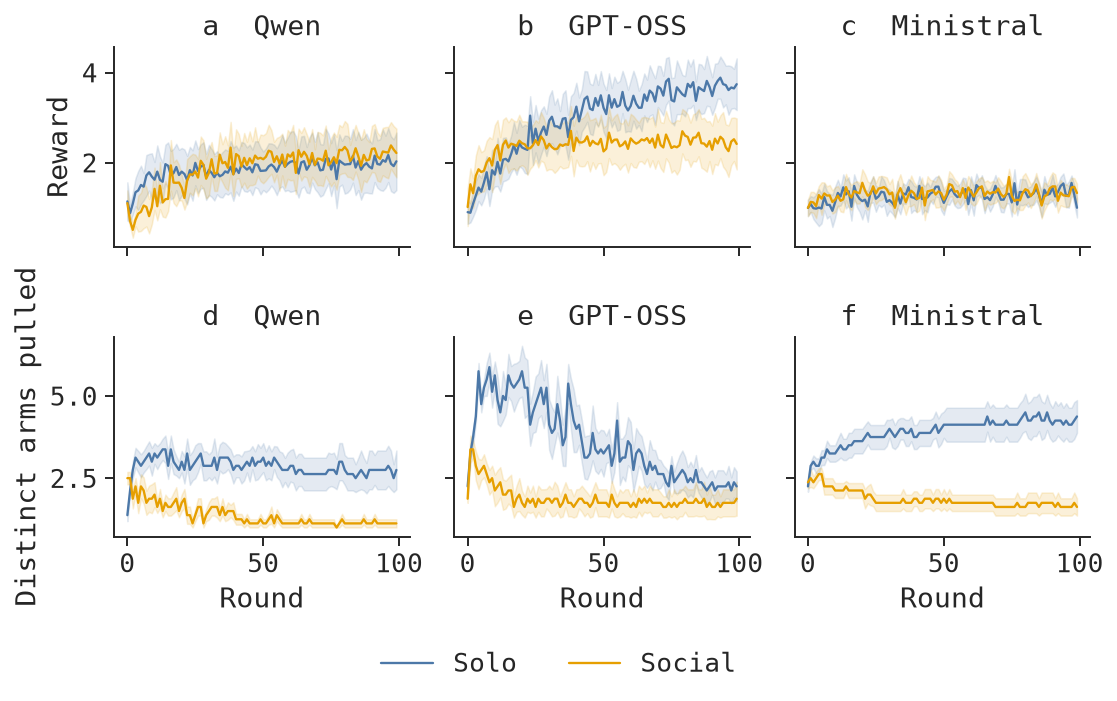}
  \caption{Finite-bandit reward and arm-use trajectories. Panels (a)--(c) show reward
  per round for Qwen, GPT-OSS-20B, and Ministral. Panels (d)--(f) show the number
  of distinct arms used by the corresponding populations. Lines are means and
  bands are seed SE.}
  \label{fig:r1-time}
\end{figure}

Figure~\ref{fig:r1-time} shows why the same copy can be individually useful and collectively insufficient. GPT-OSS-20B social populations rapidly use a narrower set of arms than solo populations, as shown in Figure~\ref{fig:r1-time}e. Qwen also selects fewer distinct arms with social access for much of the run, as shown in Figure~\ref{fig:r1-time}d. These are descriptive trajectories; the randomized IS intervention provides the causal test of adding private search.

\begin{figure}[!tbp]
  \centering
\includegraphics[width=.92\linewidth]{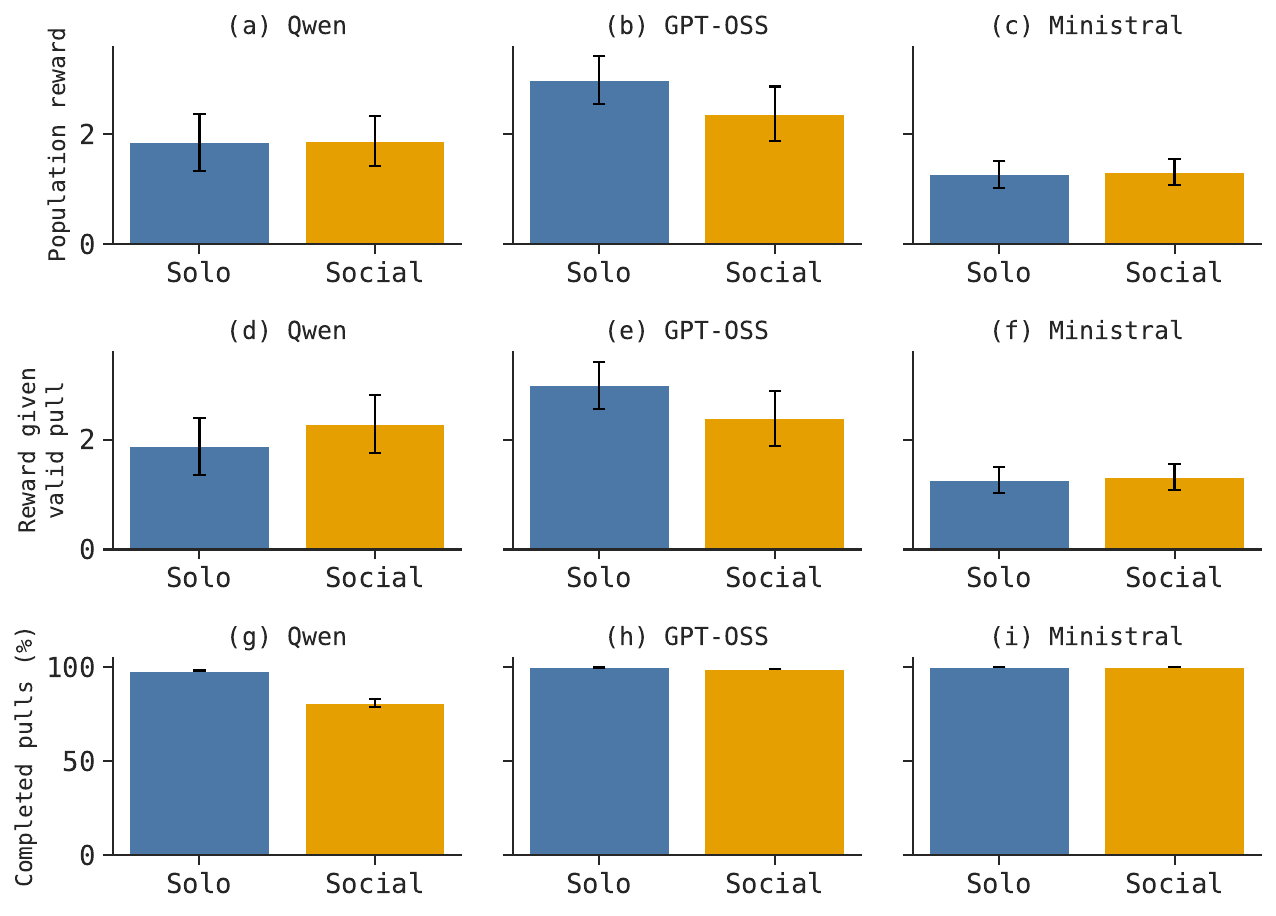}
  \caption{Primary finite-bandit outcomes under the stationary, neutral-prompt, expiring-budget condition. Columns show Qwen, GPT-OSS, and Ministral. Panels (a)--(c) show population reward; (d)--(f) reward conditional on a valid pull; (g)--(i) pull completion. Missing pulls earn zero in population reward. Bars show eight-seed means and SE.}
  \label{fig:r1-completion}
\end{figure}

Completion-aware outcomes separate earning more on a completed pull from completing more pulls (Figure~\ref{fig:r1-completion}). Conditional reward is selected by whether an action finishes and is therefore descriptive. The randomized PE intervention, rather than this conditional comparison, provides the causal test of preserving an execution reserve.

\textbf{Mechanism effects from the randomized interventions.}

For causal interventions, we distinguish an effect within social populations, the social-minus-solo difference under treatment, and the paired interaction
\begin{equation}
\Delta= (Y_{\mathrm{social,treated}}-Y_{\mathrm{solo,treated}})
       -(Y_{\mathrm{social,default}}-Y_{\mathrm{solo,default}}).
\label{eq:did}
\end{equation}
We avoid treating agent-rounds as independent observations and do not interpret associations between chosen copying and reward as causal effects.

\begin{figure}[!tbp]
  \centering
  \includegraphics[width=.84\linewidth]{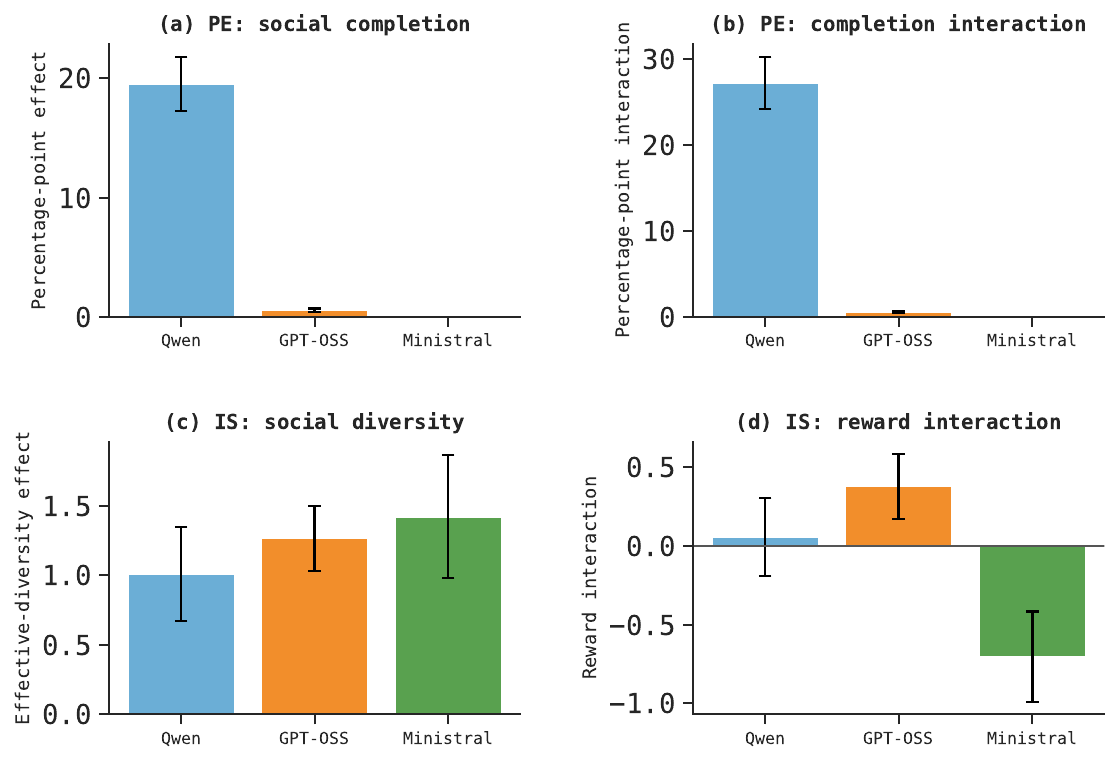}
  \caption{Mechanism estimates for the finite bandit. Panel (a) is PE's effect on completion
  within social populations. Panel (b) is its completion interaction from
  Equation~\ref{eq:did}. Panel (c) is IS's effect on effective selected-arm
  diversity within social populations. Panel (d) is its reward interaction.
  Bars show paired means with seed SE over eight seeds.}
  \label{fig:r1-mechanisms}
\end{figure}

The distinction between panels in Figure~\ref{fig:r1-mechanisms} is important. A positive effect in panel (a) or (c) shows that the intervention changes a targeted outcome within social populations. Only panels (b) and (d) ask whether the intervention changes social populations more than their matched solo controls. A positive interaction can still leave raw social performance below raw solo performance, which is why the main text begins with Figure~\ref{fig:r1-interventions}.

\textbf{Why is Ministral's independent-search interaction negative?}
\label{app:ministral-is}

\begin{figure}[!tbp]
\centering
\includegraphics[width=.94\linewidth]{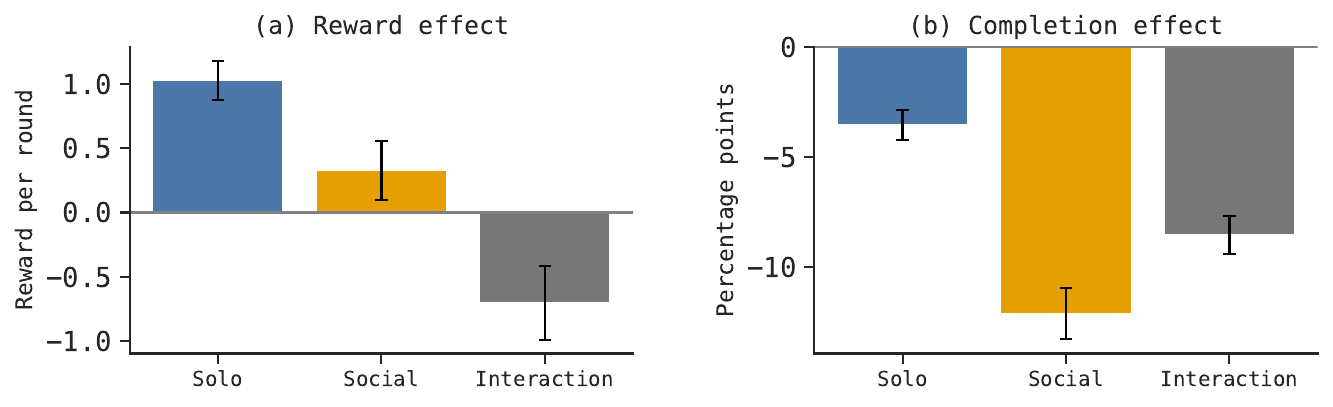}
\caption{Effects of independent search on Ministral. Panels (a)--(b) separate the treatment effect within solo, the effect within social, and their difference for reward and completion. Each uses eight matched causal-control seeds, without pooling the original default replications. Bars show paired mean effects and SE.}
\label{fig:r1-ministral-is}
\end{figure}

IS raises mean reward by $1.031\pm0.154$ in solo and $0.327\pm0.231$ in social populations, yielding a social-minus-solo interaction of $-0.704\pm0.287$ (Figure~\ref{fig:r1-ministral-is}a). The negative interaction therefore does not mean the intervention lowers social reward: solo improves more. Completion falls in both conditions, by $3.56$ percentage points in solo and $12.10$ in social (Figure~\ref{fig:r1-ministral-is}b). Forced search can improve which options are found while making it harder to finish a pull, especially in the social policy. These randomized treatment effects identify that tradeoff; they do not separately identify how much of the reward difference is mediated by completion versus arm choice.

\textbf{Hierarchical source selection.}
\label{app:source-ucb}

The source index is
\begin{equation}
I_j(t)=\frac{S_j+n_0m_j}{N_j+n_0}
+\sqrt{\frac{2\log\!\left(\max(\sum_kN_k+n_0P,2)\right)}{N_j+n_0}},
\label{eq:source-ucb}
\end{equation}
where $j$ is the agent or one of its peers, $S_j$ and $N_j$ summarize observed source payoffs, $P$ is the population size, and $n_0=1$. The prior gives the agent itself a modest initial advantage. For skill scores in $[0,1]$, the common prior mean is 0.5 and the self prior is 0.55. Observing updates the selected peer's source estimate from its reported training score; choosing oneself updates the private source from the training score of the archive-selected skill after revision. These are scores, not estimates of the marginal improvement caused by copying or revising. The lower-level archive is private to each agent. Since peers' skills change during learning, standard stationary-arm UCB guarantees do not apply to this source-selection problem.

\textbf{Discounted explore--observe--exploit.}
\label{app:deoe}

DEOE stands for \textbf{discounted explore--observe--exploit}. It is inspired by DiscountMachine, the winning strategy in the social-learning tournament of \citet{Rendell2010-ro}. It is not a reproduction of that entry. At round $t$, the implemented policy explores independently with probability $1/\sqrt{t+1}$. Otherwise, a social agent observes a random peer after a payoff that is surprising relative to its current estimate, after 20 rounds without an observation, or with the same decaying exploration probability. For an arm last sampled at $\tau_a$, it discounts the estimate toward the agent's mean:
\begin{equation}
\widetilde\mu_a(t)=\overline\mu_t+0.95^{t-\tau_a}
\left(\widehat\mu_a-\overline\mu_t\right).
\end{equation}
The agent then pulls the arm with the highest discounted estimate. Even in a stationary environment, realized Gaussian payoffs remain noisy, so a surprising draw can trigger a social check. Discounting is unnecessary for stationary means in an ideal estimator. The policy retains one rule across stationary and changing conditions.

\begin{figure}[!tbp]
\centering
\includegraphics[width=\linewidth]{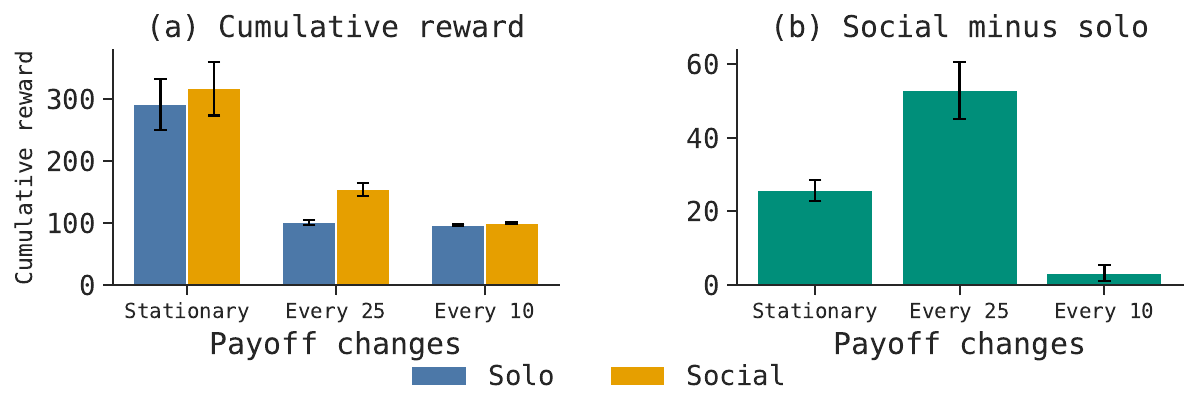}
\caption{Hierarchical social UCB versus solo UCB under stationary and changing payoffs. Panel (a) shows cumulative reward; (b) shows the paired social-minus-solo contrast. Bars show means and seed SE. Both algorithms reset their estimates at the known payoff-change boundaries, so this comparison does not test their ability to detect changes.}
\label{fig:r1-ucb-volatility}
\end{figure}

\begin{figure}[!tbp]
\centering
\includegraphics[width=\linewidth]{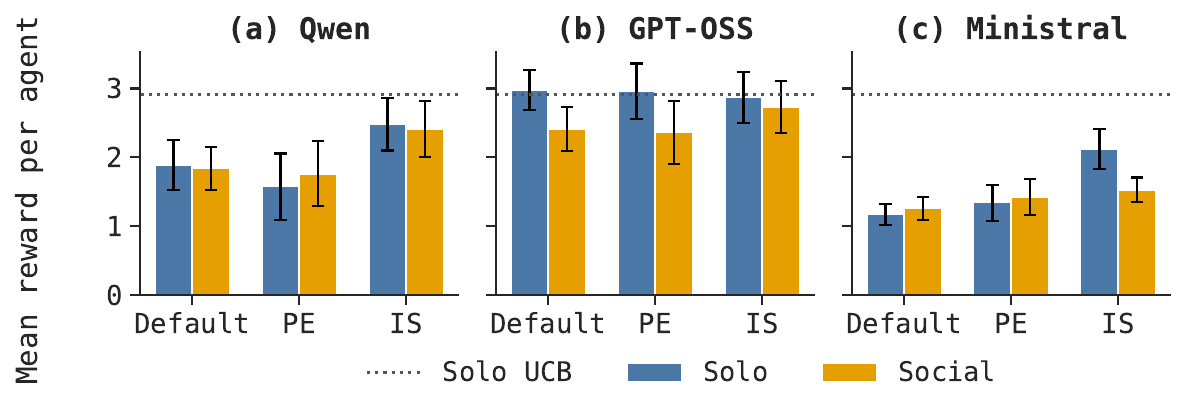}
\caption{Mean reward per round for the interventions in Figure~\ref{fig:r1-interventions}, averaged over all 100 rounds rather than measured only at the endpoint. Panels (a)--(c) show the three models. Multiplying these means and SE by 100 gives the main cumulative-reward figure; the two summaries therefore yield the same ranking.}
\label{fig:r1-intervention-mean}
\end{figure}

\begin{figure}[!tbp]
\centering
\includegraphics[width=\linewidth]{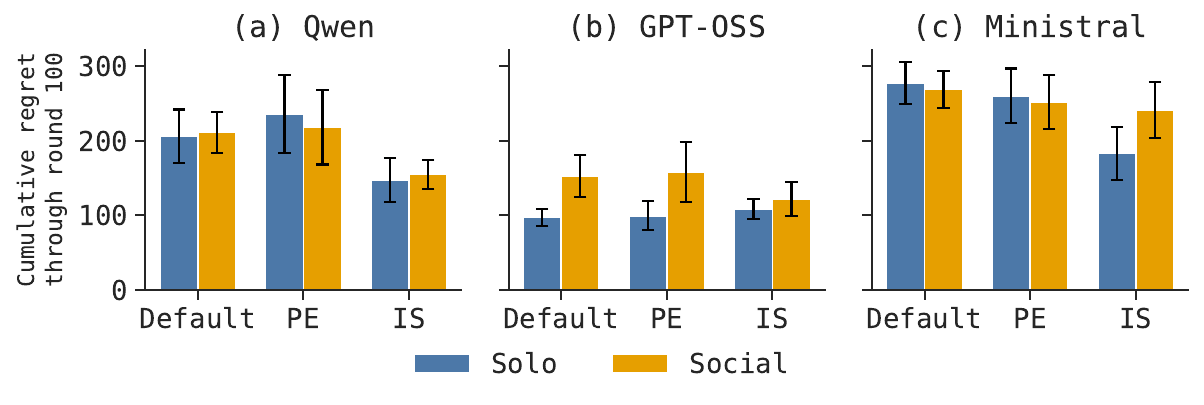}
\caption{Cumulative expected regret through round 100 for the same populations and interventions as Figure~\ref{fig:r1-interventions}. Panels (a)--(c) show the three models. Lower regret is better; bars show means and SE across run replicates.}
\label{fig:r1-intervention-regret}
\end{figure}

\begin{figure}[!tbp]
\centering
\includegraphics[width=\linewidth]{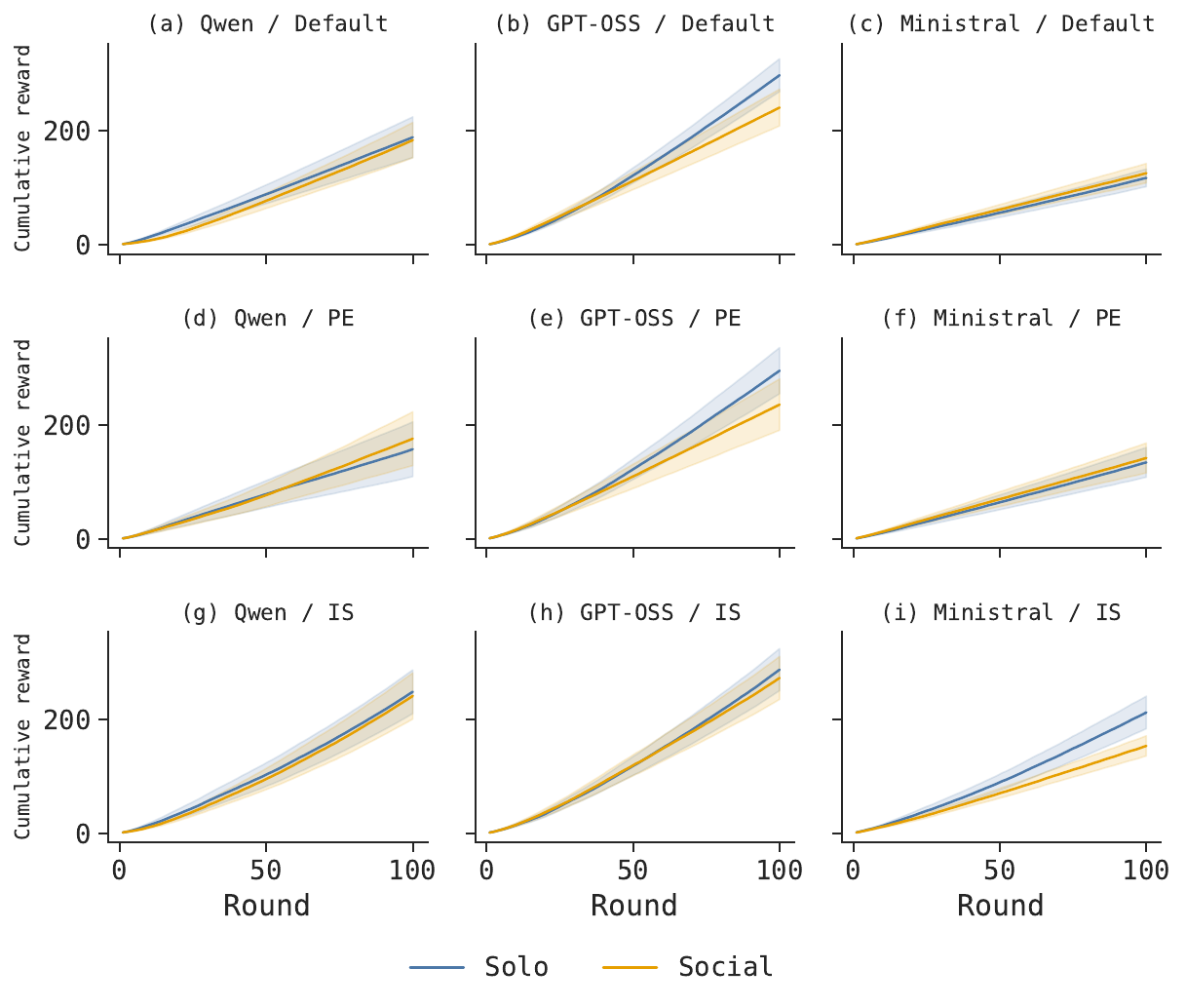}
\caption{Cumulative reward trajectories underlying Figure~\ref{fig:r1-interventions}. Columns show the three models; rows show default, protected execution, and independent search. Lines show means and shading shows SE across run replicates. These curves distinguish improvement throughout learning from an effect confined to its endpoint.}
\label{fig:r1-intervention-learning}
\end{figure}

\section{Social learning with an effectively infinite option set}
\label{app:r2}

The continuous bandit begins with no known arms. \textsc{Innovate} samples a new coordinate and reveals an informational payoff, while \textsc{Observe} reveals a peer's most recent coordinate under the access condition. Neither earns scored reward. Only \textsc{Pull} produces the round's payoff. The primary landscape is a fixed smooth random function with length scale 0.15; nearby coordinates therefore have correlated means. Ten agents interact for 100 rounds with 4,096 expiring tokens per round.

\begin{figure}[!tbp]
  \centering
  \includegraphics[width=.80\linewidth]{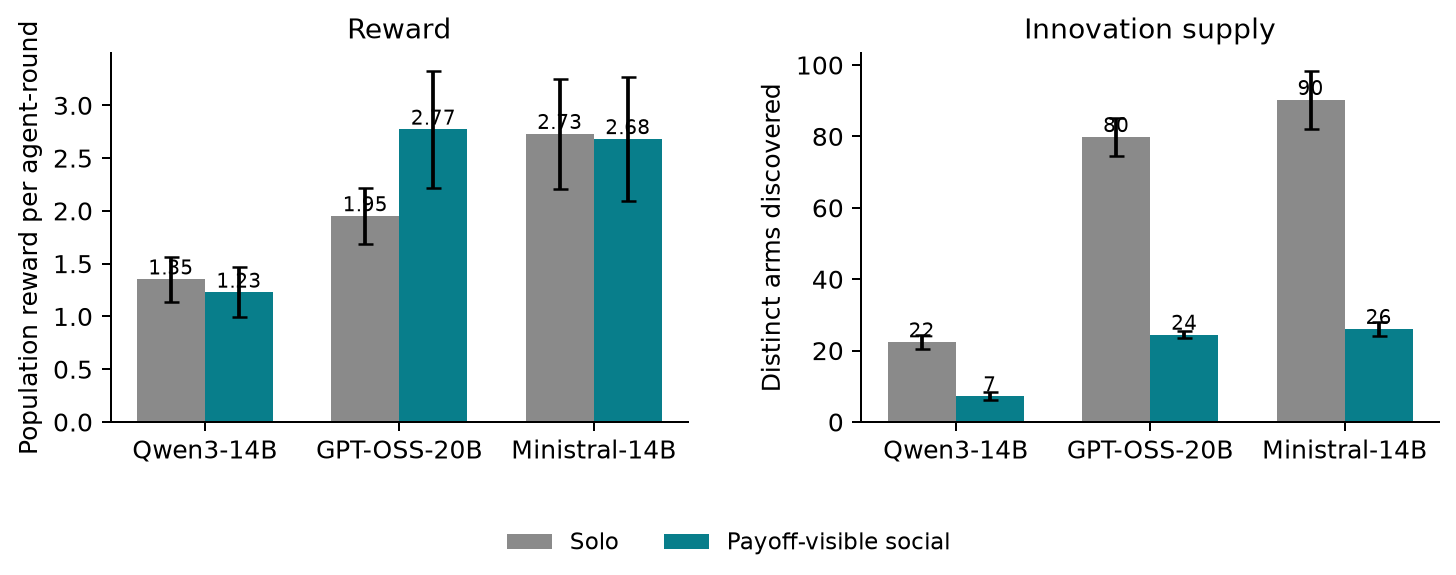}
  \caption{Continuous-bandit stationary structured environment. Panel (a) shows absolute
  reward for solo and payoff-visible social populations. Panel (b) shows the
  independent discovery supply. Error bars are seed SE over matched populations.}
  \label{fig:r2-main}
\end{figure}

Social access sharply reduces the number of independent arms generated for all three models, as shown in Figure~\ref{fig:r2-main}b. The reward consequence is model dependent: GPT-OSS-20B improves in the stationary landscape while Qwen and Ministral do not show the same clear benefit, as shown in Figure~\ref{fig:r2-main}a. This prevents a stronger claim that reduced diversity necessarily lowers short-run performance.

\begin{figure}[!tbp]
  \centering
  \includegraphics[width=.80\linewidth]{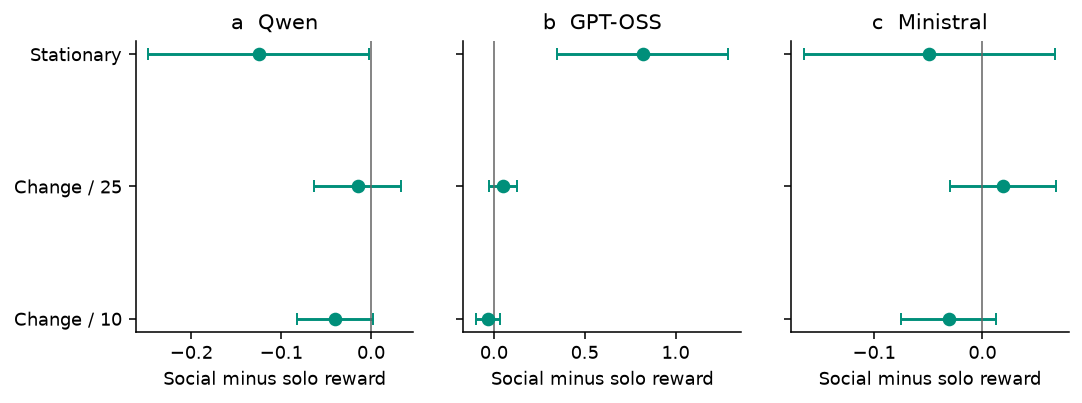}
  \caption{Continuous-bandit payoff-visible social minus solo reward when payoffs are
  stationary or change every 25 or 10 rounds. Panels (a)--(c) show Qwen,
  GPT-OSS-20B, and Ministral. Points are paired means with seed SE.}
  \label{fig:r2-volatility}
\end{figure}

The benefit becomes less stable when payoffs change. GPT-OSS-20B's social-minus-solo reward is $+0.820\pm0.471$ in the stationary condition, $+0.049\pm0.076$ when payoffs change every 25 rounds, and $-0.034\pm0.065$ when they change every 10 rounds, as shown in Figure~\ref{fig:r2-volatility}b. The same comparison for Qwen and Ministral is shown in Figure~\ref{fig:r2-volatility}a,c. Because the reward landscape is continuous, continuous-bandit regret uses the best value on a dense reference grid and is an approximation rather than an exact infinite-arm optimum.

Cumulative reward and approximate regret provide complementary summaries of these same actions (Figures~\ref{fig:r2-cumulative} and~\ref{fig:r2-regret}). The per-round trajectories in Appendix Figure~\ref{fig:r2-time}a--c show when the reward differences emerge. Panels (d)--(f) show the corresponding supply of independently discovered coordinates. They make clear that social populations can reduce discovery early even when the short-run reward effect is small or positive.

\begin{figure}[!tbp]
  \centering
  \includegraphics[width=.94\linewidth]{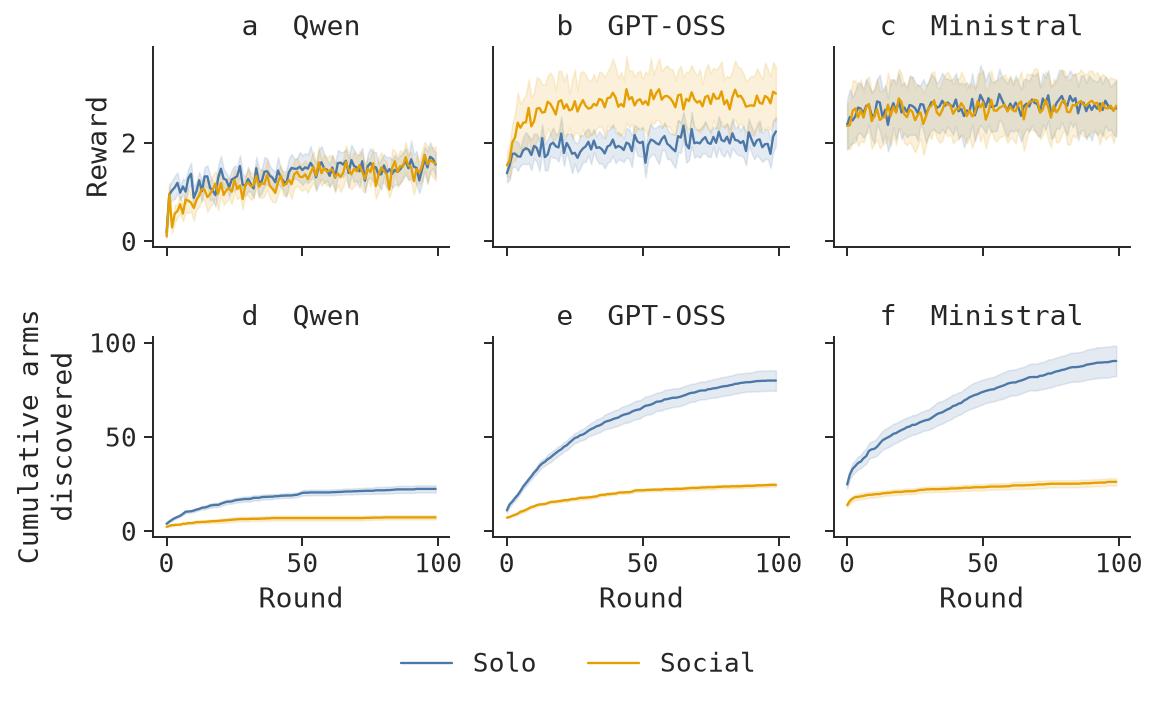}
  \caption{Continuous-bandit trajectories in the stationary structured environment.
  Panels (a)--(c) show reward per round for Qwen, GPT-OSS-20B, and Ministral.
  Panels (d)--(f) show cumulative independently discovered arms for the same
  models. Lines are means and bands are seed SE.}
  \label{fig:r2-time}
\end{figure}

\begin{figure}[!tbp]
\centering
\includegraphics[width=.94\linewidth]{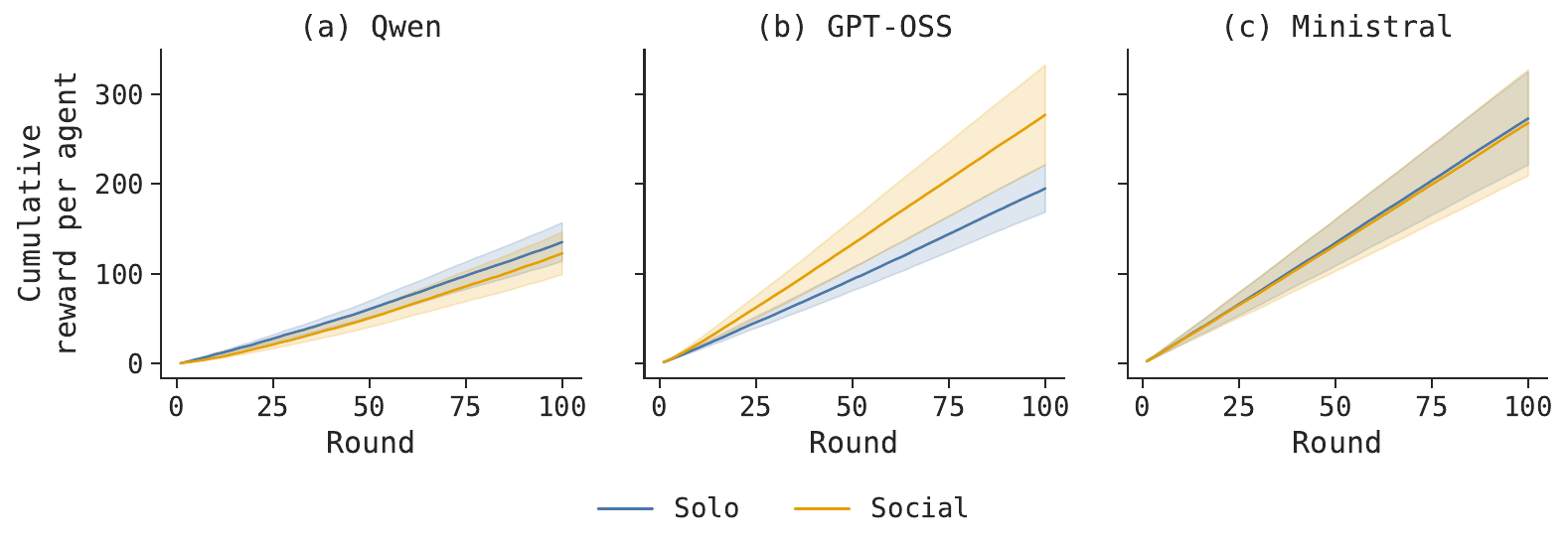}
\caption{Cumulative reward in the stationary continuous landscape. Panels (a)--(c) show Qwen, GPT-OSS, and Ministral under the primary expiring-budget condition, with population-seed means and SE.}
\label{fig:r2-cumulative}
\end{figure}

\begin{figure}[!tbp]
\centering
\includegraphics[width=.94\linewidth]{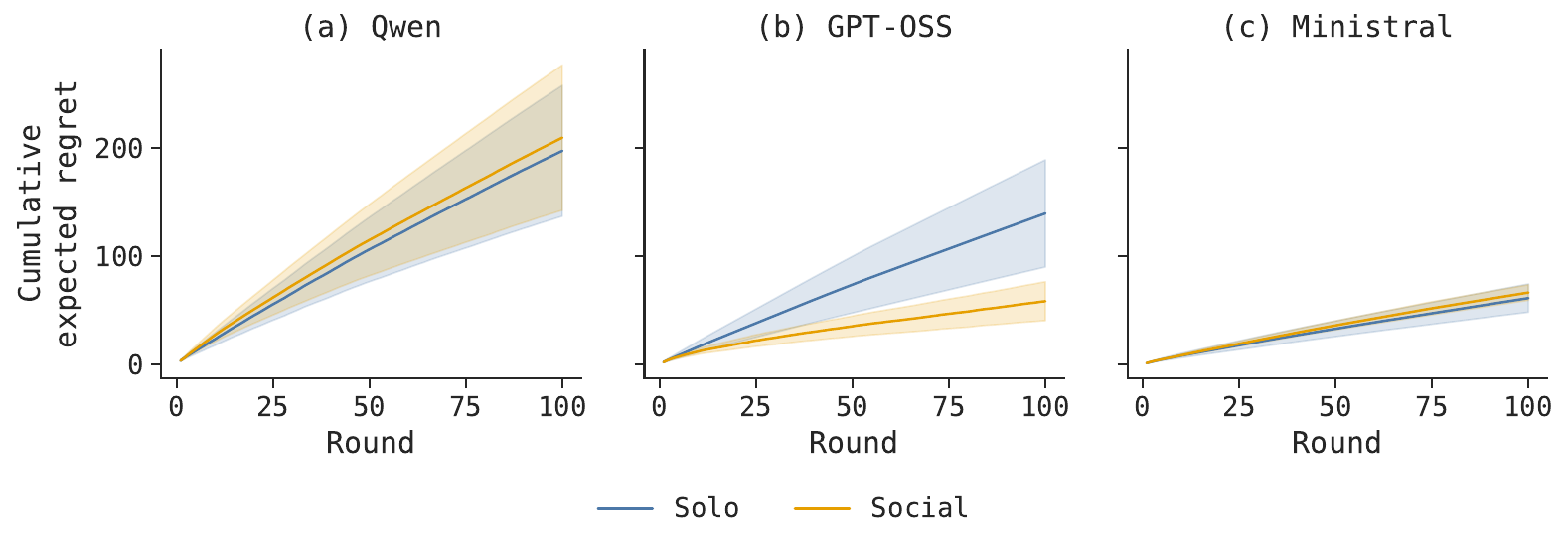}
\caption{Cumulative expected regret relative to the best value on the saved dense reference grid. Panels (a)--(c) show the same models and actions as Figure~\ref{fig:r2-cumulative}. The continuous-space oracle is approximate; these are not exact regret bounds over infinitely many arms.}
\label{fig:r2-regret}
\end{figure}

\begin{figure}[!tbp]
  \centering
  \includegraphics[width=.88\linewidth]{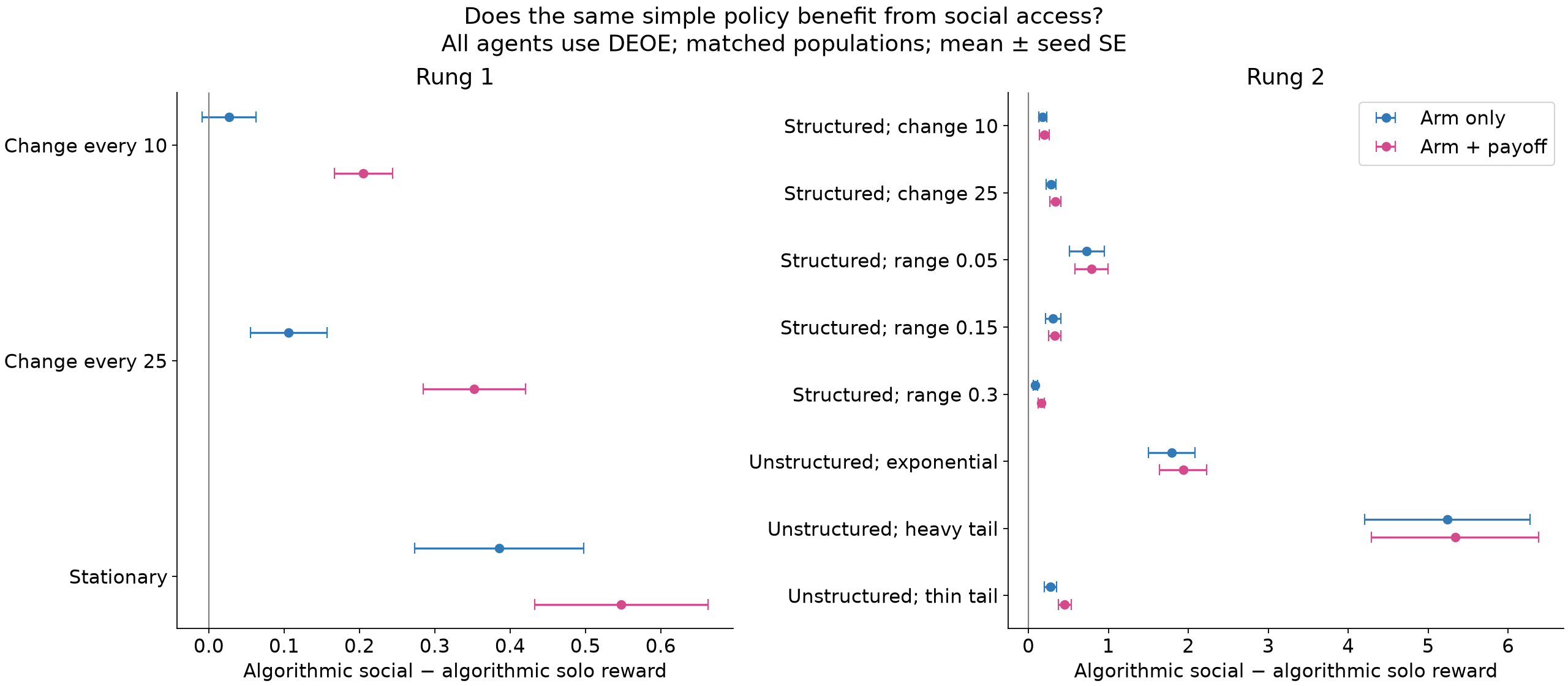}
  \caption{Paired social-minus-solo reward for DEOE across completed finite-bandit and
  continuous-bandit environments. Blue points expose actions only; pink points also expose
  payoff. Error bars are seed SE over eight matched populations.}
  \label{fig:deoe-across}
\end{figure}

DEOE benefits from social information in this setting as well. In the stationary length-scale-0.15 environment, the payoff-visible social-minus-solo effect is $+0.331\pm0.076$ reward per round, as shown in Figure~\ref{fig:deoe-across} on the continuous-bandit row. The effect establishes an available policy-level benefit under DEOE's cost model and does not imply that every reduction in innovation is optimal.

\section{Social learning with fixed reusable skills}
\label{app:r3}

\begin{figure}[!tbp]
\centering
\resizebox{\linewidth}{!}{%
\begin{tikzpicture}[font=\small,>=Stealth]
  \node[anchor=west,font=\bfseries] at (0,2.0) {Example jobs};
  \foreach \x/\id/\deadline/\profit in {0/J0/1/20,1.5/J1/2/100,3/J2/1/40,4.5/J3/2/35,6/J4/3/30}{
    \node[draw=solo,rounded corners,fill=solo!8,minimum width=1.25cm,minimum height=.85cm,align=center]
      at (\x,1.2) {\textbf{\id}\\$d=\deadline, p=\profit$};}
  \draw[->,very thick,evolve] (6.8,1.2)--(7.7,1.2);
  \node[anchor=west,font=\bfseries] at (7.8,2.0) {Valid schedule};
  \foreach \x/\slot/\id in {8.4/1/J2,9.8/2/J1,11.2/3/J4}{
    \node[draw=evolve,fill=evolve!8,minimum width=1.15cm,minimum height=.85cm,align=center]
      at (\x,1.2) {slot \slot\\\textbf{\id}};}
  \node[anchor=west] at (7.8,.4) {$40+100+30=170$; all deadlines are met.};
\end{tikzpicture}
}
\caption{Illustration of the fixed-skill scheduling task. The benchmark uses 16 jobs
and up to eight slots. Invalid schedules receive zero.}
\label{fig:r3-task}
\end{figure}
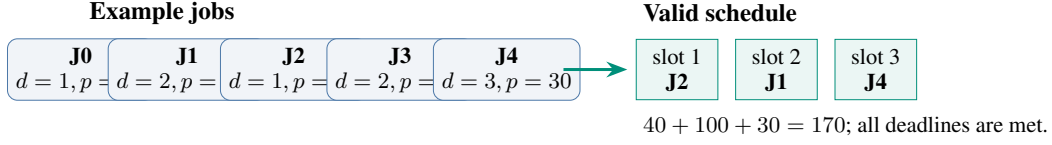

The fixed-skill experiment replaces abstract arms with eight Markdown procedures for a unit-time job-scheduling task. Each private instance contains 16 jobs with deadlines and profits. The model must return an ordered schedule in which every selected job meets its deadline; reward is selected profit divided by the optimal profit. Figure~\ref{fig:r3-task} gives a small example.

\begin{figure}[!tbp]
  \centering
  \includegraphics[width=.90\linewidth]{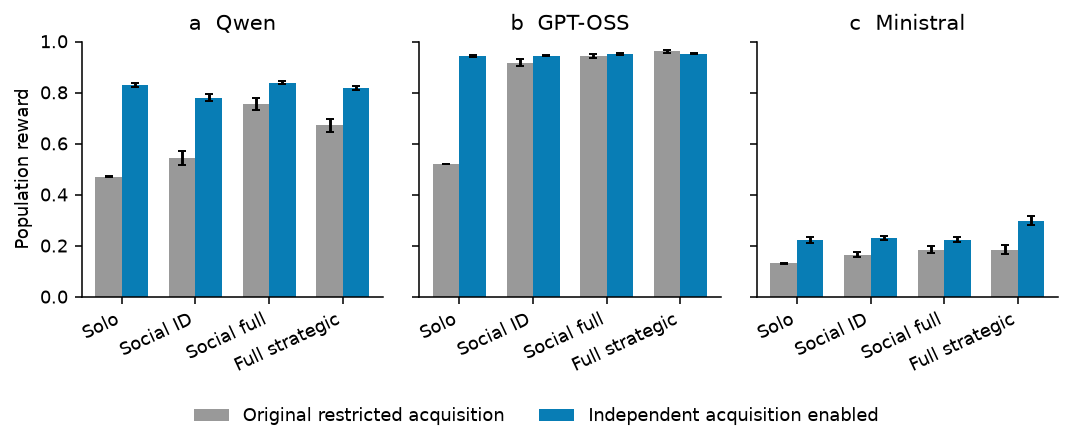}
  \caption{Fixed-skill population reward before and after equalizing access to all
  eight skills. Panels (a)--(c) show Qwen, GPT-OSS-20B, and Ministral. Gray bars
  are the original restricted-acquisition design; blue bars enable independent
  acquisition in both solo and social conditions. Error bars are seed SE.}
  \label{fig:r3-access}
\end{figure}

The original design gave social agents access to better procedures through peers while solo agents remained limited to their initial skill. The equal-access control allows both conditions to acquire any skill independently. Once access is equalized, most of the original social advantage disappears, as shown by the blue bars in Figure~\ref{fig:r3-access}a--c. This indicates that redistribution was useful when it was the only route to better procedures. It provides little average value once both conditions can reach the same menu.

In Figure~\ref{fig:r3-access}, ``Social ID'' reveals only a peer's last skill identifier. ``Social full'' additionally reveals its skill text, latest payoff, reward history, and population use and reward statistics. ``Full strategic'' supplies the same information with instructions to copy selectively, preserve independent search, and discount stale evidence. These information and guidance variants are supporting controls; they are not pooled with one another.

\begin{figure}[!tbp]
  \centering
  \includegraphics[width=.88\linewidth]{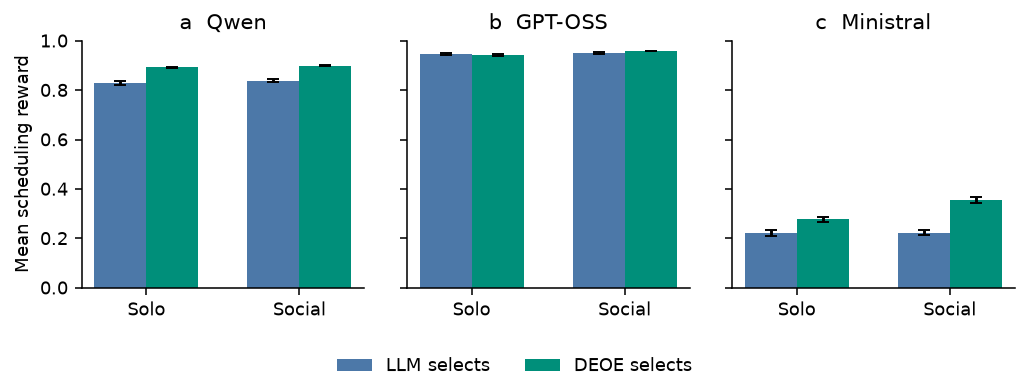}
  \caption{Fixed-skill reward when the LLM or DEOE selects a skill and the same LLM
  executes the schedule. Panels (a)--(c) show Qwen, GPT-OSS-20B, and Ministral.
  Bars are means with seed SE over eight seeds.}
  \label{fig:r3-deoe}
\end{figure}

We also separate skill selection from schedule execution with a DEOE+LLM hybrid. DEOE chooses the skill without model inference, while the same language model executes the schedule conditioned on that skill. The hybrid improves five of six model-by-access cells, with its largest increase for Ministral social execution, as shown in Figure~\ref{fig:r3-deoe}a--c. This comparison also protects execution tokens by removing the language-model selection call, so it is an intervention on the selector and its resource use together rather than a pure selector effect.

\section{Skill-evolution checks and additional results}
\label{app:r4-sanity}

\subsection{Revision and transmission in deployed skill lineages}
\label{app:lineages}

\begin{figure}[!tbp]
\centering
\includegraphics[width=\linewidth]{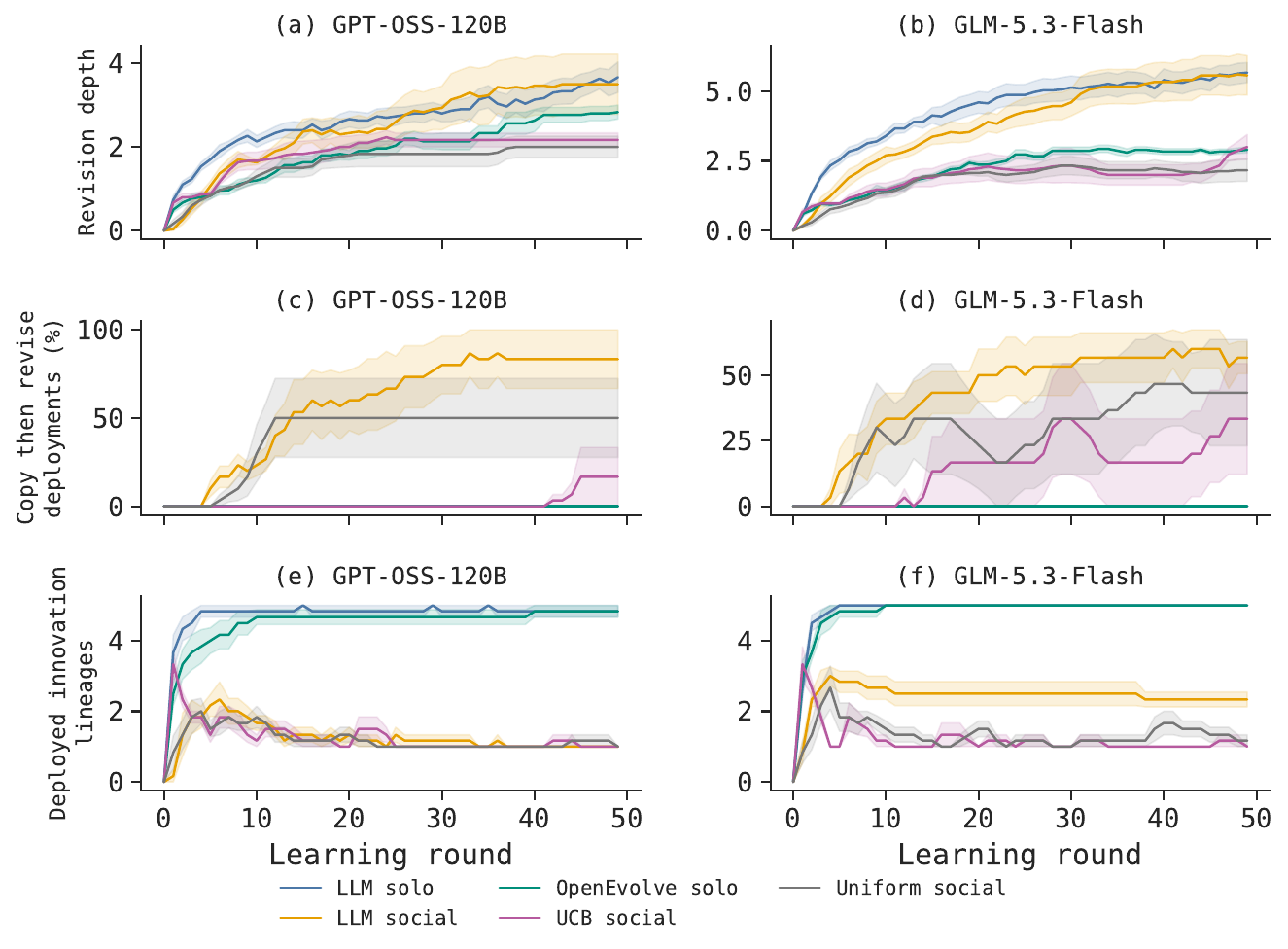}
\caption{Ancestry of deployed skills, with population means and seed SE. Columns show GPT-OSS and GLM. Panels (a)--(b) count private revision steps in a deployed file's ancestry; (c)--(d) count deployments whose ancestry includes copying a non-initial skill followed by another revision; (e)--(f) count distinct founding revision lineages currently deployed by the five agents. This is recorded first-acquisition ancestry, not all possible paths to identical file content.}
\label{fig:r4-lineages}
\end{figure}

We reconstruct each agent's acquired skill versions from the saved revision and observation events. A revision adds one to its parent's revision depth; a copy inherits the source's ancestry from the previous round. Reacquiring identical file content does not overwrite its first-acquisition history. The common initial file has depth zero and is not counted as a discovered lineage. A founding revision is the first new version descended from that initial file. Its descendants share a lineage even if their later contents differ. Figure~\ref{fig:r4-lineages} counts these lineages among the five deployed skills, not among every candidate retained in the archives, and does not measure semantic diversity.

Final LLM-social revision depth is $3.50\pm0.72$ for GPT-OSS and $5.57\pm0.71$ for GLM, compared with $3.67\pm0.36$ and $5.67\pm0.36$ in their solo populations (Figure~\ref{fig:r4-lineages}a--b). Copying followed by further revision appears in the ancestry of $83.3\pm16.7\%$ and $56.7\pm6.1\%$ of final social deployments, respectively (Figure~\ref{fig:r4-lineages}c--d). This requires transmission of a non-initial revision, followed by another private revision somewhere in its subsequent lineage; the final user need not be the agent that made that revision. Merely copying the common initial file does not qualify. GPT-OSS social populations deploy one founding lineage at the endpoint, versus $4.83\pm0.17$ in solo; GLM deploys $2.33\pm0.21$ versus five (Figure~\ref{fig:r4-lineages}e--f). Social exchange therefore supports revision chains while concentrating the ancestry of the skills in use.

\begin{figure}[!tbp]
\centering
\includegraphics[width=.94\linewidth]{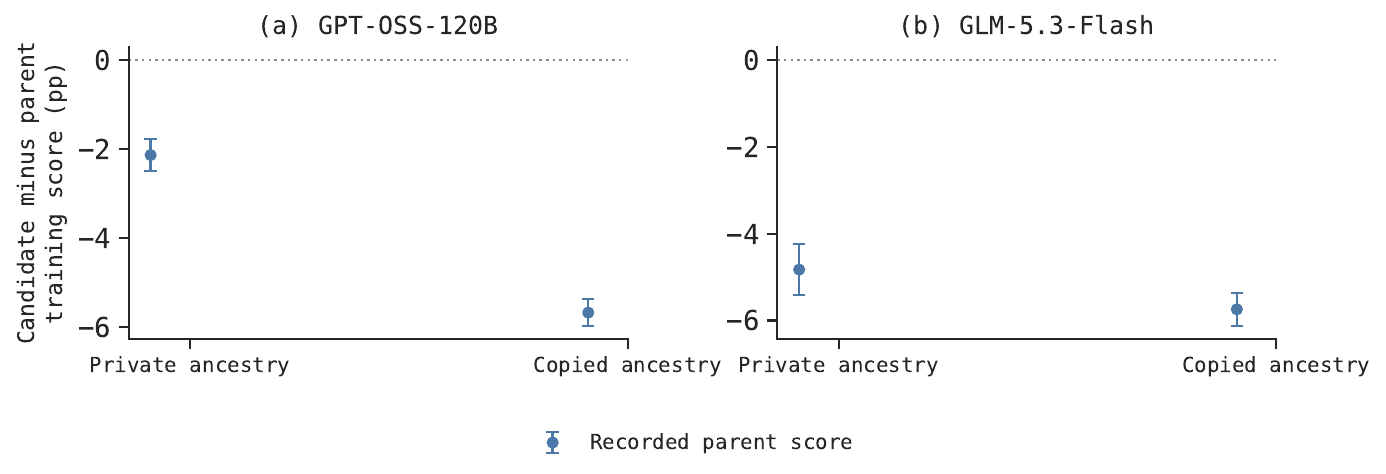}
\caption{Training-score changes from private versus socially derived parents in LLM-social populations. Panels (a)--(b) show GPT-OSS and GLM. Points and SE first average scored proposals within population seed. Parent fitness is the recorded score available before revision, not a simultaneous fresh evaluation. These are proposal gains, not gains of the best selected skill.}
\label{fig:r4-revision-gains}
\end{figure}

We also compare the training score of proposed revisions with the recorded score of their parents (Figure~\ref{fig:r4-revision-gains}). These averages include unsuccessful proposals rather than only retained winners. Revising a skill does not also reevaluate its parent: reevaluation is a separate learning action. The comparison therefore uses the parent score available to the learner, which can be optimistic after selection. Parent ancestry is chosen rather than randomized, so these contrasts describe the proposals made from each source, not the causal benefit of copying the same starting file. Held-out performance of selected descendants remains the generalization test. Mean proposed changes are below recorded parent fitness in both ancestry groups (Figure~\ref{fig:r4-revision-gains}a--b). This does not contradict improvement of the selected skill: most proposals can be worse than their parent while search retains the useful exceptions.

\subsection{Uncertainty across populations and test items}
\label{app:test-sampling}

The main error bands measure variability across six independent population seeds conditional on the same 200 held-out tasks. To examine sensitivity to those tasks, we retain task-level scores, average the five agents within each seed, and resample the 200 task identities with replacement. A task draw is shared across conditions, seeds, and checkpoints so their matching is preserved. We separately resample population seeds and tasks together. Each procedure uses 5,000 draws; the standard deviation of the bootstrap estimates is reported as SE. This assesses sampling of the existing tasks and populations, not a new task distribution or new stochastic executions.

\begin{figure}[!tbp]
\centering
\includegraphics[width=.94\linewidth]{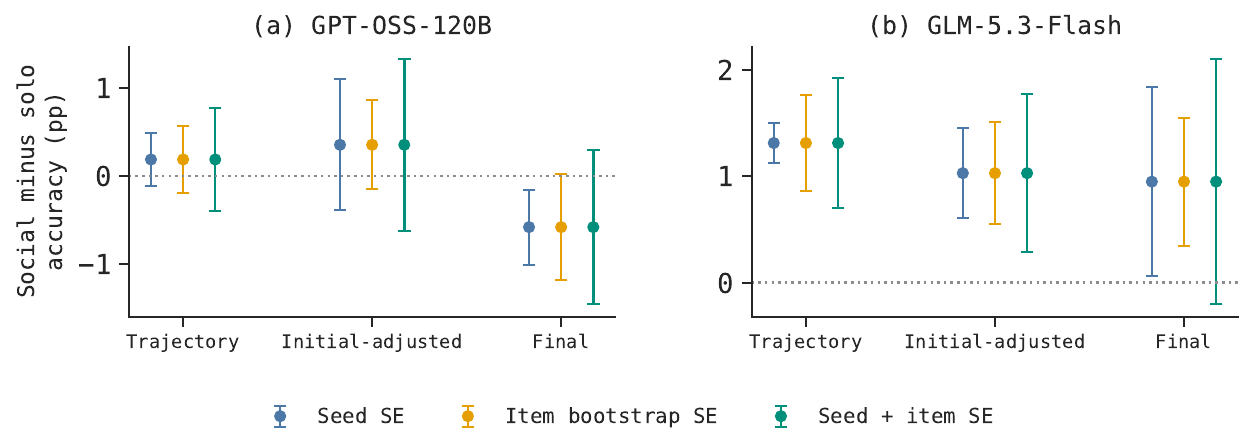}
\caption{Seed and test-item uncertainty for LLM-social minus LLM-solo whole-answer accuracy. Panels (a)--(b) show GPT-OSS and GLM. The same estimates are shown with conventional seed SE, item-only bootstrap SE, and joint seed-and-item bootstrap SE. Trajectory is the trapezoidal average over rounds 0, 10, 25, and 49; initial-adjusted subtracts each condition's starting accuracy.}
\label{fig:r4-sampling}
\end{figure}

For GLM, the social-minus-solo trajectory advantage is $1.31$ percentage points, with seed-only SE $0.19$, task-only bootstrap SE $0.45$, and joint bootstrap SE $0.61$ (Figure~\ref{fig:r4-sampling}b). After adjusting for initial accuracy, its $1.03$-point gain has SE $0.42$, $0.48$, and $0.75$, respectively (Figure~\ref{fig:r4-sampling}b). The final accuracy contrasts are $-0.58$ points for GPT-OSS and $+0.95$ for GLM, with joint SE $0.88$ and $1.15$ (Figure~\ref{fig:r4-sampling}a--b). The point estimate is unchanged, but consistent population seeds do not eliminate uncertainty from a finite test set.

\subsection{Private improvement and learning trajectories}

\begin{figure}[!tbp]
\centering
\includegraphics[width=\linewidth]{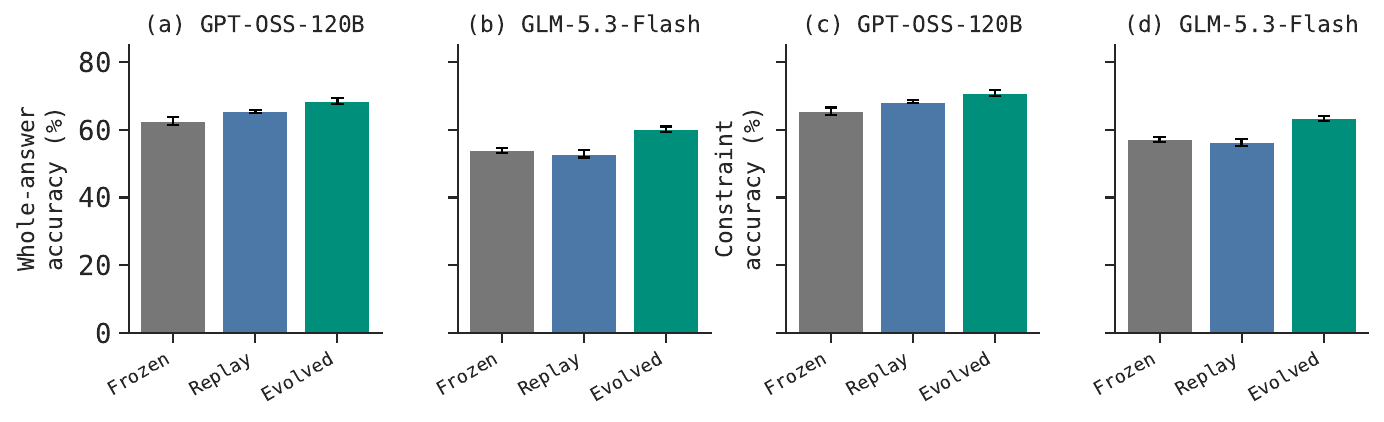}
\caption{Earlier single-agent weak-skill capability checks. Panels (a)--(b) show whole-answer accuracy for GPT-OSS and GLM; (c)--(d) show individual-constraint accuracy. Frozen and evolved use eight seeds each and the same 200 hidden tasks. Replay evaluates the starting file saved inside the evolved run: eight GPT-OSS seeds and seven complete GLM seeds. The remaining GLM replay is incomplete and excluded rather than averaged over a different task denominator. Error bars are seed SE. These runs are not pooled with the corrected population experiment.}
\label{fig:r4-offline}
\end{figure}

\textbf{Earlier standalone checks also found an improvement, but use a different baseline comparison.} In the single-agent weak-skill diagnostic, frozen and evolved GPT-OSS achieve $62.56\pm1.13\%$ and $68.44\pm0.97\%$ whole-answer accuracy; GLM achieves $54.00\pm0.74\%$ and $60.19\pm0.77\%$ (Figure~\ref{fig:r4-offline}a--b). Each condition uses eight seeds. The evolved runs train on the same fixed 100 examples; frozen and evolved skills are evaluated on the same 200 held-out tasks. They should not be read as the gain expected from every subsequent population protocol. In particular, a separately run frozen condition and a replay of the initial file inside an evolved run need not have identical measured accuracy. We show both references where complete replay is available, rather than attributing the whole difference between studies to evolution quality (Figure~\ref{fig:r4-offline}a--d).

\begin{figure}[!tbp]
\centering
\includegraphics[width=\linewidth]{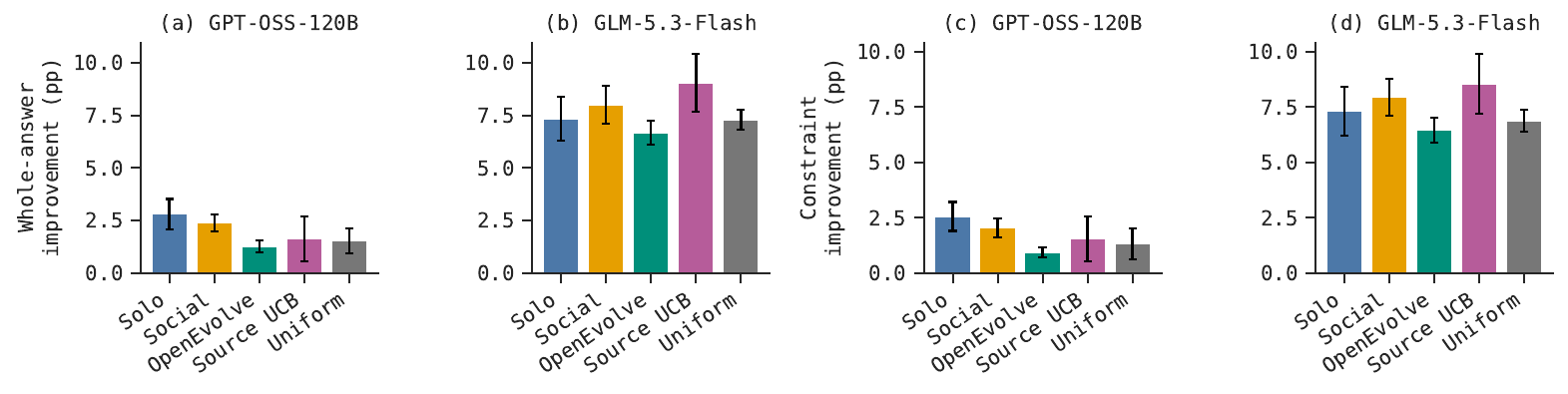}
\caption{Final minus initial held-out accuracy, paired within population seed. Panels (a)--(b) show whole-answer accuracy gains for GPT-OSS and GLM; (c)--(d) show gains on individual constraints. Bars show six-seed means with SE.}
\label{fig:r4-gains}
\end{figure}

\textbf{The corrected population study improves over its own starting skills.} Figure~\ref{fig:r4-gains} compares each condition's final selected skill with its matched initial skill, using the same hidden tasks and executor. It is a within-run improvement measure, not a comparison to a separately executed frozen condition. Whole-answer gains are positive for both models across the controllers shown, with corresponding gains on individual constraints (Figure~\ref{fig:r4-gains}a--d). The full solo OpenEvolve comparison establishes that the current population protocol can improve privately, while the LLM-solo comparison is the appropriate baseline for endogenous social learning.

\begin{figure}[!tbp]
\centering
\includegraphics[width=\linewidth]{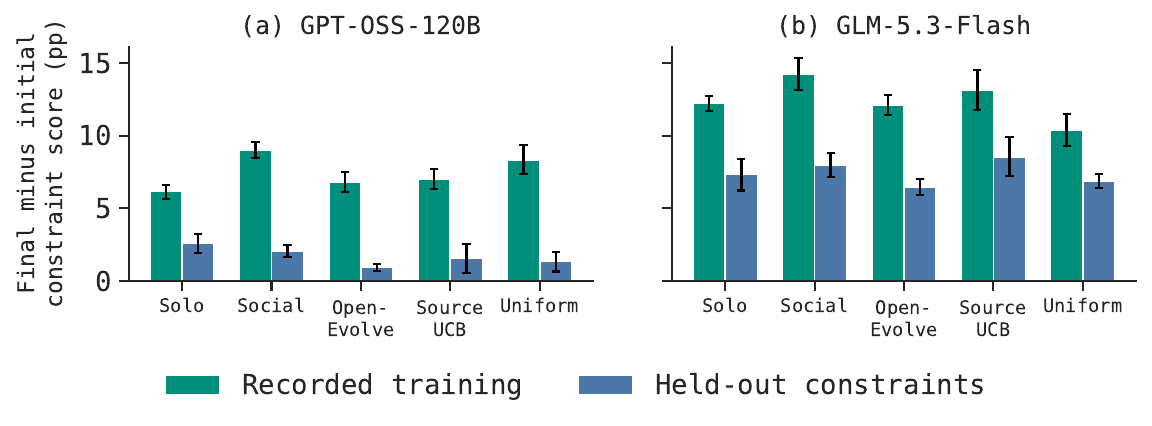}
\caption{Final-minus-initial constraint-accuracy gains on training and held-out examples. Panels (a)--(b) show GPT-OSS and GLM. Bars show matched within-population gains averaged across six seeds, with SE. Training gains are larger, but differences in token caps and reuse of selection scores prevent interpreting the gap as overfitting alone.}
\label{fig:r4-train-test}
\end{figure}

\begin{figure}[!tbp]
\centering
\includegraphics[width=.94\linewidth]{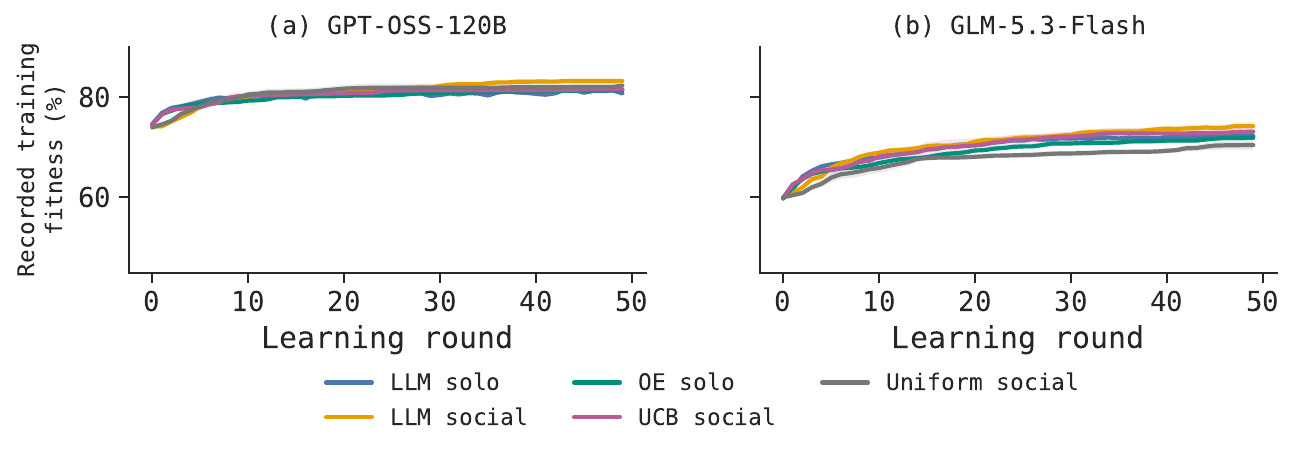}
\caption{Recorded training fitness of deployed skills over learning, for GPT-OSS (a) and GLM (b). This is the score available to the learner, not an independently re-estimated hidden score. All curves average agents within population and show six-seed mean and SE.}
\label{fig:r4-training}
\end{figure}

\begin{figure}[!tbp]
\centering
\includegraphics[width=.94\linewidth]{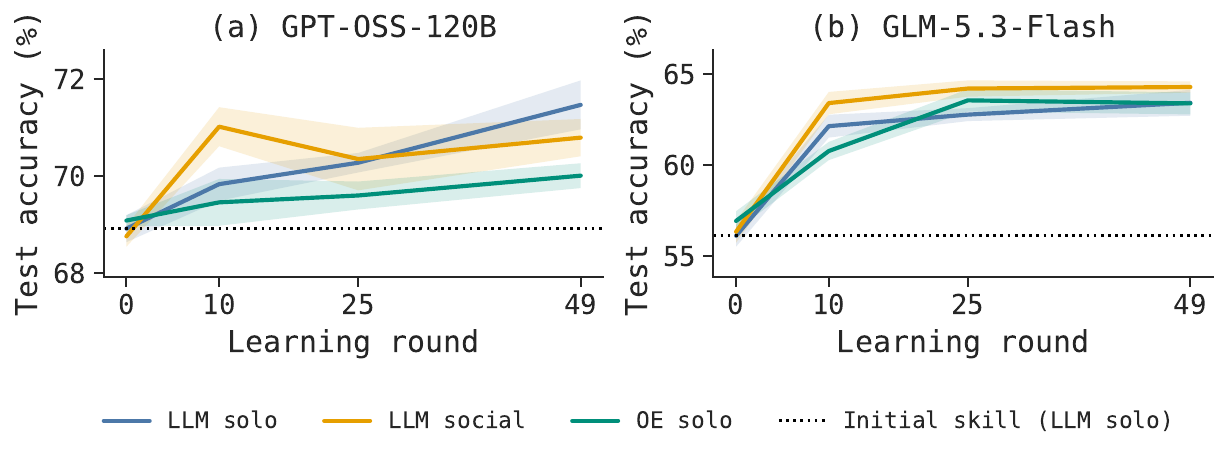}
\caption{Held-out individual-constraint accuracy at the same checkpoints as the main whole-answer metric. Panels (a)--(b) show GPT-OSS and GLM. The dotted reference is the initial LLM-solo skill, and shaded regions are population-seed SE.}
\label{fig:r4-constraints}
\end{figure}

\textbf{Training fitness and held-out performance are different outcomes.} Figure~\ref{fig:r4-training} tracks the saved training fitness of deployed skills. These values guide selection and may be optimistic because the same examples are reused throughout search. The hidden trajectories in Figure~\ref{fig:r4-constraints} test whether the selected procedures also work on examples that supplied no feedback; improved training rank alone is not evidence of better generalization. Figure~\ref{fig:r4-train-test} compares gains using the individual-constraint metric on both splits. Training evaluations allow 8,704 completion tokens per example, whereas held-out execution allows 32,768; recorded training scores are also reused rather than freshly remeasured at every checkpoint. Consequently, the contrast combines selection, data, and evaluation-protocol differences.

\begin{figure}[!tbp]
\centering
\includegraphics[width=.94\linewidth]{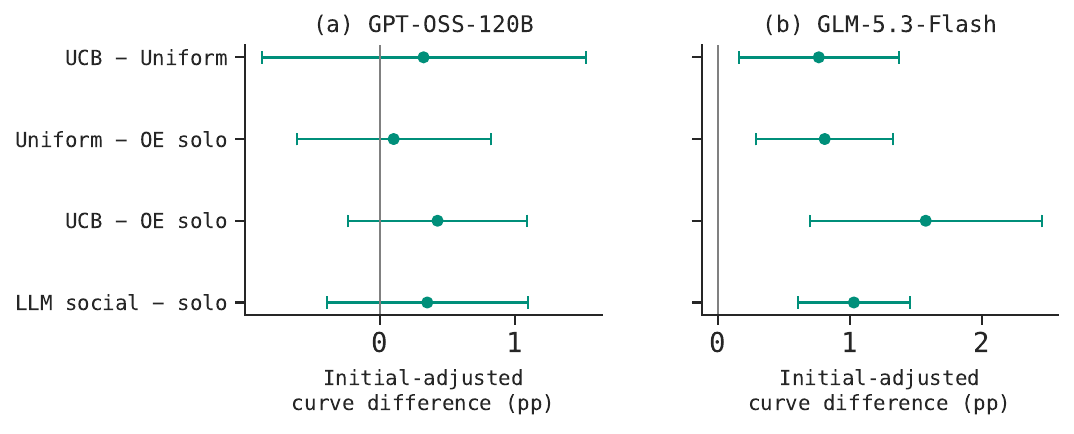}
\caption{Trajectory-average contrasts after subtracting each run's initial whole-answer accuracy. Panels (a)--(b) show GPT-OSS and GLM. Points and error bars show the paired mean and SE across six populations. This controls the starting offset, not every possible difference in subsequent search.}
\label{fig:r4-initial-adjusted}
\end{figure}

An initially stronger social population could create an apparent speed advantage without learning more. To check this, we subtract each population's starting accuracy before comparing its trajectory with the matched solo run. GLM retains an average improvement advantage of $1.03\pm0.42$ percentage points (Appendix Figure~\ref{fig:r4-initial-adjusted}b), indicating that the pattern is not simply an initial offset.

Training gains are larger than held-out gains across controllers (Appendix Figure~\ref{fig:r4-train-test}a--b). For GPT-OSS, LLM-social training constraint accuracy rises by $9.02\pm0.55$ points, versus $2.03\pm0.42$ on held-out constraints; GLM gains $14.25\pm1.11$ and $7.95\pm0.84$ points, respectively (Appendix Figure~\ref{fig:r4-train-test}a--b). Thus, improving the score used for selection does not translate one-for-one into better test performance. Training scores are reused during selection and evaluated under a different per-example token cap, so this difference is not a clean estimate of overfitting alone.

\begin{figure}[!tbp]
\centering
\includegraphics[width=\linewidth]{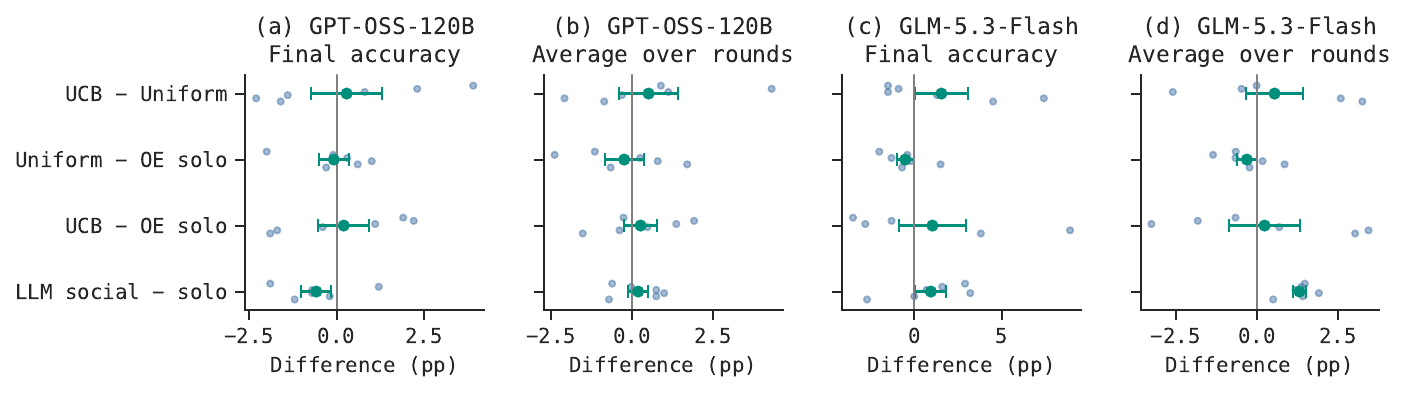}
\caption{Paired controller differences in percentage points. Panels (a)--(b) show GPT-OSS final and trajectory-average accuracy; (c)--(d) show the corresponding GLM contrasts. Small points are matched population seeds; larger points and whiskers are the paired mean and SE. UCB and uniform are compared with full solo OpenEvolve, and with each other.}
\label{fig:r4-paired}
\end{figure}

\textbf{Learning speed and final accuracy answer different questions.} We integrate held-out accuracy with the trapezoidal rule between rounds 0, 10, 25, and 49, then divide by 49 to retain accuracy units. Figure~\ref{fig:r4-paired} presents paired final and integrated differences rather than asking the reader to infer uncertainty from overlapping marginal error bars. GLM's LLM-social trajectory is higher on average in every population seed, even though the final contrast is uncertain (Figure~\ref{fig:r4-paired}c--d). This integral includes initial performance differences and assumes linear interpolation between measured checkpoints; it does not locate the exact round when a target accuracy was first reached. The constraint-level trajectories provide a complementary outcome rather than an independent replication (Figure~\ref{fig:r4-constraints}).

\begin{figure}[!tbp]
\centering
\includegraphics[width=\linewidth]{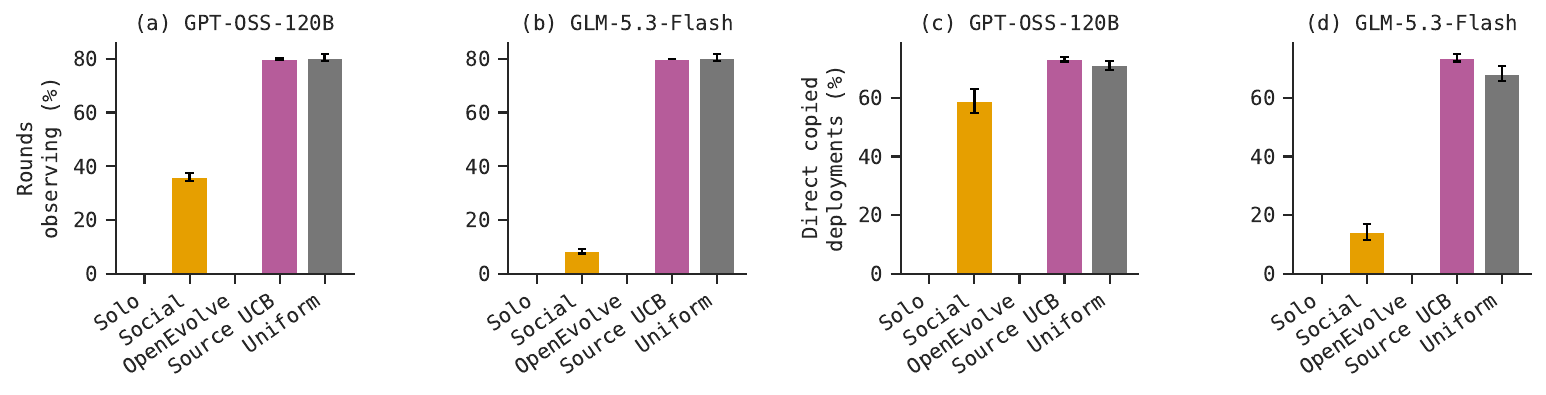}
\caption{Overall observation and copied-file deployment rates. Panels (a)--(b) show observation as a fraction of learning rounds; (c)--(d) show deployments of directly acquired social files. Denominators include all post-initialization agent-rounds. Bars show population-seed mean and SE.}
\label{fig:r4-behavior-summary}
\end{figure}

\textbf{Initial differences do not explain the entire GLM curve advantage.} Subtracting each condition's own initial accuracy before integrating leaves a GLM social-minus-solo difference of $1.03\pm0.42$ percentage points (Figure~\ref{fig:r4-initial-adjusted}b). This sensitivity check measures improvement above the starting point, whereas the main curve comparison measures attained accuracy. Both are paired across population seeds.

\begin{figure}[!tbp]
\centering
\includegraphics[width=\linewidth]{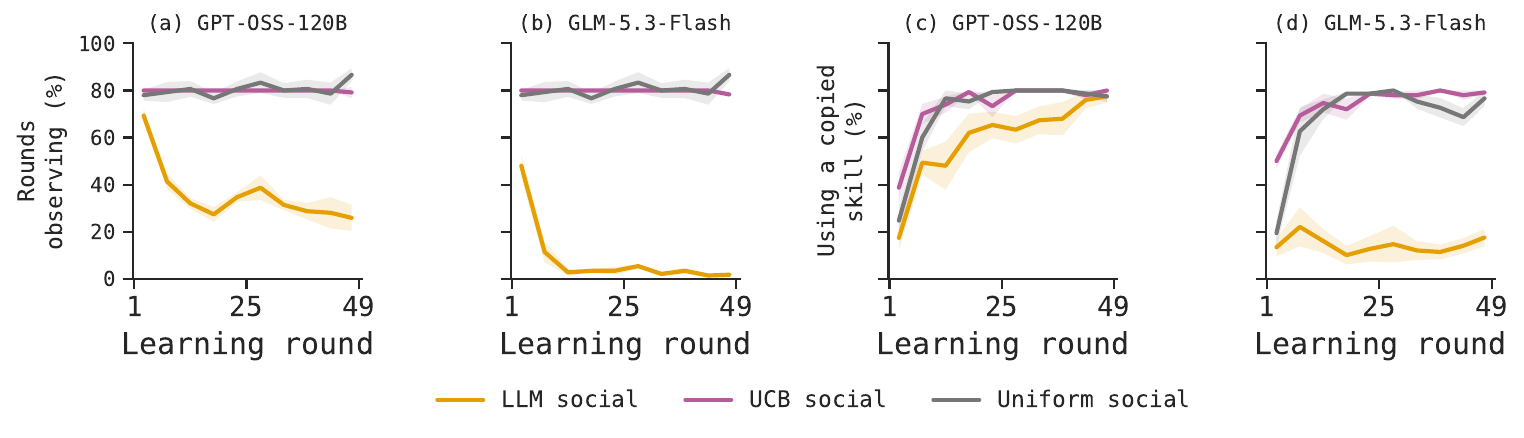}
\caption{Observation and use of copied skills over learning. Panels (a)--(b) show the fraction of rounds spent observing; (c)--(d) show deployment of a file first acquired socially. Curves average non-overlapping five-round blocks within seed (four rounds in the last block), then show seed mean and SE. Revising a copied file creates a locally acquired descendant, which is not counted as direct copied-file deployment.}
\label{fig:r4-behavior}
\end{figure}

\begin{figure}[!tbp]
\centering
\includegraphics[width=\linewidth]{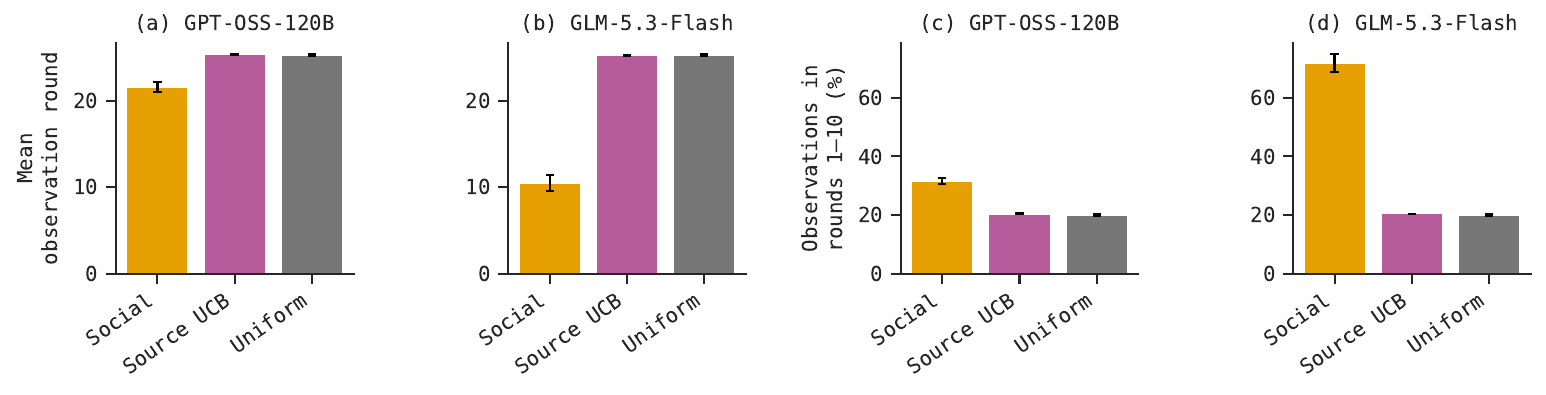}
\caption{When observations occur. Panels (a)--(b) show the mean observation round; (c)--(d) show the percentage of all observations occurring in rounds 1--10. Each statistic is computed within a population before averaging across six seeds. Error bars are seed SE.}
\label{fig:r4-timing}
\end{figure}

\begin{figure}[!tbp]
\centering
\includegraphics[width=\linewidth]{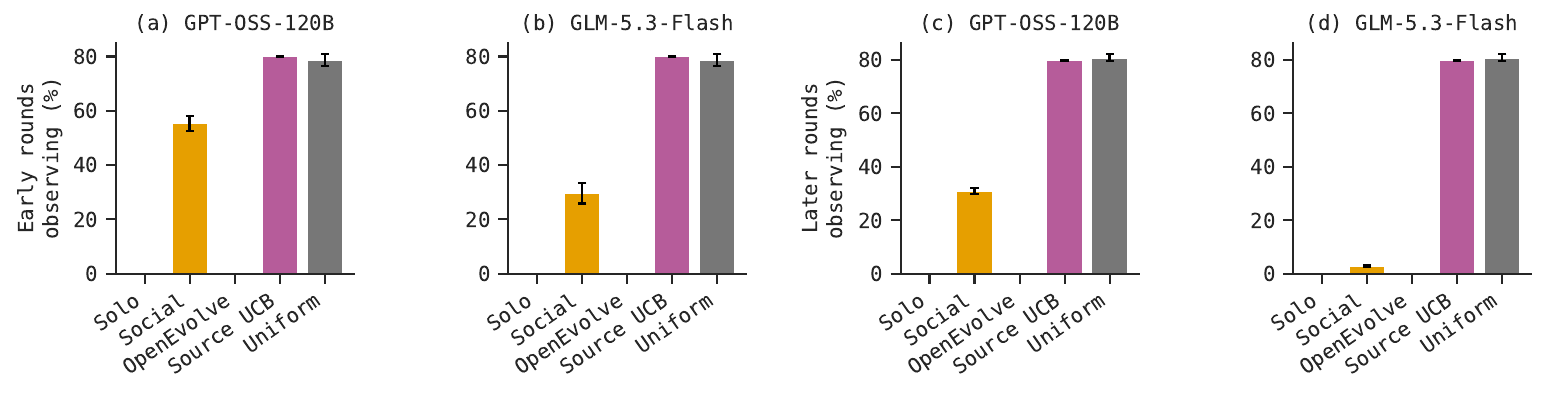}
\caption{How often agents observe within a time interval. Panels (a)--(b) divide observations by the available agent-rounds in rounds 1--10; (c)--(d) use rounds 11--49. Unlike Figure~\ref{fig:r4-timing}c--d, the denominator includes rounds with no observation. Bars show mean and SE across population seeds.}
\label{fig:r4-early-late}
\end{figure}

\textbf{GLM's observations are concentrated earlier, not merely less frequent.} Conditional on an observation occurring, its mean round is $10.45\pm0.92$ for GLM and $21.55\pm0.56$ for GPT-OSS (Figure~\ref{fig:r4-timing}a--b). This conditions on observations: an agent that never observes does not contribute an event to this timing statistic. The complementary early/late rate instead divides observations by all available agent-rounds in each interval (Figure~\ref{fig:r4-early-late}). These denominators answer different questions. A high fraction of observations occurring early does not imply that most early rounds are spent observing.

\begin{figure}[!tbp]
\centering
\includegraphics[width=\linewidth]{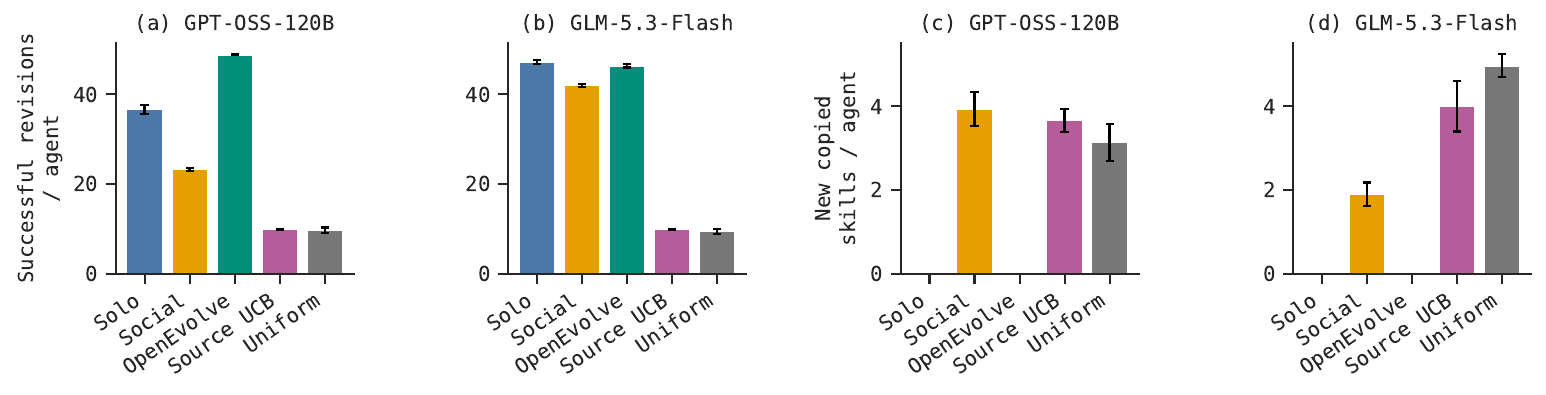}
\caption{Private and social additions to skill libraries. Panels (a)--(b) show successful revisions per agent; (c)--(d) show newly acquired social skill versions per agent. Repeated observations of an already acquired file do not count as new acquisitions. Agents are averaged within population before computing six-seed means and SE.}
\label{fig:r4-search}
\end{figure}

\textbf{Frequent observation reduces the supply of private revisions.} The single learning-action budget means that observation can replace a revision. Figure~\ref{fig:r4-search} separates successful private revisions from newly imported skill versions, rather than treating every repeated observation as a new discovery. This helps explain why a controller can spend less without improving more. It does not establish that every displaced revision would have produced a useful skill, or that the observed differences between model families are caused only by copying. UCB and uniform selection also produce nearly the same overall observation frequency in this horizon (Figure~\ref{fig:r4-behavior-summary}a--b). Their uncertain accuracy difference therefore does not establish that directing social attention is generally useless; it evaluates this particular source rule and the behavior it actually produces.

\begin{figure}[!tbp]
\centering
\includegraphics[width=\linewidth]{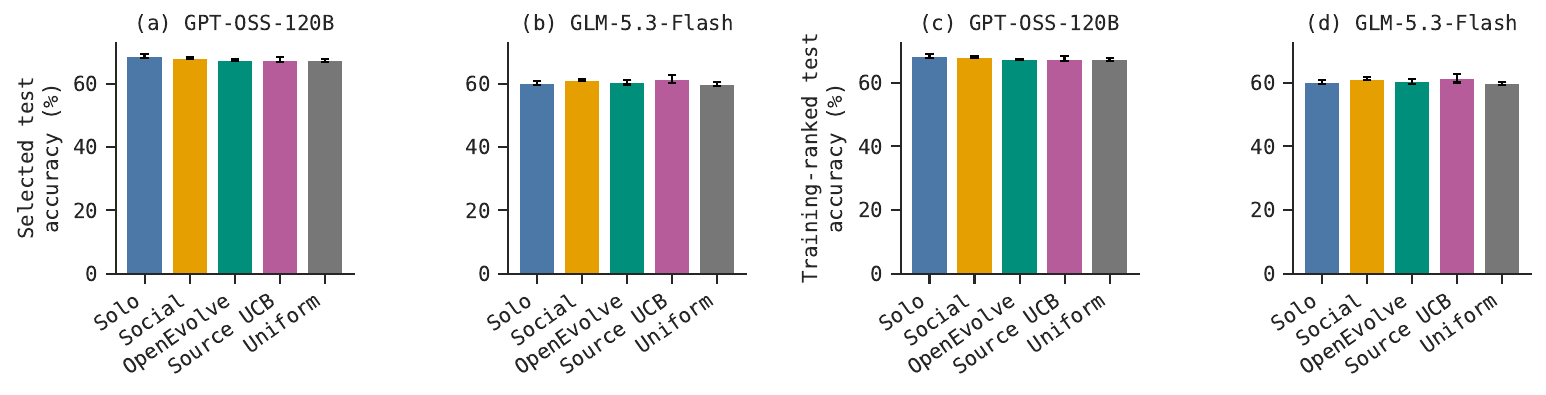}
\caption{Saved selected skills versus training-ranked alternatives. Panels (a)--(b) show selected-skill test accuracy; (c)--(d) show alternative-skill accuracy. Both use the same held-out tasks. Bars are population means with SE; these marginal bars do not by themselves estimate a paired-effect SE.}
\label{fig:r4-selection}
\end{figure}

\textbf{A training-ranked alternative does not reveal a large missed deployment gain.} We replay each agent's training-ranked alternative on the same hidden tasks and executor used for its selected skill. Selected and alternative performance are close for the LLM-controlled policies (Figure~\ref{fig:r4-selection}a--d). For algorithmic archive policies, the selected skill already follows training rank, so equality is expected rather than a separate validation. This diagnostic is limited to the saved, training-selected alternative: it does not test every peer file, select a hindsight winner using test results, or prove that no beneficial social transfer was possible.

\begin{figure}[!tbp]
\centering
\includegraphics[width=.94\linewidth]{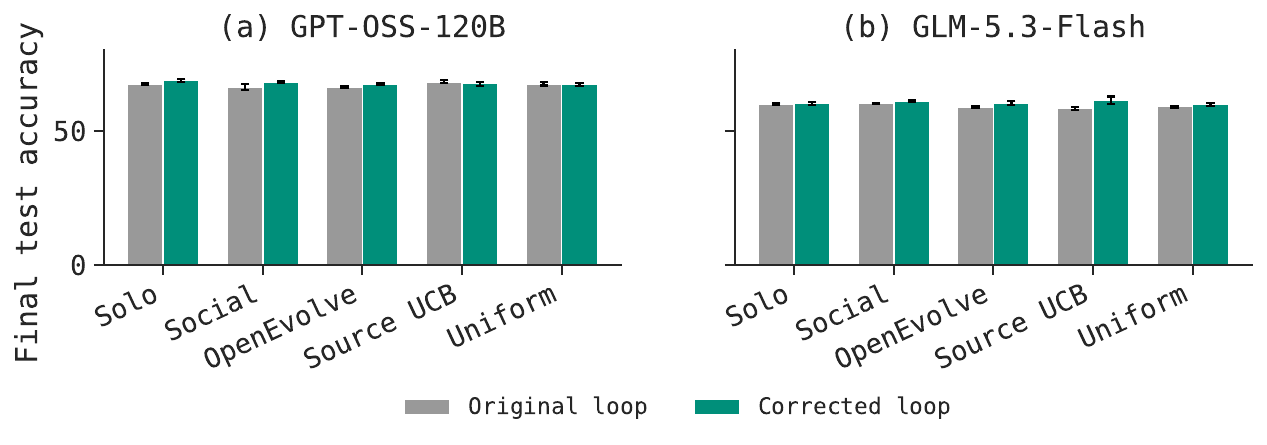}
\caption{Original versus corrected population experiments, displayed separately. Panels (a)--(b) show GPT-OSS and GLM final whole-answer accuracy. Each bar uses six population seeds with SE. These are different implementations, not additional interchangeable replicates.}
\label{fig:r4-versions}
\end{figure}

\textbf{Earlier and corrected implementations must not be pooled.} The corrected loop repaired parent-fitness feedback, archive lineage and random-state restoration, and seed/task-order handling. The original population loop also included one additional post-initialization decision. Figure~\ref{fig:r4-versions} preserves both sets of endpoints to make this change visible. Because several implementation details changed together, their difference is not a controlled estimate of any one fix. Main-text claims use only the corrected experiment.

\textbf{Decoding and selection.} Both skill-evolution models use temperature 0.6 and top-$p$ 0.95, with up to 32,768 tokens per execution call and medium reasoning effort where supported by the configured provider. The training set is fixed throughout learning. Parent selection, mutation, evaluation, and deployment use only training feedback. OpenEvolve's archive and island migrations operate within an agent; cross-agent transmission is permitted only by the specified social action. Thus diversity maintained inside an archive is distinct from the decision to seek another agent's work. In full solo OpenEvolve, the archive chooses which skills to revise, keeps alternative skills in separate subpopulations, and executes the highest-scoring skill.

\section{Learning expense and accounting}
\label{app:performance-cost}

\begin{table*}[!tbp]
\centering
\small
\setlength{\tabcolsep}{5pt}
\caption{Resource use for LLM-social minus LLM-solo populations. Tokens include both prompts and completions and are reported in millions. ``Learning'' includes search, revision, selection, and training-time execution; ``all'' additionally includes every held-out evaluation. USD uses the provider-reported cost for the same calls. Effects are paired means $\pm$ SE across six seeds.}
\label{tab:r4-resources}
\begin{tabular}{lrrrr}
\toprule
Model & Learning tokens & All tokens & Learning USD & All USD \\
\midrule
GPT-OSS-120B & $-17.34\pm2.08^{***}$ & $-22.80\pm9.04$ n.s. & $-2.25\pm0.15^{***}$ & $-2.13\pm0.19^{***}$ \\
GLM-5.3-Flash & $-1.74\pm3.32$ n.s. & $-4.64\pm11.67$ n.s. & $-0.51\pm0.23$ n.s. & $-0.30\pm0.27$ n.s. \\
\bottomrule
\multicolumn{5}{l}{\footnotesize $^{***}p<0.001$; n.s.: $p\geq0.05$. Negative values mean social uses fewer resources.}
\end{tabular}
\end{table*}

\subsection{Comparison at matched learning tokens}
\label{app:matched-spending}

We compare LLM-social and LLM-solo at matched cumulative learning tokens separately within every model and population seed. For each pair, we use the smaller final token expenditure, provided it lies in both observed ranges. We linearly interpolate accuracy between each run's neighboring checkpoints and subtract solo from social. This includes all six seeds for each model, without extrapolation. The target can differ between seeds: this is a within-pair comparison of realized learning curves, not a prospectively assigned common budget. Tokens include prompts and completions across all five agents, including initialization and the private search that produced copied files, but exclude hidden evaluation. The reconstructed final totals agree exactly with the learning ledgers for all 60 runs. Tokens are our accounting unit, not a hardware-independent measure of FLOPs.

\begin{figure}[!tbp]
\centering
\includegraphics[width=.94\linewidth]{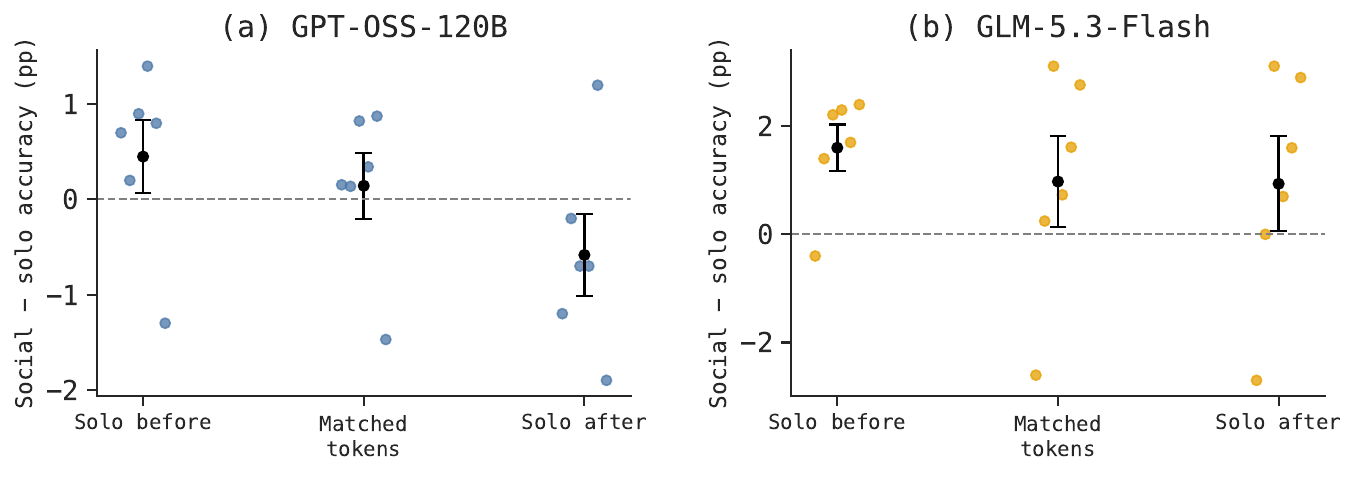}
\caption{Matched learning-token comparison from saved checkpoints. Panels (a)--(b) show GPT-OSS and GLM. Colored points are paired population-seed contrasts; black points and whiskers are means and SE. The middle comparison interpolates both policies at their common token endpoint. The neighboring comparisons retain social at that target and use solo's preceding or following checkpoint instead. Tokens include prompts and completions across the population, excluding hidden evaluation. No curve is extrapolated.}
\label{fig:r4-matched-spending}
\end{figure}

At each pair's common endpoint, the estimated social advantage $\Delta J(C)$ in whole-answer accuracy is $+0.14\pm0.35$ percentage points for GPT-OSS and $+0.98\pm0.85$ for GLM (Figure~\ref{fig:r4-matched-spending}a--b). Neither difference is statistically significant. Figure~\ref{fig:r4-matched-spending} also substitutes the solo checkpoints immediately before and after the target to show sensitivity to interpolation. These are alternative observed comparisons, not statistical bounds: accuracy need not change monotonically between checkpoints. These estimates do not erase the observed savings; they distinguish saving computation over a completed run from outperforming independent learning at the same expenditure. SE and paired tests for this comparison are across six population seeds, conditional on the fixed held-out tasks.

We apply the same within-seed interpolation to each algorithmic social controller and full solo OpenEvolve. At equal learning tokens, UCB minus OpenEvolve accuracy is $+0.79\pm0.76$ points for GPT-OSS and $+3.27\pm1.77$ for GLM; uniform minus OpenEvolve is $+0.58\pm0.90$ and $+1.56\pm0.36$, respectively (Figure~\ref{fig:r4-efficiency-tokens}). The GLM uniform contrast has nominal paired $p<0.01$; the other three contrasts are not significant. These comparisons concern the complete source policies, including their different rates of observing rather than revising. They do not isolate the value of selecting one peer rather than another.

\subsection{Why compare performance at matched computation?}
\label{app:metric-motivation}

Consider two students preparing for the same exam. A student studying alone scores 80\% after five hours and 82\% after ten. A student learning from a friend stops after five hours and scores 80\%. Dividing final score by study time makes the second student look almost twice as efficient, yet both reach 80\% at five hours. Avoiding the extra study is useful, but it does not establish better learning at the same budget. The same distinction applies to cumulative reward per token: a ratio compares average returns over the work each policy chose to do, whereas a matched-budget comparison asks what each could reach after equal expenditure. Cumulative performance also measures success throughout learning rather than performance at a particular endpoint; we retain both views without treating them as interchangeable.

In our experiments, the relevant comparison is an interacting population against an equally sized population of independent learners. All of their learning tokens count, including the search that makes a copied skill available. Thus, parallel search is available to both sides, and copying does not receive discoveries for free. A positive $\Delta J(C)$ supports the social learning procedure over the tested independent baseline; it does not imply that no individual algorithm could achieve the same result. Held-out execution costs remain a separate consideration, since equal learning expenditure need not imply equally short test-time answers.

The bandit baselines provide a concrete example of effective social learning. Solo UCB and hierarchical social UCB use no LLM inference tokens, so their token expenditure is identical: zero. We therefore compare cumulative reward directly over the same rounds and environment seeds, rather than dividing by zero to compute reward per token. Hierarchical UCB earns $25.68\pm2.87$ more cumulative reward per agent over 100 rounds (Appendix Figure~\ref{fig:r1-cumulative}d). Social observations let it use information from its peers' pulls instead of having to gather every payoff estimate through its own search. This illustrates the advantage we seek: access to others' discoveries improves performance without additional LLM inference expenditure. The algorithms still perform ordinary computation; zero here refers specifically to our token accounting, not to zero computational work.

When skills can evolve, recursive social improvement requires a further step: agents must build on what they copy. One agent might discover a useful procedure, another copy and revise it, and a third improve it again. Spreading discoveries cheaply is valuable, but our criterion asks whether these chains of revision and reuse let the population reach better performance than equally resourced solo learners. The goal is not merely to share what has already been found, but to make further progress by using others' discoveries as starting points.

\subsection{Token allowances and realized use}
\label{app:budget-audit}

Every learning action, including \textsc{Keep}, costs the tokens used to make the decision, but \textsc{Keep} avoids the cost of revising, evaluating, or observing. The library summary shown at each decision also counts against the budget. Only the most recently observed skill is shown in full, which matches the social information available in the bandit and keeps prompts from growing with every copy.

The per-agent round allowance is $1{,}001{,}472=100\times8{,}704+4\times32{,}768$ completion tokens: room for one full training evaluation and up to four calls for learning selection, revision, deployment selection, and execution. Controllers share this allowance, but need not use every call. Each training example has an 8,704-token cap; revision, controller, and execution calls have separate caps of up to 32,768. Hidden tests use 32,768 tokens per example outside the learning allowance. This generous aggregate limit allows full-batch search; it is not intended to reproduce the tight finite-bandit budget.

\begin{figure}[!tbp]
\centering
\includegraphics[width=.94\linewidth]{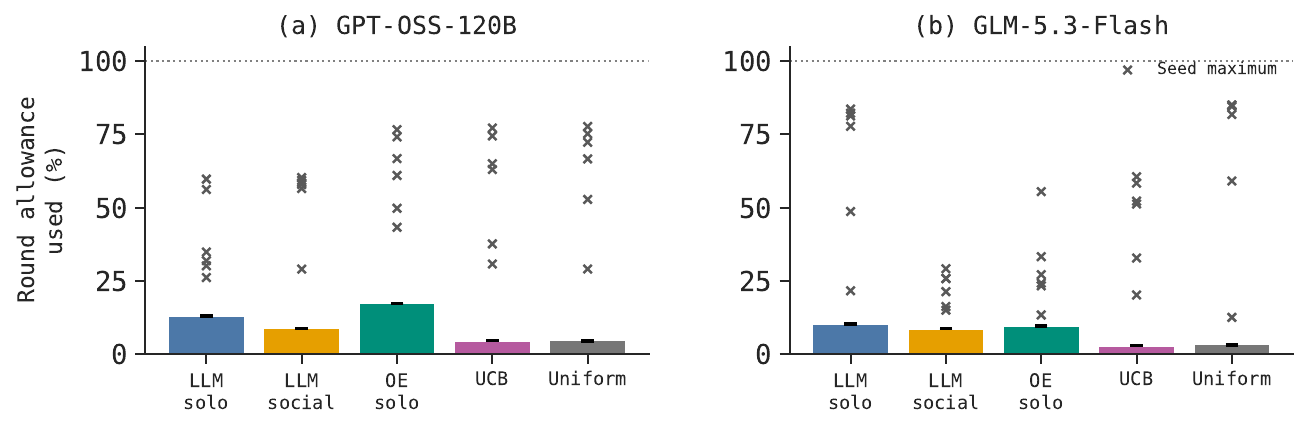}
\caption{Realized completion-token use as a percentage of the round allowance. Panels (a)--(b) show GPT-OSS and GLM. Bars average all initialization and learning agent-rounds within seed, then show seed means and SE. Crosses show each seed's maximum agent-round fraction, all below the dotted aggregate limit. This does not measure individual-call cap hits or the tokens reserved before a call.}
\label{fig:r4-budget}
\end{figure}

No completed agent-round exhausts the aggregate allowance (Figure~\ref{fig:r4-budget}). Mean use is $12.93\pm0.23\%$ in GPT-OSS LLM-solo and $8.75\pm0.12\%$ in LLM-social; GLM uses $10.22\pm0.32\%$ and $8.53\pm0.13\%$, respectively (Figure~\ref{fig:r4-budget}a--b). Individual calls can still reach their own caps or return invalid output, and budget reservations can constrain which calls may begin. Thus an unused aggregate balance does not prove that every attempted revision or answer had sufficient tokens. These records cover completed rounds, not all abandoned attempts or provider retries in the billing ledger.

\subsection{Reading learning costs}

We report performance together with the population's cumulative token use, $C_T(\boldsymbol\pi)$, and distinguish efficiency, $E(\boldsymbol\pi)$, from the matched-cost social advantage, $\Delta J(C)$ (Section~\ref{sec:formulation}). The comparison is direct: does a policy perform better, use fewer resources, or do both? Agents receive the same token limits but may use very different amounts because observing a saved skill is cheaper than running another private search. Showing performance and computation separately prevents a ratio from hiding that distinction.

For evolving skills, the optimum over possible revisions is unknown, so we compare held-out accuracy at saved checkpoints, its average over learning rounds, and its relationship to token use and API expense. Learning-only measures include every call that can affect the evolving skill; end-to-end measures add the held-out generations. Both sum across the full population, including the private search performed by agents whose skills are later copied. The learning curves below use learning-only resources, while Table~\ref{tab:r4-resources} reports both definitions from the persistent provider ledgers.

Prompt and completion tokens must be counted together. Completion-only accounting suggests that both social policies are cheaper, but Table~\ref{tab:r4-token-components} shows why this is misleading for GLM: fewer generated tokens are offset by longer prompts. USD is another complementary measure because providers price prompt and completion tokens differently. We therefore use total tokens and reported cost for the main resource claim rather than choosing whichever measure produces the cleanest contrast.

\begin{table*}[!tbp]
\centering
\small
\setlength{\tabcolsep}{4pt}
\caption{Learning-token decomposition for LLM-social minus LLM-solo populations, in millions of tokens. Prompt and completion columns sum to the total column. Values are paired means $\pm$ SE across six seeds.}
\label{tab:r4-token-components}
\begin{tabular}{lrrr}
\toprule
Model & Prompt & Completion & Total \\
\midrule
GPT-OSS-120B & $-6.87\pm1.55^{**}$ & $-10.46\pm0.74^{***}$ & $-17.34\pm2.08^{***}$ \\
GLM-5.3-Flash & $+2.51\pm2.75$ n.s. & $-4.25\pm0.76^{**}$ & $-1.74\pm3.32$ n.s. \\
\bottomrule
\multicolumn{4}{l}{\footnotesize $^{**}p<0.01$, $^{***}p<0.001$; n.s.: $p\geq0.05$.}
\end{tabular}
\end{table*}

\begin{figure}[!tbp]
\centering
\includegraphics[width=.94\linewidth]{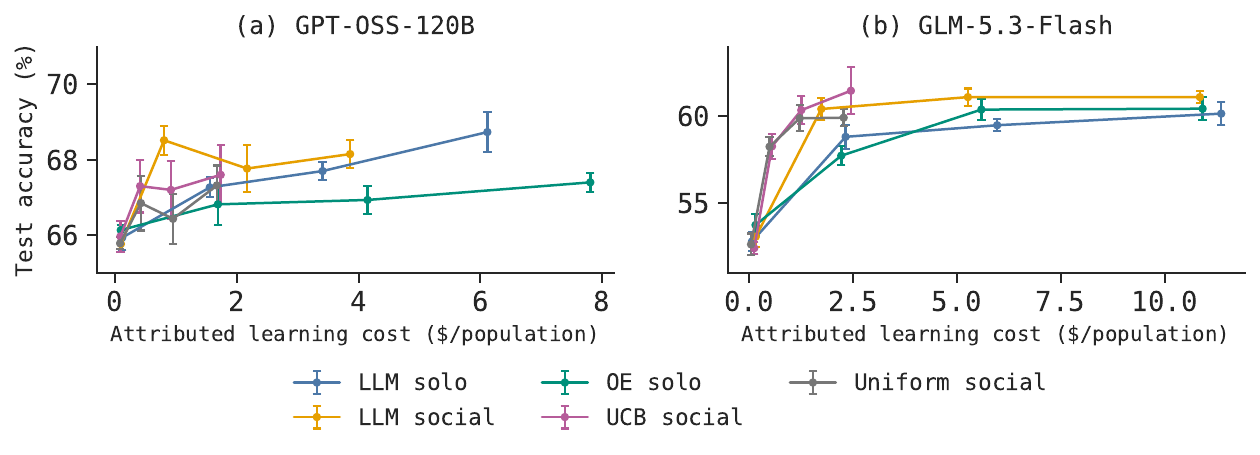}
\caption{Performance and spending during skill evolution. Panels (a)--(b) show GPT-OSS and GLM on separate vertical scales for comparison with each model's own solo policy. Each point is a saved checkpoint at mean cumulative learning expense per five-agent population, with vertical accuracy SE across six seeds. Higher and further left is preferable. Costs include private discovery throughout the population and exclude hidden evaluation. OE denotes OpenEvolve.}
\label{fig:r4-efficiency-cost}
\end{figure}

\begin{figure}[!tbp]
\centering
\includegraphics[width=.94\linewidth]{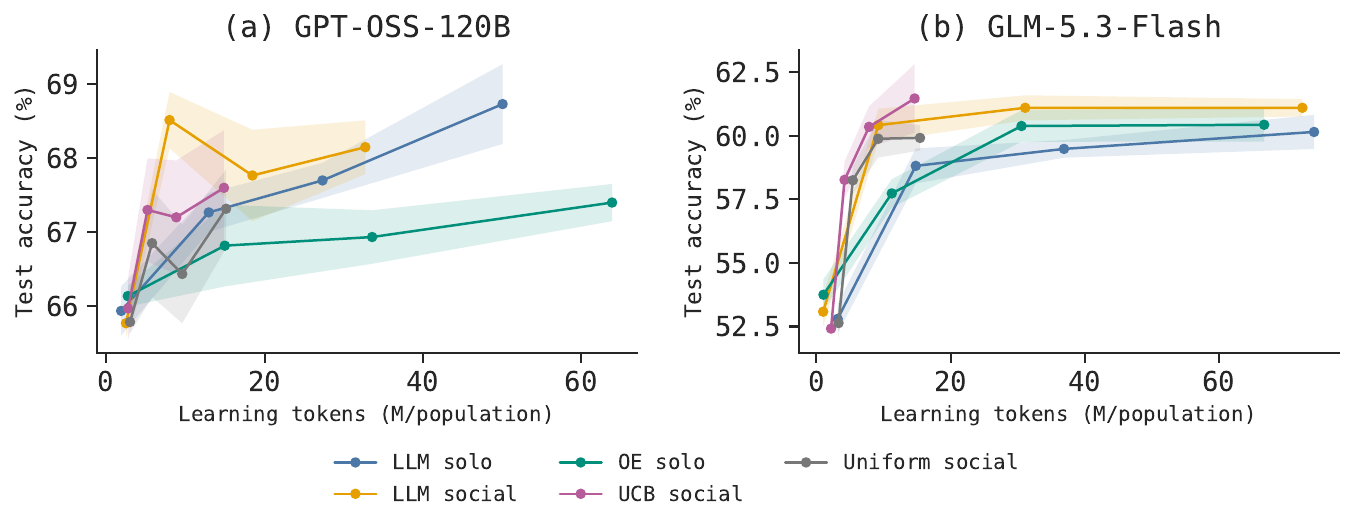}
\caption{Held-out accuracy against cumulative learning tokens, including prompts and completions across all five agents and excluding hidden evaluation. Panels (a)--(b) show GPT-OSS and GLM, with separate vertical scales. Points are saved checkpoints at mean token expenditure; shading shows accuracy SE across six population seeds. Matched-token contrasts are computed within each seed, not from these averaged curves.}
\label{fig:r4-efficiency-tokens}
\end{figure}

Figure~\ref{fig:r4-efficiency-cost} places the saved checkpoints on performance--cost axes; Appendix Figure~\ref{fig:r4-efficiency-tokens} gives the corresponding total-learning-token view. These comparisons characterize how the models divide effort between improvement and reuse. They do not show that social learning universally improves the tradeoff, or that either model is on the optimal frontier. Rather, they identify different outcomes that a binary classification of social learning as helpful or harmful would obscure.

\textbf{Reading the performance--spending plot.} Follow a single color from left to right: its four points summarize the skills selected at rounds 0, 10, 25, and 49, evaluated on the same held-out task set. The horizontal position is the mean total learning expense accumulated by the population; the vertical position is its mean test accuracy. Compare social and solo within each model's panel. Further left means less spending; higher means better accuracy. Separate vertical scales make within-model changes visible, so slopes should not be compared across panels. The main-text bars provide the simpler final-accuracy and token-use comparison.
\label{app:costs}

\begin{figure}[!tbp]
\centering
\includegraphics[width=.94\linewidth]{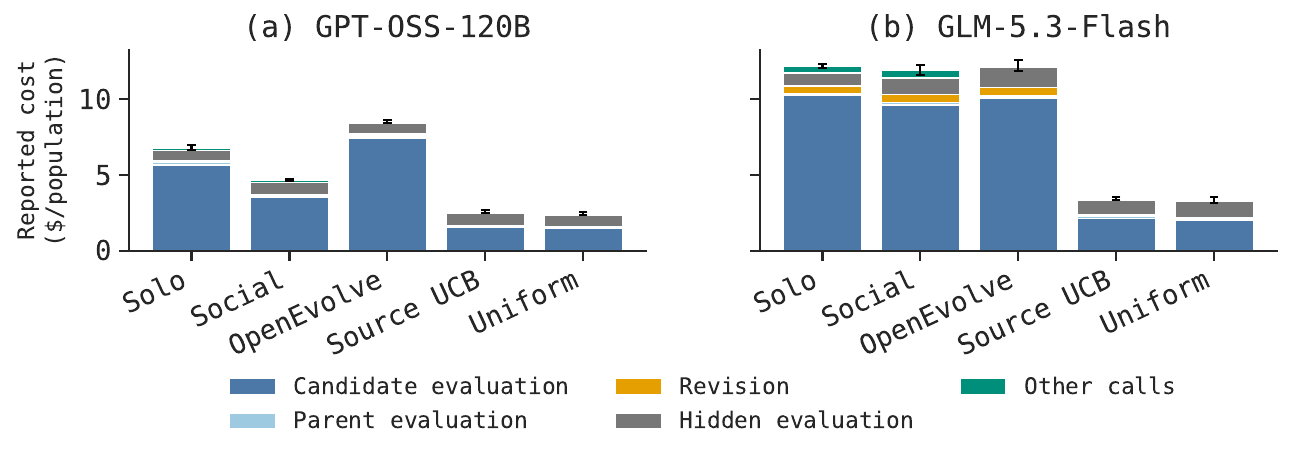}
\caption{Reported API expense by phase, averaged over six five-agent populations per condition; error bars show SE of the total. Panels (a)--(b) show GPT-OSS and GLM. Hidden evaluation is included here but excluded from the matched-learning-token comparison in Appendix~\ref{app:matched-spending}. Unconfirmed usage reservations are not attributed to these reported-cost components.}
\label{fig:r4-costs}
\end{figure}

Figure~\ref{fig:r4-costs} separates candidate evaluation, parent evaluation, revision proposals, hidden evaluation, and remaining calls. Candidate evaluation dominates the expensive private-search conditions, while source policies often replace it with observation. The learning-cost comparison excludes hidden evaluation because it supplies no learning feedback; the end-to-end comparison includes it because evaluation-time verbosity still consumes resources. Costs are reported API charges, not local cluster compute costs or an estimate of local GPU expenditure. Calls whose usage was not returned retain conservative reservations and are not treated as free.

\textbf{A round is not a fixed amount of computation.} Figures~\ref{fig:r4-efficiency-cost} and~\ref{fig:r4-efficiency-tokens} re-express the hidden checkpoints against recorded learning expense and total learning tokens. These views complement accuracy versus round: a revision can require many task evaluations, whereas observing a saved skill need not. They describe the realized policies, not a randomized experiment giving each policy the same dollar budget. Costs attributed to completed learning rounds can differ from ledger totals because retries and unresolved charges are not fully assigned to an individual round.

\begin{figure}[!tbp]
\centering
\includegraphics[width=\linewidth]{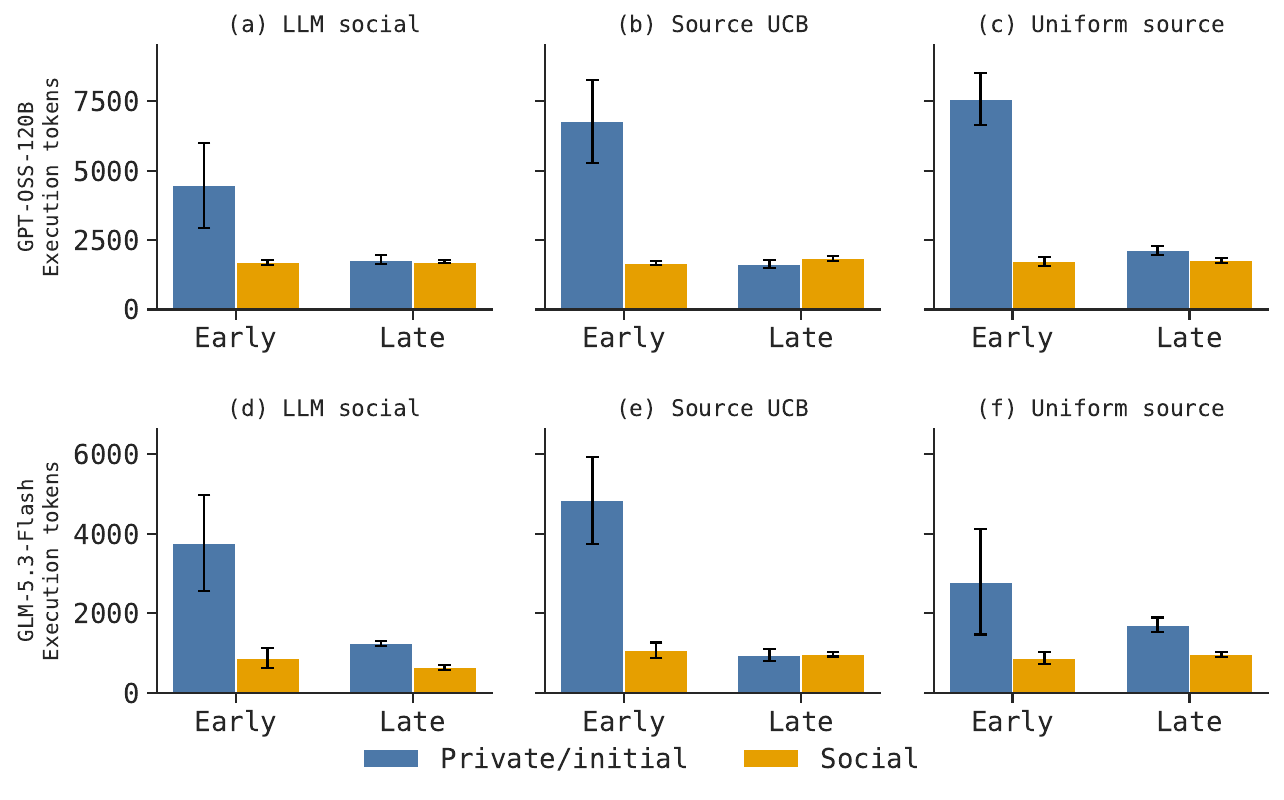}
\caption{Completion tokens used to execute the selected skill, by acquisition origin and learning period. Early is rounds 1--10; late is 11--49. Columns compare LLM social, source UCB, and uniform source selection; rows show GPT-OSS and GLM. Conditional means are computed within population before taking six-seed means and SE. All cells have six contributing populations; their numbers of deployments differ. These are execution costs, not tokens spent deciding to observe.}
\label{fig:r4-origin-execution}
\end{figure}

\textbf{Executing copied versus privately acquired skills.} We classify the deployed file by how the agent first acquired it, rather than by whether it observed in the current round. In early LLM-social rounds, GPT-OSS uses $1{,}694\pm104$ execution tokens with directly copied files versus $4{,}473\pm1{,}525$ with private or initial files; late in learning these means are $1{,}724\pm62$ and $1{,}787\pm163$ (Figure~\ref{fig:r4-origin-execution}a). GLM uses $872\pm253$ versus $3{,}764\pm1{,}205$ early, and $644\pm66$ versus $1{,}244\pm59$ late (Figure~\ref{fig:r4-origin-execution}d). UCB and uniform copying likewise associate with lower early execution costs, but their late patterns differ (Figure~\ref{fig:r4-origin-execution}b--c,e--f). These are conditional means, not effects of assigning a copied skill: file quality, task difficulty, and selection history may differ. The private category includes initial files and local revisions of copied files, so it does not imply an entirely private ancestry.

\begin{figure}[!tbp]
\centering
\includegraphics[width=\linewidth]{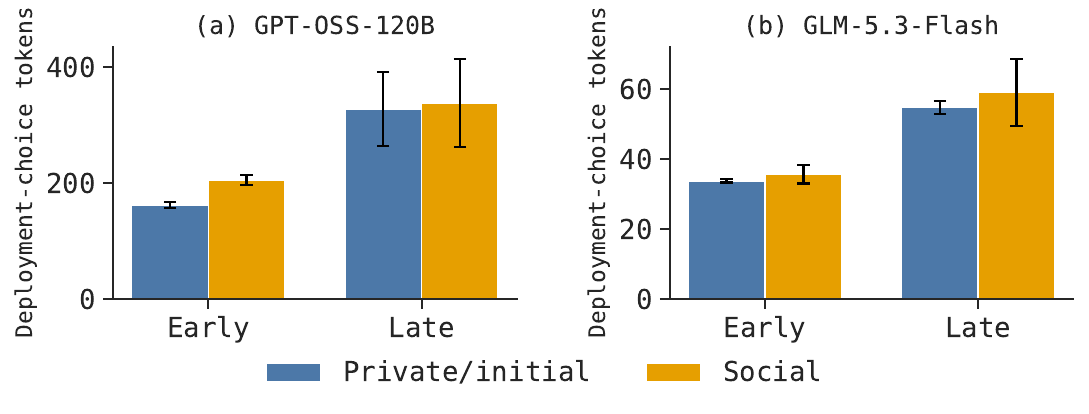}
\caption{LLM-social tokens spent choosing a skill to deploy, conditional on the selected file's acquisition origin. Panels (a)--(b) show GPT-OSS and GLM; periods and uncertainty match Figure~\ref{fig:r4-origin-execution}. This decision considers a library, so its tokens cannot be attributed solely to reading the selected file. UCB and uniform controllers choose deployment algorithmically and make no corresponding LLM decision call.}
\label{fig:r4-origin-deployment}
\end{figure}

\section{Who observes whom, and how much computation do those decisions use?}
\label{app:social-attention}

\begin{figure}[!tbp]
\centering
\includegraphics[width=.94\linewidth]{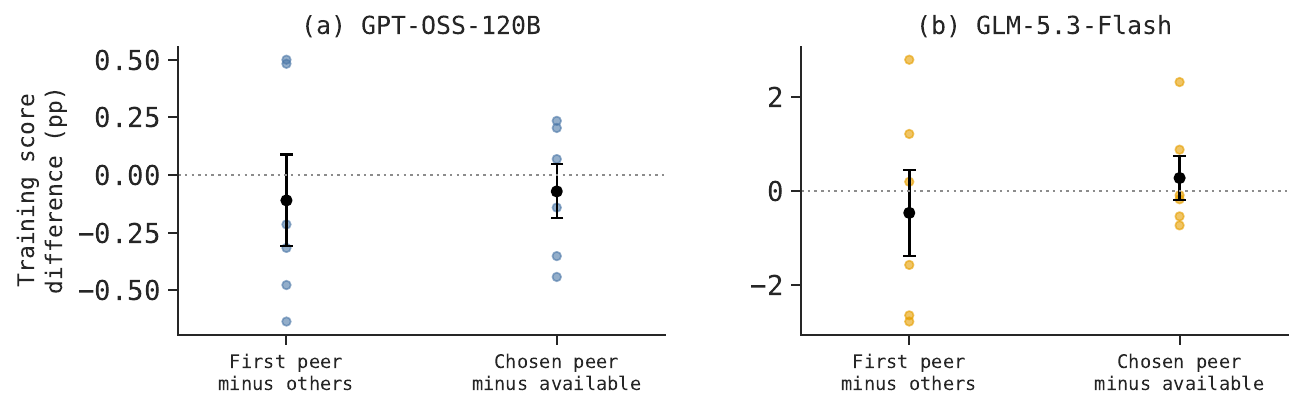}
\caption{Source quality at observed rounds in LLM-social populations. Panels (a)--(b) show GPT-OSS and GLM. First peer minus others compares the first eligible ID with the remaining peers; chosen minus available compares the selected peer with the mean across all four eligible peers. Small points are population-seed averages; black points show their mean and SE. These are retrospective diagnostics, not additional information revealed before source selection.}
\label{fig:r4-peer-quality}
\end{figure}

\textbf{Preferred list positions do not have an established score advantage.} At each realized observation, we compare the first eligible peer's previous training score with the mean of the other eligible peers. The within-population average difference is $-0.11\pm0.20$ percentage points for GPT-OSS and $-0.47\pm0.92$ for GLM (Figure~\ref{fig:r4-peer-quality}a--b). The chosen peer's score advantage over the available-peer mean is also small (Figure~\ref{fig:r4-peer-quality}a--b). These comparisons do not explain the concentration of attention by superior recorded scores, but neither identify a causal position bias: source choice, observation timing, and peer quality develop together. Shuffled-order runs would be needed to separate position from identity.

\begin{figure}[!tbp]
\centering
\includegraphics[width=.94\linewidth]{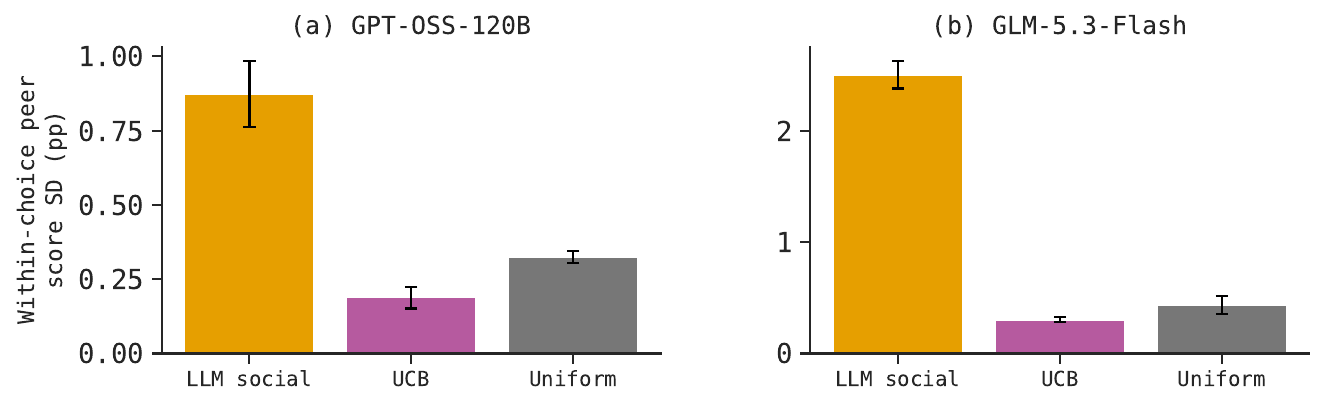}
\caption{Separation of peer training scores at realized observations. Panels (a)--(b) show GPT-OSS and GLM. For each observation, we compute the standard deviation across four available scores, then average observations within seed. Bars show seed means and SE. Different controllers generate different observation times and candidate pools.}
\label{fig:r4-peer-spread}
\end{figure}

\textbf{Algorithmic populations offer little score separation between peers.} The within-choice standard deviation of the four available training scores averages $0.19\pm0.04$ percentage points for GPT-OSS UCB and $0.30\pm0.02$ for GLM UCB, versus $0.87\pm0.11$ and $2.51\pm0.12$ in LLM-social populations (Figure~\ref{fig:r4-peer-spread}a--b). UCB's sources therefore often supply similar scores. This is consistent with copying spreading a small set of highly ranked files; it does not show that all undiscovered skills have similar value or establish why UCB and uniform have similar test accuracy. The score spread is an outcome of each policy, not a randomized environmental property.

\begin{figure}[!tbp]
\centering
\includegraphics[width=\linewidth]{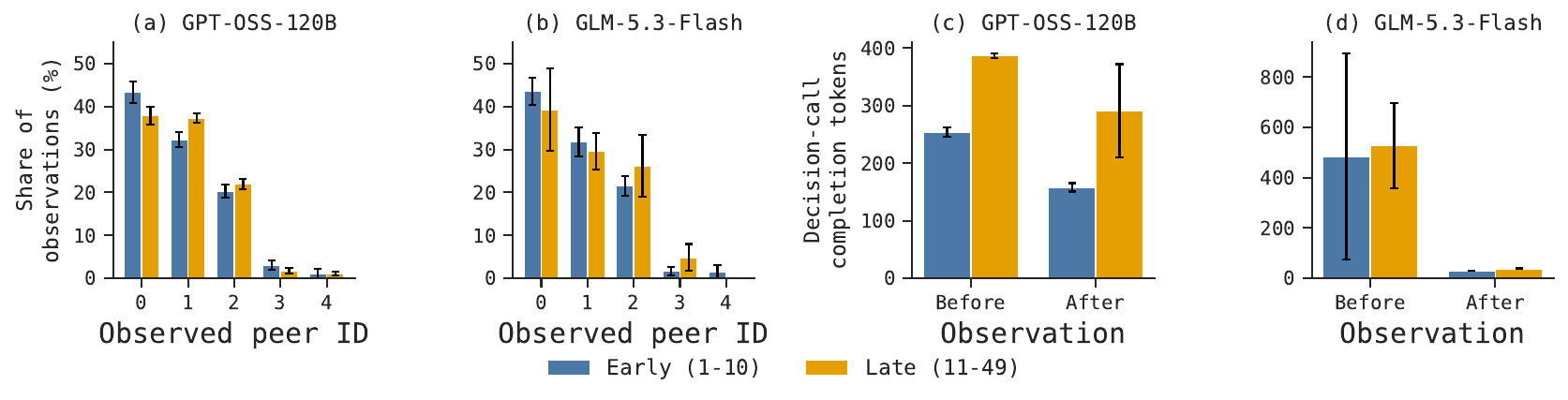}
\caption{Whom LLM agents observe and how much computation their decisions use. Panels (a)--(b) show each peer's share of observations in rounds 1--10 and 11--49. IDs follow the order of the available-peer list; they do not encode skill quality. Panels (c)--(d) show completion tokens conditional on observation: the learning decision before receiving the file and the deployment choice afterward. Bars show population means with seed SE; decision-cost panels have separate vertical scales. All calls, including unusually long ones, contribute to the means.}
\label{fig:r4-attention}
\end{figure}

\begin{figure}[!tbp]
\centering
\includegraphics[width=\linewidth]{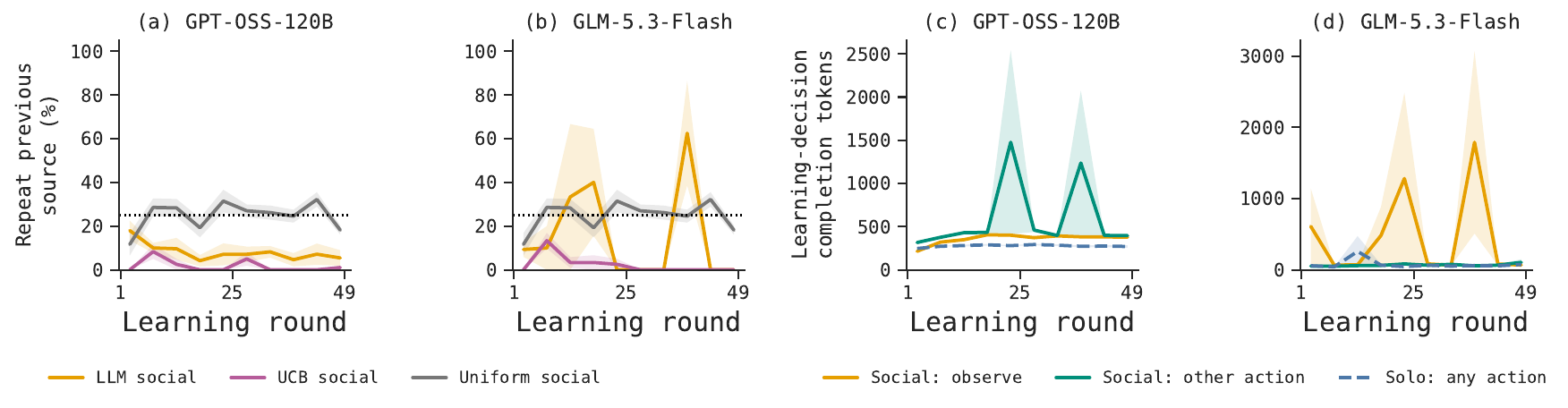}
\caption{Source repetition and learning-decision cost over time. Panels (a)--(b) condition on an observation with an earlier source to compare against; the dotted line is the uniform-repeat probability. Panels (c)--(d) compare learning-selection completion tokens for social observation, other social actions, and solo actions. Curves show means with SE across eligible population seeds in five-round blocks. Late GLM observations are sparse, so conditional estimates can use fewer seeds and should be read with the coverage saved in the accompanying data.}
\label{fig:r4-source-persistence}
\end{figure}

\begin{figure}[!tbp]
\centering
\includegraphics[width=\linewidth]{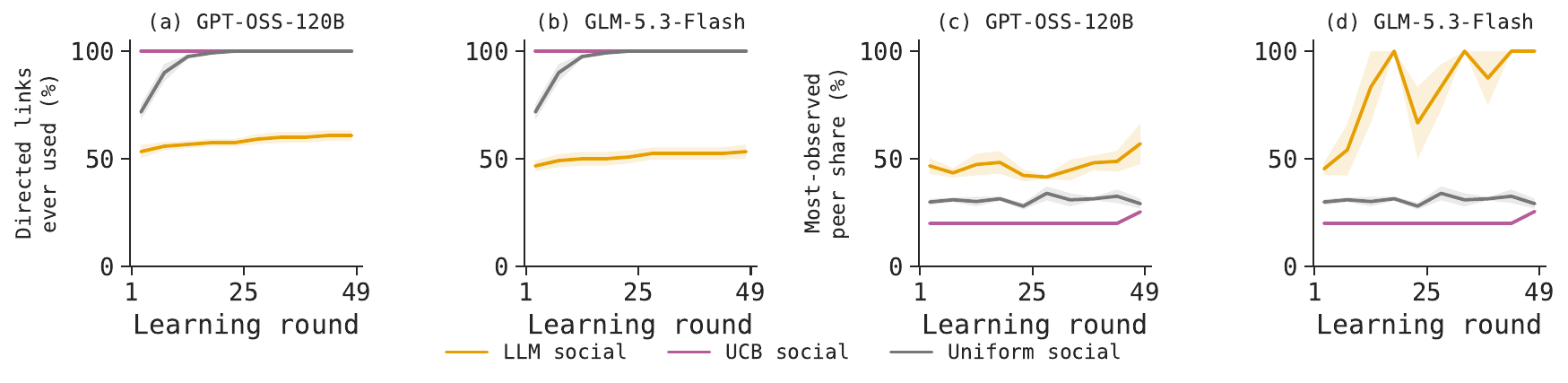}
\caption{Breadth and concentration of observation. Panels (a)--(b) show the cumulative fraction of possible directed links used; (c)--(d) show the largest incoming observation share within each five-round block. Curves show population-seed means with SE shading. Conditional concentration is undefined, rather than zero, in blocks without observations. Sparse observation can produce high concentration even under random choice, so the realized uniform baseline is more informative than a concentration threshold alone.}
\label{fig:r4-network-coverage}
\end{figure}

\begin{figure}[!tbp]
\centering
\includegraphics[width=\linewidth]{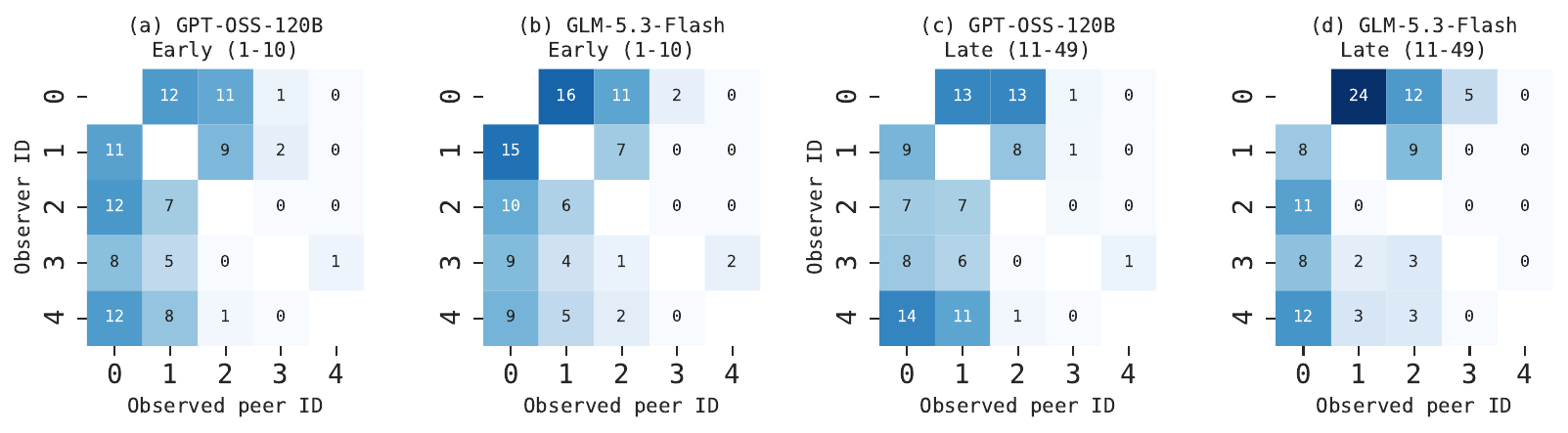}
\caption{Directed observation graphs for LLM-controlled social learning. Panels (a)--(b) show rounds 1--10; (c)--(d) show rounds 11--49. Each annotated cell is the mean percentage of all observations assigned to that observer--peer pair, with identical color scales across panels. Cells are descriptive edge weights, not independent samples. Late GLM graphs summarize substantially fewer observations; the observation-rate comparison is shown in Figure~\ref{fig:r4-early-late}.}
\label{fig:r4-observation-graphs}
\end{figure}

\textbf{Social attention concentrates on a few peers, without becoming consistently cheaper (RQ3).} Copying rates alone do not show whether an agent consults a broad range of peers or repeatedly draws from a small part of the population. Both LLMs direct much of their attention toward the first-listed peer IDs, and this preference is visible in both early and late learning (Figure~\ref{fig:r4-attention}a--b). The directed graphs show the individual relationships behind this concentration (Appendix Figure~\ref{fig:r4-observation-graphs}a--d). Concentration does not mean returning to the same peer on every observation: GPT-OSS usually changes sources, but does so within a restricted set of relationships (Appendix Figures~\ref{fig:r4-source-persistence}a and~\ref{fig:r4-network-coverage}a). Because peer IDs appear in a fixed order, we cannot attribute these preferences solely to learning which peers produce better skills.

\begin{figure}[!tbp]
\centering
\includegraphics[width=\linewidth]{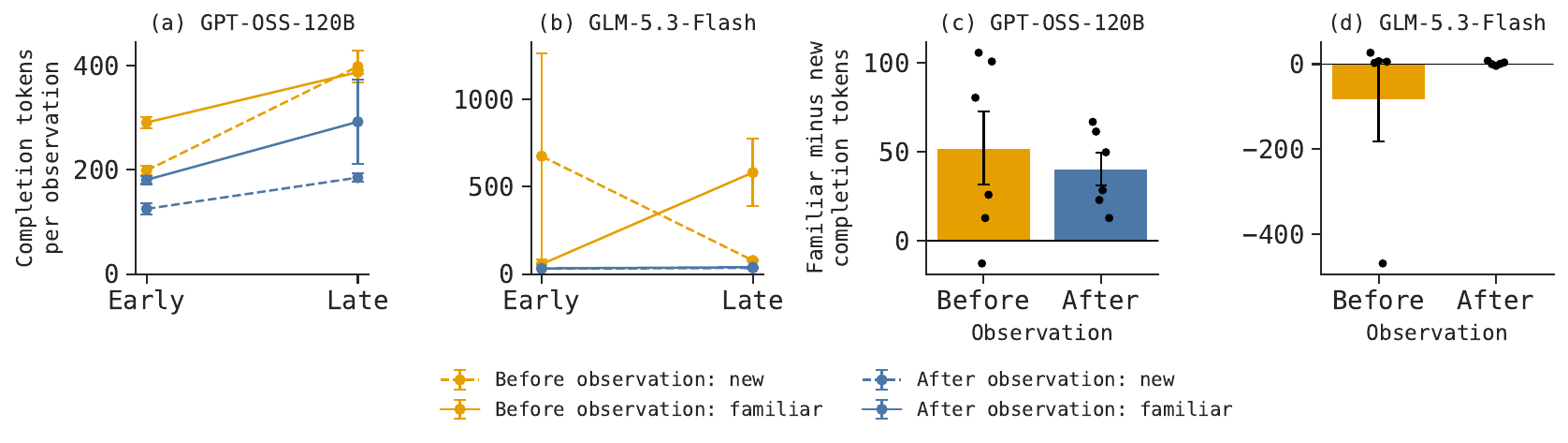}
\caption{Decision costs when observing familiar and new sources. Panels (a)--(b) show conditional means with seed SE, separately for learning selection before observation and deployment selection afterward; solid lines denote familiar sources and dashed lines new ones. Panels (c)--(d) show familiar-minus-new differences matched within observer and five-round block, averaged within seed. Black points are population estimates and bars show their mean with SE. No unobserved cell is imputed as zero.}
\label{fig:r4-familiarity}
\end{figure}

We can also ask whether observing becomes a cheaper decision as these relationships develop. GPT-OSS uses $254\pm8$ completion tokens to choose an observation in early learning and $387\pm4$ later, rather than spending less as it gains experience with peers (Figure~\ref{fig:r4-attention}c). GLM's mean is much more variable, so there is no similarly clear temporal pattern (Figure~\ref{fig:r4-attention}d). A comparison of familiar and new sources within the same observer and short time window likewise does not show a consistent token saving from familiarity (Appendix Figure~\ref{fig:r4-familiarity}c--d). These costs describe the learning decision before observation and the deployment choice afterward, not just reading the copied file. We therefore find structured social attention, but not evidence that relationships automatically make social reasoning more efficient.

Once a file is acquired, using it has a separate cost. Directly copied skills are associated with fewer execution tokens early in learning under all three social controllers, but this difference narrows or changes later (Appendix Figure~\ref{fig:r4-origin-execution}a--f). We also report deployment-choice tokens by selected-file origin (Appendix Figure~\ref{fig:r4-origin-deployment}a--b). Neither comparison isolates a causal effect of copying: agents choose which skills to acquire and use.

\label{app:observation-networks}

We reconstruct the observation graph from completed learning rounds in the corrected experiment. A directed edge from agent $i$ to agent $j$ means that $i$ observed $j$'s previously deployed skill. Repeating that observation increases the edge's weight; it does not necessarily add a new skill, because the peer may still be using the same file. Graphs describe actual observation, not permission to observe or later execution of the copied file. Initialization is excluded because no observation decision is made then.

\textbf{Persistence and breadth answer different questions.} The source-persistence statistic asks whether an observer returns to the same peer as on its immediately preceding observation, even if several rounds have passed (Figure~\ref{fig:r4-source-persistence}a--b). Each observer's first observation has no previous source and is excluded from this denominator. Uniform sampling from the other four agents repeats a previous source with probability one quarter. In Figure~\ref{fig:r4-network-coverage}a--b, we instead count how many of the 20 possible directed links have ever been used. Panels (c)--(d) ask what share of observations in a five-round block went to the most-observed peer. This peer can change from block to block. A population can explore many relationships over its lifetime while repeatedly consulting only a few of them at a particular stage.

Figure~\ref{fig:r4-observation-graphs} shows which identities receive attention in the LLM-controlled populations. Within each population and time interval, an edge's weight is its fraction of all observations; we then average these fractions across seeds. Rows are observers and columns are observed peers. The diagonal is unavailable because agents cannot observe themselves. Agent IDs reflect their positions in the experiment, not externally assigned reputations or fixed differences in ability. Shared patterns across seeds can therefore reflect index preferences or prompt order as well as responses to skill quality. The matrices describe behavior; they are not evidence that the most-observed peer is objectively the best teacher.

\textbf{Familiarity is not randomly assigned.} A source is familiar if the recipient has observed it before, whether or not it was the most recent source. Figure~\ref{fig:r4-familiarity}a--b separates completion tokens used before observation, when choosing the learning action and source, from those used afterward, when choosing a skill for deployment. It compares observations of familiar and previously unobserved sources in early and late learning. These are averages conditional on observation, not population-wide token costs. The before-observation call may consider private revision as well as copying, and the afterward call may consider any acquired skill; neither is a pure measure of effort spent reading the peer's file.

To reduce differences in who is deciding and when, Figure~\ref{fig:r4-familiarity}c--d compares familiar and new sources only within the same observer and five-round block, then averages those contrasts within population seed. Only blocks containing both types contribute: six GPT-OSS populations and five GLM populations meet this requirement (Figure~\ref{fig:r4-familiarity}c--d). This does not remove differences in candidate quality, prompt contents, or reasons for observing. In particular, later observations of new sources may be rare. We save the number of contributing blocks and populations with the estimates, rather than interpreting a small conditional sample as a randomized familiarity effect. Realized completion-token counts include whatever reasoning the provider bills as completion, but do not tell us whether the reasoning was useful.

\section{Randomized tests of copying time}
\label{app:timing-plan}

\begin{figure}[!tbp]
\centering
\includegraphics[width=\linewidth]{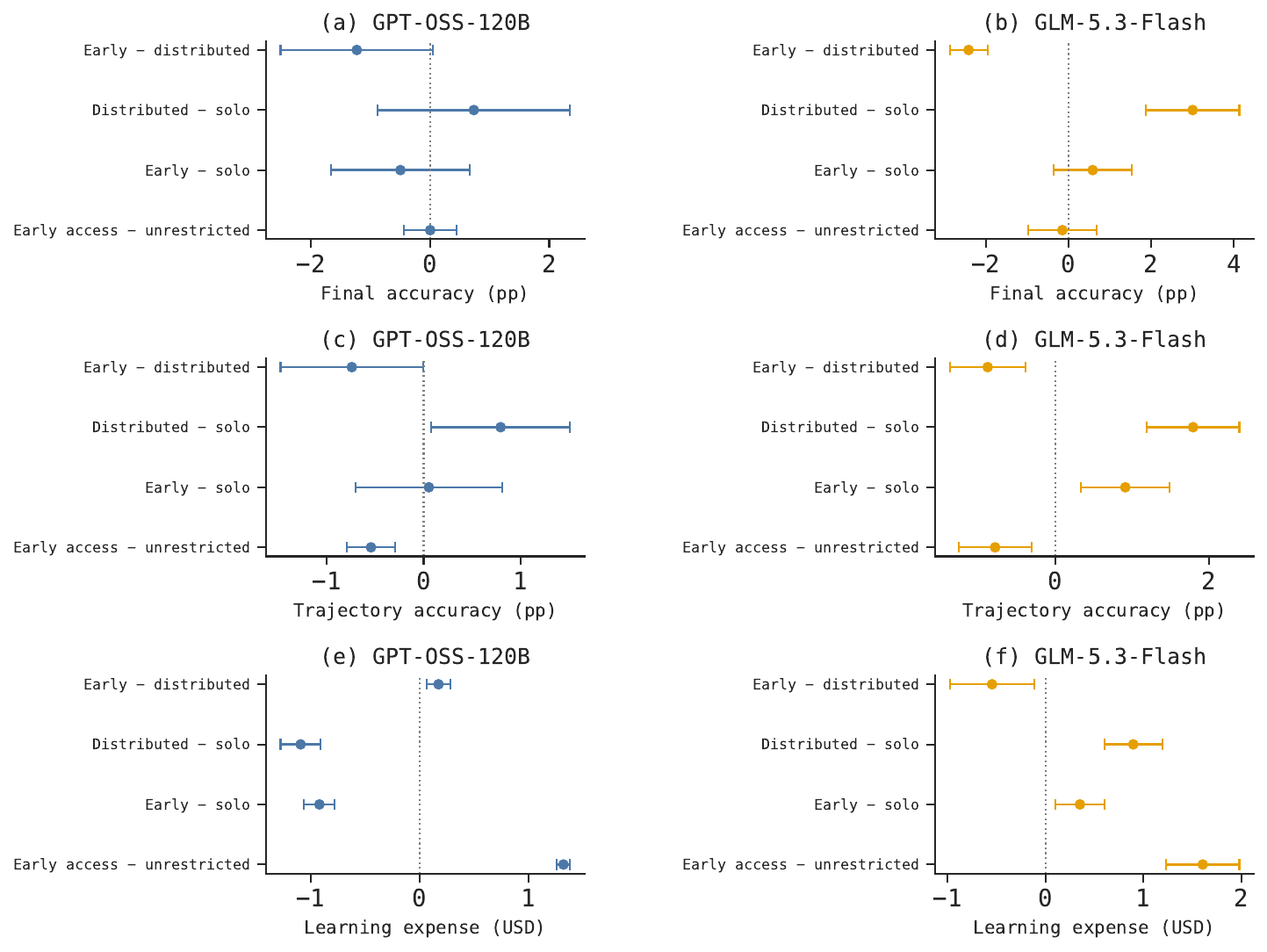}
\caption{Paired effects of copying schedules, with means and seed SE. Columns show GPT-OSS and GLM; rows show final accuracy, trajectory-average accuracy, and learning expense excluding hidden tests. Early and distributed both assign four observations; early access only permits optional observation in rounds 1--10. Solo and unrestricted controls are reused, not new independent replications.}
\label{fig:r4-timing-paired}
\end{figure}

Figure~\ref{fig:r4-timing-paired} reports the paired accuracy and expense contrasts behind the trajectory plots. Panels (a)--(b) compare final accuracy; (c)--(d) compare the trapezoidal trajectory average; (e)--(f) compare attributed learning expense. A timing policy can alter how much private search follows an observation, so fixing the observation count need not fix total expense. All contrasts average agents within population and pair the same six seed identifiers.

The descriptive results above raise a causal question: does GLM learn faster because it copies early? We test this in two ways, using six matched population seeds per condition and the same static IFBench protocol. In the assigned-timing experiment, every agent observes exactly four times. The early condition places all four observations at randomly assigned rounds within 1--10; the distributed condition places one in each quarter of learning. The model still chooses whom to observe and which acquired skill to deploy. This comparison changes timing while holding the observation count fixed. In the access-window experiment, observation is available only in rounds 1--10 but remains optional. We compare this condition with the reused unrestricted-social and solo populations. Restricting the window can change both when and how often agents copy, so it is a practical policy comparison rather than a timing-only effect.

\begin{figure}[!tbp]
\centering
\includegraphics[width=0.9\linewidth]{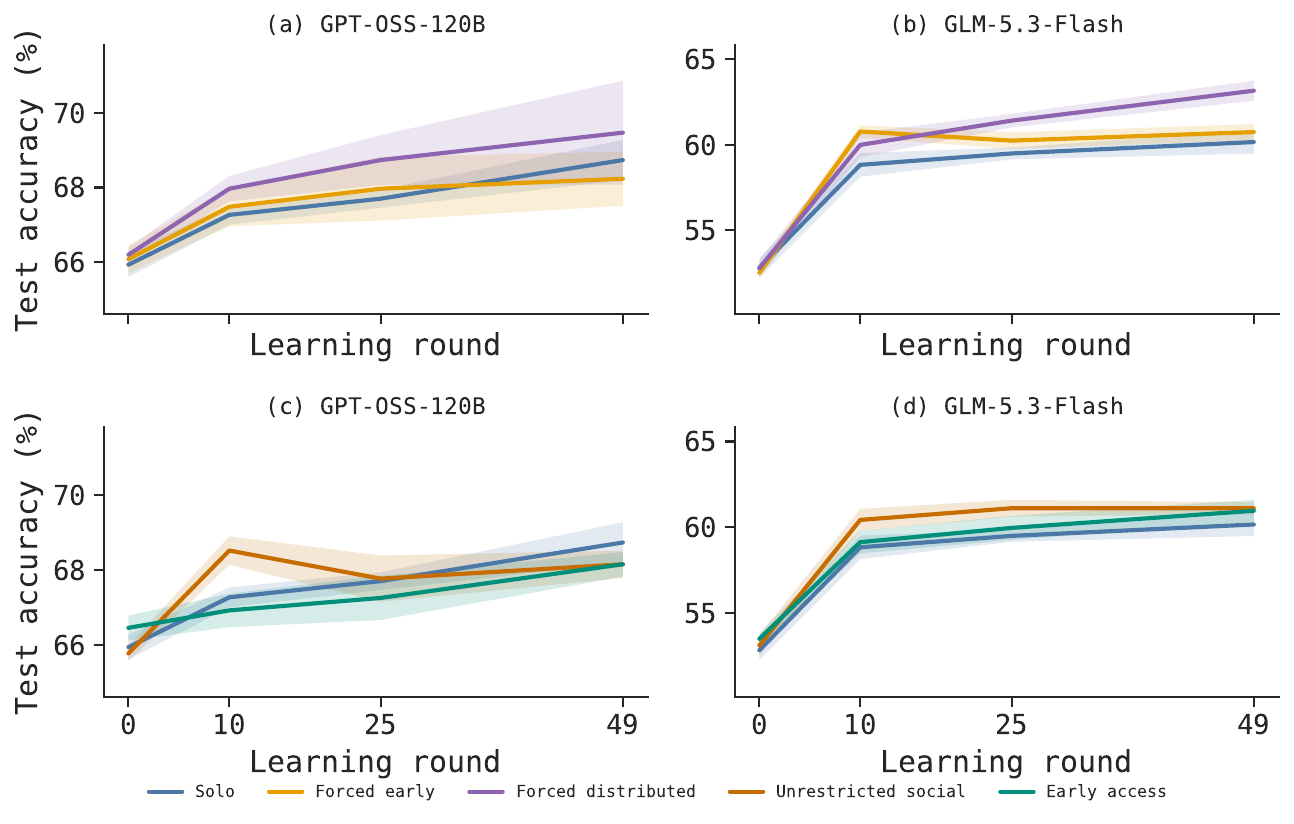}
\caption{Held-out whole-answer accuracy under randomized copying schedules. Panels (a)--(b) compare four forced observations in rounds 1--10 with four observations distributed across learning; solo is shown as a reference. Panels (c)--(d) compare optional access only in rounds 1--10 with unrestricted social access and solo learning. Lines show six-seed means and shaded SE. All comparisons use the common checkpoints at rounds 0, 10, 25, and 49; test feedback never enters learning.}
\label{fig:r4-copy-timing}
\end{figure}

\textbf{Concentrating observations early does not reproduce GLM's endogenous social advantage.} For GLM, forced early observation lowers final whole-answer accuracy by $2.42\pm0.46$ percentage points relative to the same four observations distributed across learning; its average accuracy over the common checkpoints is $0.88\pm0.50$ points lower (Figure~\ref{fig:r4-copy-timing}b). The trajectory difference remains negative after subtracting each population's starting accuracy ($-0.62\pm0.44$ points). Distributed observation finishes $3.00\pm1.13$ points above matched solo learning, whereas forced early observation finishes only $0.58\pm0.95$ points above solo. GPT-OSS has the same early-minus-distributed direction at the endpoint ($-1.23\pm1.28$ points), but the estimate is much less precise (Figure~\ref{fig:r4-copy-timing}a). Thus, the intervention provides no evidence that moving copies earlier is sufficient; for GLM it does the opposite.

\begin{figure}[!tbp]
\centering
\includegraphics[width=\linewidth]{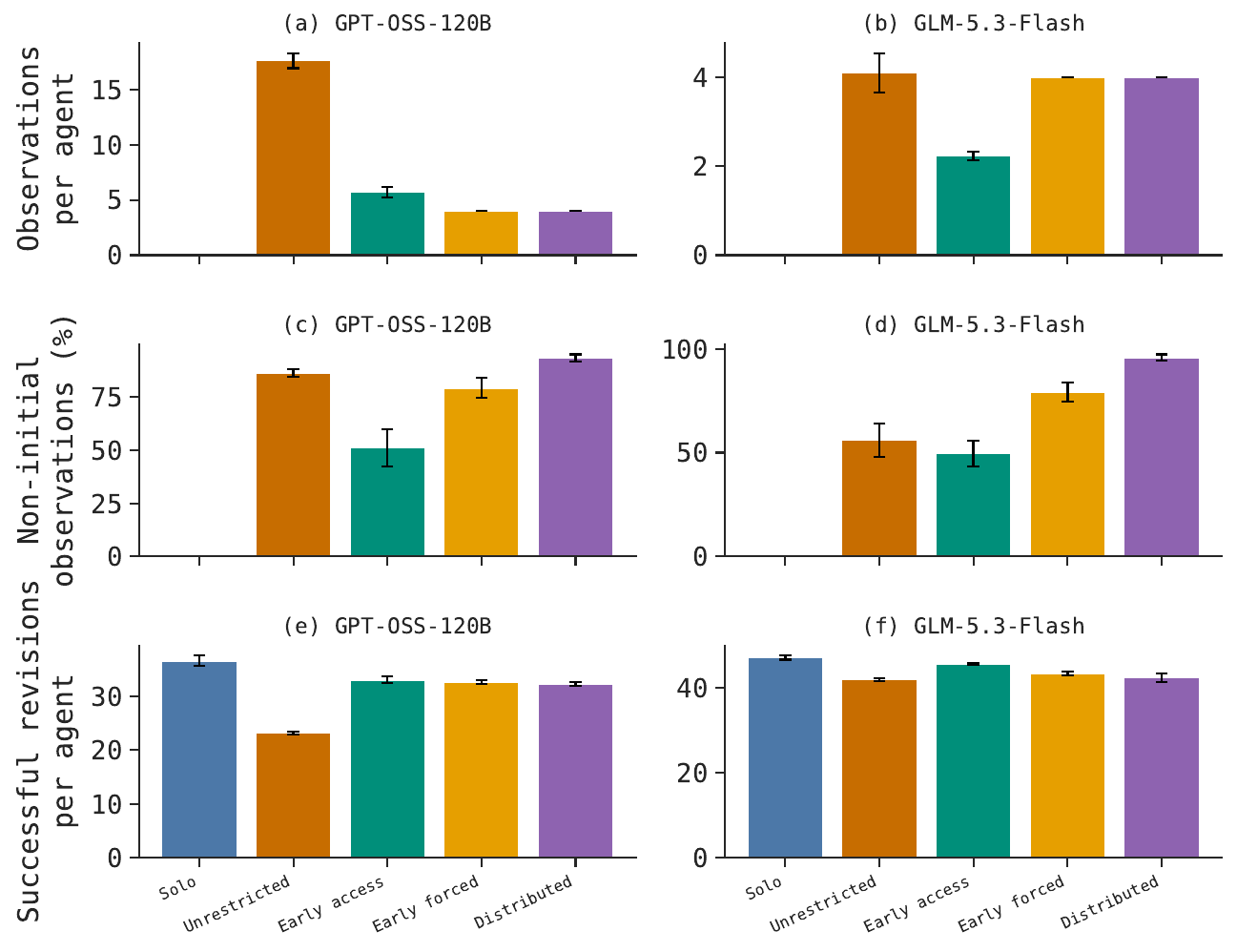}
\caption{How the timing interventions change learning behavior. Panels (a)--(b) show observations per agent; panels (c)--(d) show the share of observations that copy a non-initial skill version; panels (e)--(f) show successful private revisions. Each statistic is first computed within a five-agent population, then averaged over six seeds with SE. Forced conditions contain exactly four observations per agent by design. ``Early access'' is optional rounds 1--10; ``Unrestricted'' permits observation throughout learning.}
\label{fig:r4-copy-timing-behavior}
\end{figure}

\textbf{Early-only access reduces copying, but does not improve performance.} GPT-OSS makes $5.70\pm0.46$ observations per agent with early-only access, compared with $17.60\pm0.69$ under unrestricted access; GLM falls from $4.10\pm0.44$ to $2.23\pm0.10$ (Figure~\ref{fig:r4-copy-timing-behavior}a--b). The displaced observations are largely replaced by private revisions (Figure~\ref{fig:r4-copy-timing-behavior}e--f). Nevertheless, early-only access has lower average trajectory accuracy than unrestricted access by $0.54\pm0.25$ points for GPT-OSS and $0.79\pm0.47$ for GLM, while their final differences are $0.00\pm0.44$ and $-0.15\pm0.83$ points (Figure~\ref{fig:r4-copy-timing}c--d). This rules out the simple claim that limiting both models to an early burst of social learning improves the balance between copying and innovation.

These experiments reverse the most tempting interpretation of the observational timing pattern. GLM's early endogenous copying may mark selective behavior---copying when a useful opportunity appears---rather than a generally superior early schedule. Under forced early observation, $79.17\pm4.73\%$ of copied versions differ from the common initial skill for both models. When observations are distributed, that share rises to $93.33\pm1.67\%$ for GPT-OSS and $95.83\pm1.54\%$ for GLM (Figure~\ref{fig:r4-copy-timing-behavior}c--d). This is consistent with assigned early observations arriving before peers have produced differentiated skills, while distributed observations allow later discoveries to enter the population. It is not a separate randomization of artifact quality: timing also changes the population's subsequent search and the skills available later. The experiment therefore estimates the practical effect of the schedule rather than the value of copying a fixed artifact at two different times, and does not imply that one schedule will be optimal for other tasks or model families.

\section{Interpretation and reporting safeguards}
\label{app:reporting}

When we describe a difference as statistically significant, we use a two-sided paired $t$-test across matched population seeds and report $p<0.001$, $p<0.01$, or $p<0.05$. The ten finite-bandit comparisons in Table~\ref{tab:r1-key} use Bonferroni correction across the full table; the three primary matched-token contrasts in Figure~\ref{fig:r1-matched-tokens} form a separate Bonferroni-corrected family. Other reported tests, including the observation-timing contrasts, are nominal and unadjusted unless stated otherwise. Each agent is averaged within its population before testing. Seed-based tests condition on the fixed evaluation tasks, with test-item uncertainty examined separately in Appendix~\ref{app:test-sampling}. A nonsignificant difference does not establish equivalence. Effect sizes and SE remain the primary summaries. The cumulative-reward percentages in Section~\ref{sec:framework} compare the seed-averaged social and solo rewards, whereas efficiency percentages average the relative change within each matched seed.

\begin{table*}[!b]
\centering
\small
\setlength{\tabcolsep}{5pt}
\begin{tabular}{llrr}
\toprule
Comparison & Outcome & Effect & Significance \\
\midrule
Hierarchical vs. solo UCB & Cumulative reward & $+25.68\pm2.87$ & $p<0.001$ \\
Qwen social vs. solo & Cumulative reward & $+2.8\pm20.3$ & n.s. \\
GPT-OSS social vs. solo & Cumulative reward & $-61.6\pm39.2$ & n.s. \\
Ministral social vs. solo & Cumulative reward & $+5.0\pm27.1$ & n.s. \\
Qwen social vs. solo & Efficiency relative to solo & $-29.6\pm13.4\%$ & n.s. \\
GPT-OSS social vs. solo & Efficiency relative to solo & $-51.9\pm7.8\%$ & $p<0.01$ \\
Ministral social vs. solo & Efficiency relative to solo & $-41.8\pm11.6\%$ & n.s. \\
Qwen PE vs. default & Pull completion & $+19.53\pm2.24$ pp & $p<0.001$ \\
Qwen IS vs. default & Selected-arm diversity & $+1.01\pm0.34$ & n.s. \\
GPT-OSS IS vs. default & Selected-arm diversity & $+1.27\pm0.23$ & $p<0.01$ \\
\bottomrule
\end{tabular}
\caption{Key paired results in the finite-option environment. Effects are treatment minus comparison, reported as mean $\pm$ SE across population seeds. Reward is cumulative through round 100. Efficiency is the within-seed percentage change in $E(\boldsymbol\pi)$ (Equation~\ref{eq:efficiency}) relative to solo, estimated using realized rewards and charged completion tokens. Intervention effects compare treated and default social populations. Significance uses two-sided paired $t$-tests with Bonferroni correction across all ten comparisons; n.s. denotes adjusted $p\geq0.05$.}
\label{tab:r1-key}
\end{table*}

For the finite-bandit social-minus-solo contrasts, we also report an approximate retrospective minimum detectable effect at 80\% power and two-sided $\alpha=0.05$. We multiply the observed paired SE by $t_{0.975,7}+z_{0.8}\approx3.21$. This calculation describes the scale of effects the eight-seed design could reliably detect under its observed variability; it is not an equivalence test and does not turn a nonsignificant result into evidence of no effect.

Every copied-option comparison conditions on the recipient having a personal alternative, and its coverage is reported. Hidden evaluation is performed only after training and never feeds back into skill evolution. Candidate selection uses training evidence, not hidden outcomes. OpenEvolve's islands migrate only within an agent's private archive; all cross-agent transfer must pass through the tested social mechanism. The uniform source control changes both how often the agent chooses itself and which peer it sees. It therefore tests adaptive source allocation as a package. A future fixed-self/social-schedule replay would be needed to isolate peer identity alone.

\begin{figure}[!tbp]
  \centering
\includegraphics[width=.94\linewidth]{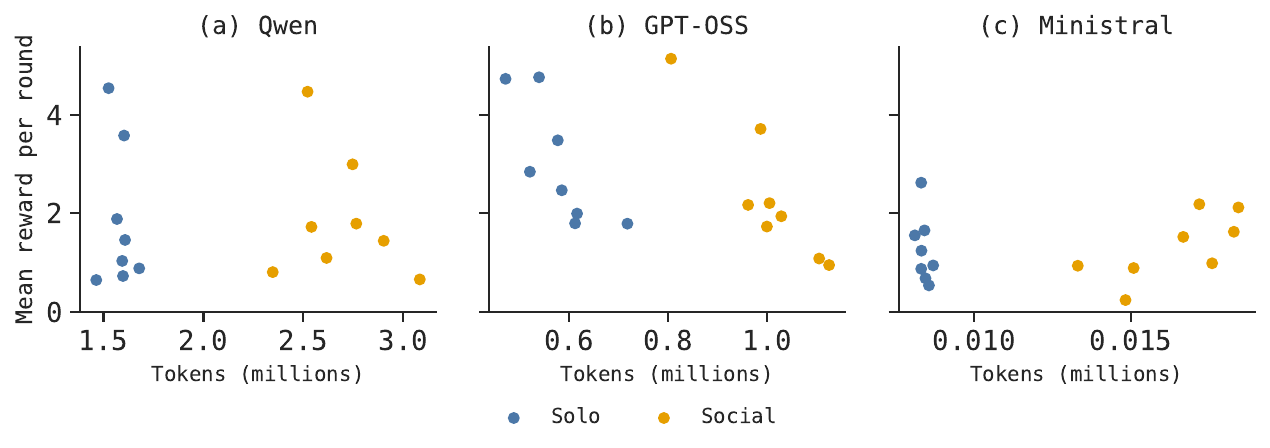}
  \caption{Finite-bandit mean reward against completion tokens in the primary stationary, neutral-prompt, expiring-budget comparison. Panels (a)--(c) show Qwen, GPT-OSS, and Ministral. Each point is one population seed. These are realized outcomes, not effects of assigning a token budget.}
  \label{fig:r1-token-efficiency}
\end{figure}

Compute is part of the behavior. The skill-evolution study logs observation events and records tokens and calls separately for learning-action selection, deployment choice, revision, candidate evaluation, task execution, and the final hidden audit. These records distinguish the decision preceding an observation from the deployment decision afterward; they do not isolate the tokens spent interpreting the copied file from all other reasoning in that decision. Hidden-audit expense is excluded from the agent's learning budget. We report both absolute spend and performance against cumulative learning tokens. A controller that saves language-model decision calls is allowed to use fewer tokens; we record that efficiency rather than adding dummy calls. For the completed finite-bandit sweep, Figure~\ref{fig:r1-token-efficiency} plots reward against charged completion tokens at the population-seed level. It uses only the primary stationary, neutral-prompt, expiring-budget comparison and describes realized expense rather than a causal return to extra tokens.

\end{document}